\documentclass[a4paper,oneside,12pt]{article}

\usepackage{epsfig,array,amsmath,amssymb,psfrag,graphicx,tensor,color,mathrsfs}
\usepackage{amsthm}

\definecolor{blue2}{RGB}{0, 30, 255}
\definecolor{dblue}{RGB}{0, 0, 150}

\usepackage[colorlinks=true, linkcolor=blue2, citecolor=blue2, urlcolor=dblue]{hyperref}

\newcommand{\noi}{\noindent}
\newcommand{\beq}{\begin{equation}}
\newcommand{\eeq}{\end{equation}}
\newcommand{\bea}{\begin{eqnarray}}
\newcommand{\eea}{\end{eqnarray}}

\newcommand{\mc}[1]{\mathcal{#1}}

\newcommand{\mbb}[1]{\mathbb{#1}}
\newcommand{\mr}[1]{\mathrm{#1}}

\newcommand{\Tc}{\hat{T}}
\newcommand{\Jc}{\hat{J}}
\newcommand{\Sc}{\hat{S}}
\newcommand{\bbTc}{\hat{\mbb{T}}}
\newcommand{\bbCc}{\hat{\mbb{C}}}
\newcommand{\Mc}{\hat{M}}
\newcommand{\Dc}{\hat{D}}
\newcommand{\Cc}{\hat{C}}

\newcommand{\JN}{\mathscr{J}}

\newtheorem{thrm}{Theorem}
\newtheorem{prop}{Proposition}[section]

\newtheorem{defin}{Definition}

\makeatletter \@addtoreset{equation}{section} \makeatother 
\renewcommand{\theequation}{\arabic{section}.\arabic{equation}}

\begin{document}

\title{\vspace{-1cm} \bf Conserved currents associated with conformal symmetry
in linearized gravity}

\author{G\'abor Zsolt T\'oth\\[4mm]
\small \textit{HUN-REN Wigner Research Centre for Physics,} \\[-1mm]
\small \textit{Konkoly Thege Mikl\'os \'ut 29-33, 1121 Budapest, Hungary} \\
\small \texttt{toth.gabor.zsolt@wigner.hun-ren.hu}}
\date{}
\maketitle

\begin{abstract}
Extending a previous work on the energy-momentum tensor of the
linearized gravitational field
in Minkowski background, conserved currents in linearized gravity
associated with Lorentz transformations,
dilatations and special conformal transformations are explored
in the framework of Lagrangian field theory.
Both outstanding special cases and general parametric families
of currents are considered.
Regarding the former,
the distinguished energy-momentum tensor $T_{\mathrm{lg}}^{ab}$
we found earlier,
guided by similarities between linearized gravity and electrodynamics,
is supplemented with an angular momentum tensor $M_{\mathrm{lg}}^{abc}$
and a dilatation current $D_{\mathrm{lg}}^a$,
associated with Lorentz and dilatation symmetries.
Concerning special conformal transformations,
a generalization of Noether's theorem that is useful
if additional conditions
(such as gauge fixing conditions) are imposed is formulated,
it is shown that special conformal transformations are
Noether symmetries of linearized gravity
in the generalized harmonic gauges,
and a special conformal tensor $C_{\mathrm{lg}}^{ab}$,
accompanying $T_{\mathrm{lg}}^{ab}$,
is obtained using the generalized Noether's theorem.
The angular momentum tensor, dilatation current
and special conformal tensor accompanying two other notable
energy-momentum tensors are discussed as well.
General multi-parameter families of $M^{abc}$, $D^a$ and $C^{ab}$
tensors accompanying a general family of energy-momentum tensors,
which was found earlier and
includes canonical energy-momentum tensors,
are also determined,
and local balance equations that are relevant
if matter is present are derived for them.
These multi-parameter families contain
$M_{\mathrm{lg}}^{abc}$, $D_{\mathrm{lg}}^a$ and $C_{\mathrm{lg}}^{ab}$
as well as the special
$M^{abc}$, $D^a$ and $C^{ab}$ tensors
that accompany the other notable energy-momentum tensors mentioned above.
Their members
do not depend on higher than first derivatives of the linearized metric.
\end{abstract}

\noi

\thispagestyle{empty}

\newpage



\section{Introduction}
\label{sec.intr}

Linearized gravity in Minkowski background is
similar to electrodynamics in several respects---in particular,
both theories have abelian gauge symmetry,
and the linearized gravitational field
and the electromagnetic field are both massless.
It is well known
that electrodynamics has conformal symmetry in the absence of sources
\cite{Bateman1, Bateman2, Cunningham, BesselHagen, Kastrup, AP1},
therefore it is natural to ask
whether linearized gravity also has conformal symmetry,
and if it does, what the corresponding conservation laws are.
The main aim of this paper is to address these fundamental questions
in the framework of classical Lagrangian field theory.
As the conformal group contains the Poincar\'e group,
the exploration of the conservation laws
corresponding to Poincar\'e symmetry
is also included in the investigation.

We have already discussed translation symmetry
and the energy-momentum tensor of the linearized gravitational field in \cite{TG},
therefore in the present paper we focus on the Lorentz,
dilatation and special conformal symmetries.
In \cite{TG}, the principal aim was to find an energy-momentum tensor
for the linearized gravitational field that is
similar to the usual energy-momentum tensor of the electromagnetic field
(which will be denoted by $T_{\mathrm{em}}^{ab}$; see (\ref{eq.Tem})),
in view of the numerous favourable properties of the latter.

The main result in \cite{TG}
was the identification of the energy-momentum tensor
\begin{equation}
\label{eq.Tlg}
T_{\mathrm{lg}}^{ab} =  2\left(F^{acd}\mathring{F}^b{}_{cd}
- \frac{1}{4} \eta^{ab} F^{ecd}\mathring{F}_{ecd}\right),
\end{equation}
where $\eta_{ab}$ denotes the Minkowski metric,
$F_{abc}$ is the \emph{Fierz tensor} (see (\ref{eq.lg.x6})
and \cite{Fierz, FP2, NN, NN2}),
and $\mathring{F}_{abc}$ is a closely related tensor given by equation (\ref{eq.circx}).
$F_{abc}$ and $\mathring{F}_{abc}$ are constructed from
the first derivatives of the linearized metric $h_{ab}$
and are antisymmetric in the first two indices.
They are partially gauge invariant.
They can be regarded as field strength tensors
in linearized gravity that are similar, to some extent,
to the electromagnetic field strength tensor.
In finding $T_{\mathrm{lg}}^{ab}$, a formulation of linearized gravity
based on the Fierz tensor had a key role.
(This formulation is also important for the present paper,
therefore it is briefly reviewed in Section \ref{sec.lingrav}.
A more detailed account can be found in \cite{TG}.)

The properties of $T_{\mathrm{lg}}^{ab}$ have been discussed in \cite{TG};
here we recall that
\begin{itemize}
\setlength\itemsep{0em}
\item[(a)] $T_{\mathrm{lg}}^{ab}$ is traceless,
\item[(b)] it is symmetric in the gauge $F_{ab}{}^b=0$, which is a particular generalized harmonic gauge
and contains the transverse traceless (TT) gauge,
\item[(c)] it satisfies the dominant energy condition (d.e.c.)
in the $F_{ab}{}^b=0$, $F_{ab0}=0$ gauge,
which contains the TT gauge,
\item[(d)] it is associated by Noether's theorem\footnote{Here we refer
to a variant of Noether's theorem;
see Section \ref{sec.nthr} or \cite{TG} for further details.}
with spacetime translations
combined with certain field-dependent gauge transformations,
\item[(e)] it differs from the canonical energy-momentum tensor following from
the Lagrangian density $L_1$ of linearized gravity (see (\ref{eq.lg.x7}))
by a trivially conserved tensor\footnote{The term `trivially conserved'
and further related terms
are explained in the last part of this section.} related to
the gauge symmetry of linearized gravity,
\item[(f)] it is invariant under those gauge transformations
whose parameter takes the form $\phi_a=\partial_a\phi$,
where $\phi$ is an arbitrary scalar function (we refer to these
gauge transformations as `scalar' gauge transformations;
the scalar gauge invariance of $T_{\mathrm{lg}}^{ab}$
follows from the fact that it can be expressed in terms of
$F_{abc}$ and $\mathring{F}_{abc}$,
which are scalar gauge invariant),
\item[(g)] it changes by a trivially conserved tensor under
general gauge transformations,
\item[(h)] it gives the same total energy and momentum
as other known energy-momentum tensors
(assuming that their normalization is properly adjusted),
if $h_{ab}$ and $\partial_ch_{ab}$ fall off
sufficiently rapidly at spatial infinity,
\item[(i)] it is invariant with respect to the duality transformation
$F_{abc}\to\tilde{F}_{abc}$ if $F_{ab}{}^b=0$
(the tilde denotes the Hodge dual on the first two indices).
\end{itemize}
Of these properties, the positivity (i.e., property (c))
was the most interesting for us.

$T_{\mathrm{em}}^{ab}$ is known to have the same properties,
with the difference that some of them are valid in the case of $T_{\mathrm{em}}^{ab}$
in more perfect form: instead of having the property (g),
$T_{\mathrm{em}}^{ab}$ is fully gauge-invariant,
(f) is also replaced by full gauge-invariance,
and (b), (c) and (i) hold for $T_{\mathrm{em}}^{ab}$ without any condition.
The form of expression (\ref{eq.Tlg}) for $T_{\mathrm{lg}}^{ab}$
is also similar to that of the usual expression
for $T_{\mathrm{em}}^{ab}$ (see (\ref{eq.Tem})),
although (\ref{eq.Tlg}) is less symmetric
(nevertheless, the asymmetry of $T_{\mathrm{lg}}^{ab}$ disappears in
the $F_{ab}{}^b=0$ gauge; see (\ref{eq.Tlg-sym})).
The counterpart of the Lagrangian density $L_1$ mentioned in property (e)
is the standard Lagrangian density of electrodynamics (see (\ref{eq.eml2})).
Although $L_1$ is not strictly gauge invariant,
it is strictly scalar gauge invariant
(as it can be expressed in terms of $F_{abc}$ and $\mathring{F}_{abc}$).

The lack of complete gauge-invariance of $T_{\mathrm{lg}}^{ab}$ is
in agreement with known results showing that
the linearized gravitational field
does not have a gauge-invariant energy-momentum tensor \cite{DesMcC,DesHen,MagSok,HoBaLa}.
Although gauge fixing is important for several of
the favourable properties of $T_{\mathrm{lg}}^{ab}$,
the gauges that are well suited to $T_{\mathrm{lg}}^{ab}$ in this respect
(the TT gauge, or the somewhat more general gauge
$F_{ab}{}^b=0$, $F_{ab0}=0$)
are not strange or inconvenient.

Overall, the properties of $T_{\mathrm{lg}}^{ab}$ are
less perfect than those of $T_{\mathrm{em}}^{ab}$,
but are still quite favourable,
and allow $T_{\mathrm{lg}}^{ab}$ to be regarded as a proper
counterpart of $T_{\mathrm{em}}^{ab}$.
It is not obvious whether linearized gravity has a single most appropriate
energy-momentum tensor,
nevertheless, $T_{\mathrm{lg}}^{ab}$ appears to be among the preferable ones.

It should be noted that
another remarkable energy-momentum tensor (denoted by $\tau^{ab}$)
was found in \cite{BHLemt,BHLthe} by Butcher, Hobson and Lasenby;
it is symmetric in harmonic gauge and
satisfies the dominant energy condition in TT gauge,
but it is not traceless even in the latter gauge.
As is mentioned in Section 6.4 of \cite{TG},
$\tau^{ab}$ also has a counterpart in electrodynamics,
which is different from $T_{\mathrm{em}}^{ab}$.
The linearized Landau--Lifshitz pseudotensor
$T_{\mathrm{LL}}^{ab}$ is also worth mentioning;
it is symmetric without any gauge fixing,
but it is also not traceless in TT gauge.

In addition to finding $T_{\mathrm{lg}}^{ab}$,
general families of energy-momentum tensors
satisfying only minimal requirements were also discussed in \cite{TG}.
In particular, a $3$-parameter\footnote{In contrast with \cite{Bicak,BS,TG},
in the present paper the overall scale of the energy-momentum tensor is not counted
as a parameter.}
family of energy-momentum tensors obtained previously
in \cite{Bicak,BS} by Bi\v{c}\'ak and Schmidt
was examined and extended into a $5$-parameter family
(the latter family arises as a result of allowing terms that contain
the completely antisymmetric tensor $\epsilon^{abcd}$).
A common feature of the energy-momentum tensors in these families
is that they do not depend on higher than first derivatives of $h_{ab}$.
The $3$-parameter family contains only one traceless tensor,
which is $T_{\mathrm{lg}}^{ab}$,
and one symmetric tensor, which, up to a normalization factor, is $T_{\mathrm{LL}}^{ab}$.
We mention that the results of \cite{Bicak,BS} were extended
in \cite{Baker} by Baker,
allowing terms that contain second derivatives
(more specifically, terms of the type $h\partial\partial h$).
Further developments, constituting the first part of a more complete investigation,
can be found in \cite{BaTa}.

Turning to the subject of the present paper,
namely the local conservation laws associated with
Lorentz transformations, dilatations and special conformal transformations,
we first note that
the angular momentum tensor $M^{abc}$,
the dilatation current $D^a$ and the special conformal tensor $C^{ab}$
(associated with the aforementioned transformations)
can be split into `orbital' and `intrinsic' parts\footnote{The intrinsic part of $M^{abc}$ is also called \textit{spin tensor}.},
which we denote by $M_{\mathrm{orb}}^{abc}$, $D_{\mathrm{orb}}^a$, $C_{\mathrm{orb}}^{ab}$
and $M_{\mathrm{int}}^{abc}$, $D_{\mathrm{int}}^a$, $C_{\mathrm{int}}^{ab}$
(see Sections \ref{sec.eads}, \ref{sec.orbint}, \ref{sec.cft} for more details).
The orbital parts are completely determined by the energy-momentum tensor:\footnote{Concerning the normalization of $M_{\mathrm{orb}}^{abc}$
relative to $T^{ab}$, see Sections \ref{sec.eads} and \ref{sec.orbint}.}
\begin{equation}
\label{eq.1.MDorb}
M_{\mathrm{orb}}^{abc} = \frac{1}{2}(T^{ab}x^c - T^{ac}x^b),\qquad
D_{\mathrm{orb}}^a = T^{ab}x_b\, ,
\end{equation}
\begin{equation}
\label{eq.1.Corb}
C_{\mathrm{orb}}^{ab} = 2T^{ac}x_c x^b - T^{ab}x_c x^c.
\end{equation}
It can be said, therefore, that after an energy-momentum tensor has been chosen,
the problem is to find the intrinsic quantities that accompany it.

$C_{\mathrm{int}}^{ab}$ can be further split into an `extrinsic'
and a `fully intrinsic' part
(see Section \ref{sec.orbint}).
The `extrinsic' part is determined by $M_{\mathrm{int}}^{abc}$ and $D_{\mathrm{int}}^a$,
therefore it can also be said that generally
it is the energy-momentum tensor,
the intrinsic parts of $M^{abc}$ and $D^a$,
and the `fully intrinsic' part of $C^{ab}$ that one aims to determine.

The orbital and intrinsic parts of $M^{abc}$, $D^a$ and $C^{ab}$
could be disregarded
or other tensors instead of the designated energy-momentum tensor
could be allowed to appear in (\ref{eq.1.MDorb}) and (\ref{eq.1.Corb}),
but from the physical point of view it is natural
to affirm the significance of the orbital and intrinsic parts,
and to accept (\ref{eq.1.MDorb}) and (\ref{eq.1.Corb})
as valid relations between
$M_{\mathrm{orb}}^{abc}$, $D_{\mathrm{orb}}^a$, $C_{\mathrm{orb}}^{ab}$
and the energy-momentum tensor.

It is known that if $T^{ab}$ is symmetric,
then $M_{\mathrm{orb}}^{abc}$ is conserved, if $T^{ab}$ is traceless,
then $D_{\mathrm{orb}}^a$ is conserved,
and if $T^{ab}$ is both symmetric and traceless,
then $M_{\mathrm{orb}}^{abc}$, $D_{\mathrm{orb}}^a$
and $C_{\mathrm{orb}}^{ab}$ are conserved.
Although the respective intrinsic quantities
may be zero in these cases, this is not necessary;
they may still need to be nonzero,
as is also emphasized and explained
in \cite{BHLang} in relation to the spin tensor.

Dilatations are known to be
Noether symmetries of linearized gravity\footnote{In other words,
the action of linearized gravity has dilatation symmetry.} \cite{FarHinHul},
as are Lorentz transformations,
therefore Noether's theorem can be applied
in finding conserved currents associated with them.
On the other hand, special conformal transformations
are symmetries of linearized gravity only in a less obvious way:
they are not Noether symmetries of linearized gravity \cite{FarHinHul},
but the field equations for the linearized Weyl tensor
do have full conformal symmetry \cite{AP3,AP4},
and in the quantized theory the correlation functions
of the Weyl tensor operator
are those of a conformal primary \cite{FarHinHul}.
For this reason it is not immediately obvious how conserved currents
associated with special conformal transformations
can be found in linearized gravity.

In the present paper we show that special conformal transformations are
also Noether symmetries of linearized gravity
if suitable gauge fixing conditions are imposed,
and therefore conserved currents
associated with them can be obtained.
More precisely,
we formulate a generalization of Noether's theorem in which
additional conditions that one may have imposed on the
allowed field configurations\footnote{A field configuration
is a particular assignment of values of a field throughout spacetime.
Unless explicitly stated, field configurations
are not assumed to satisfy the field equations.},
such as gauge fixing conditions,
are taken into account,
show that special conformal transformations are
Noether symmetries of linearized gravity in the generalized harmonic gauges,
and obtain associated conserved currents using the generalized Noether's theorem.

It is important to note that the gauge symmetry of linearized gravity
allows the action of the spacetime symmetry transformations
on $h_{ab}$ to be modified by
suitable accompanying gauge transformations,
similarly to electrodynamics and Yang--Mills theory
\cite{BesselHagen,Jackiw1,JM,Jackiw2,ErLei,BanRey,BakLinSme}
(see also Appendix \ref{sec.em}).
Taking into account this possibility, we allow the various transformations,
which act on $h_{ab}$ and with which $M^{abc}$, $D^a$ and $C^{ab}$
are associated by Noether's theorem,
to be combinations of the `simple' spacetime transformations
(see Section \ref{sec.pdsym})
with suitable, generally field-dependent, gauge transformations.
We also allowed such combined transformations in \cite{TG},
and they appear in the context of linearized gravity
in \cite{BaTa,HoBaLa} (see also \cite{BaKu,HoBaLa2}) as well.
For brevity, they will be referred to as modified transformations.

We apply a general and well-known variant of Noether's theorem
(see Section \ref{sec.nthr})
that is well suited for the present investigation.
This variant and its generalization mentioned above
give a class of conserved currents, rather than a unique one,
for a given symmetry;
therefore, additional considerations are necessary
to obtain definite currents.
Regarding these considerations, we mention here that
we restrict our attention to Lorentz covariant currents that
do not depend on higher than first derivatives of $h_{ab}$
and are quadratic;
the full details will be given in the subsequent sections.

Although we are primarily interested in finding
special conserved currents having favourable properties,
we also consider general multi-parameter families of conserved currents,
as in \cite{TG}.
Multi-parameter families are worth considering
because they provide a broader perspective
and a useful background for the investigation of the distinguished currents.

Regarding the distinguished special conserved quantities,
we look for suitable special $M^{abc}$, $D^a$, $C^{ab}$ tensors
to accompany $T_{\mathrm{lg}}^{ab}$, $T_{\mathrm{LL}}^{ab}$, and $\tau^{ab}$.
The principal aim of this paper is to deal with the case of $T_{\mathrm{lg}}^{ab}$,
but the cases of $T_{\mathrm{LL}}^{ab}$ and $\tau^{ab}$ are also interesting,
both in themselves and for comparison with the case of $T_{\mathrm{lg}}^{ab}$.
A distinguished spin tensor
accompanying $\tau^{ab}$ has already been proposed in \cite{BHLang,BHLthe};
our aim regarding this tensor is to reproduce it
within the framework we adopt in the present paper.

Analogy with electrodynamics was an important guiding principle
in the identification of $T_{\mathrm{lg}}^{ab}$ in \cite{TG},
and it will also be important in the
construction of the $M^{abc}$, $D^a$ and $C^{ab}$
tensors accompanying $T_{\mathrm{lg}}^{ab}$,
therefore an appendix (Appendix \ref{sec.em})
on the construction of $M^{abc}$, $D^a$ and $C^{ab}$
tensors in electrodynamics is included in the paper.
If the energy-momentum tensor of the electromagnetic field
is taken to be the standard one (i.e., $T_{\mathrm{em}}^{ab}$;
see (\ref{eq.Tem})),
it is possible and reasonable to take
$M^{abc} = M_{\mathrm{orb}}^{abc}$,
$D^a = D_{\mathrm{orb}}^a$ and
$C^{ab} = C_{\mathrm{orb}}^{ab}$,
as was shown already in \cite{BesselHagen},
therefore analogy with electrodynamics suggests that
in linearized gravity one should also aim to choose
the $M^{abc}$, $D^a$ and $C^{ab}$ tensors
accompanying $T_{\mathrm{lg}}^{ab}$ in such a way
that their intrinsic parts are zero
or have special favourable properties.
If some of these properties require gauge fixing, then
the existence of at least one gauge in which all favourable properties
become manifest is also desirable.
In addition, this gauge (or at least one of these gauges) should be
well known and useful in other respects as well, if possible.
While aiming to have zero or almost zero intrinsic parts,
other requirements, such as that
$T^{ab}$, $M^{abc}$, $D^a$ and $C^{ab}$
should not depend on higher than first derivatives of $h_{ab}$,
must also be respected.

As for general multi-parameter families of conserved currents,
the $5$-parameter family of energy-momentum tensors
mentioned above will be supplemented with
accompanying multi-parameter families of
$M^{abc}$, $D^a$ and $C^{ab}$ tensors.
The $5$-parameter family of energy-momentum tensors contains
a $2$-parameter family of canonical energy-momentum tensors;
those will also be supplemented with
$M^{abc}$, $D^a$ and $C^{ab}$ tensors
that can be considered `canonical'.

In the literature on linearized gravity
the energy-momentum received the most attention
\cite{MagSok,NN,BHLemt,BHLthe,Butcher,Bicak,BS,Barnett,Baker,AghAndBha,Maggiore,HEL,HoBaLa},
the angular momentum is less studied \cite{BHLang,BHLthe,Butcher,Barnett,AghAndBha,Maggiore},
whereas the dilatation current and the special conformal tensor
do not seem to have been investigated in detail.
It should be noted, on the other hand,
that the conserved currents of the spin-two field\footnote{Here, the spin-two field is, in effect, the linearized Weyl tensor.}
were completely classified, up to equivalence, in \cite{AP2}
(see also \cite{AP3,AP4}).

\subsection*{Organization of the paper}

In Section \ref{sec.conf}, the general framework
we use for the discussion of the conserved currents
associated with conformal symmetry is introduced.
The generalization of Noether's theorem,
which can be regarded as a part of the framework,
is also presented in this section.
In Section \ref{sec.lingrav},
the formulation of linearized gravity in the framework of
Lagrangian field theory,
making use of the Fierz tensor, is recalled.
The gauges that are important for the subsequent considerations are introduced.
In Section \ref{sec.lg.em}, conserved currents associated with
the Poincar\'e and dilatation symmetries of linearized gravity are constructed and discussed.
In Section \ref{sec.gharm},
the special conformal symmetry of linearized gravity in generalized harmonic gauges
is discussed,
and the conservation laws
associated with Poincar\'e and dilatation symmetry found in Section \ref{sec.lg.em}
are extended to full conformal symmetry.
In Section \ref{sec.concl}, the main results of the paper
are summarized and concluding remarks are made.
In Appendix \ref{sec.em}, the conformal symmetry and
the associated conserved currents in electrodynamics are discussed briefly,
in order to allow comparison with linearized gravity.

\subsection*{Notation and conventions}

We mostly apply the same notation and conventions as in \cite{TG}.
Spacetime tensor indices will be denoted by $a,b,\dots$.
The dimension of the spacetime is mostly $4$ throughout the paper,
but in Section \ref{sec.conf}
it can have any value greater than $1$
(moreover, in Section \ref{sec.nthr}, a one-dimensional base manifold is also allowed).
The dimension of spacetime minus $1$ will be denoted by $d$;
the possible values of $a,b,\dots$ are $0,1,2,\dots,d$.
The Minkowski
metric \hfill (with \hfill signature \hfill $(+,-,\dots ,-)$) \hfill will \hfill be \hfill denoted \hfill by \hfill $\eta_{ab}$; \hfill
coordinates \hfill in \hfill which\\
$\eta_{ab} = \mathrm{diag}(1,-1,-1,\dots,-1)$ will be used
(although throughout the paper the particular form of the matrix of $\eta_{ab}$
is mostly not relevant;
in most cases it is sufficient for the matrix of $\eta_{ab}$ to be constant).
Spatial tensor indices will be denoted by $i, j, \dots$;
they can take the values \hfill $1, 2, \dots, d$. \hfill
They \hfill are \hfill raised \hfill and \hfill lowered \hfill by \hfill the \hfill spatial \hfill Euclidean \hfill metric\\
$\eta_{ij} = \mathrm{diag}(-1,-1,\dots,-1)$.
$\Box$ will denote the wave operator $\partial_a\partial^a$.
The conservation of a tensor (i.e., of a tensorial current)
$U^{ab\dots}$ will mean that
its divergence on the first index is zero,
i.e.\ $\partial_aU^{ab\dots} = 0$.

Occasionally square brackets will be used to emphasize that
the enclosed expression is an antisymmetric tensor.
The brackets $(\, )$ and $[\, ]$ will be used on indices
to denote symmetrization and antisymmetrization, respectively.
Combinatorial normalization factors will be understood to be included
where this notation is used.
The notation $\mc{A}_{\dots}[\dots]$ will also be used to denote antisymmetrization;
the indices to be antisymmetrized will be written in the subscript.
For example,
$\mc{A}_{ab}[T^{\dots a \dots b \dots}] =
(1/2)(T^{\dots a \dots b \dots}-T^{\dots b \dots a \dots})$.
$\epsilon^{abcd}$ will denote the totally antisymmetric tensor in $4$ dimensions,
normalized so that $\epsilon^{0123}=1$.
The Hodge dual of a tensor $T^{ab\dots}$ that is antisymmetric
in the first two indices will be denoted by a tilde:
$\tilde{T}^{ab\dots} = (1/2)\epsilon^{abcd}T_{cd}{}^{\dots}$.

Similarly to \cite{TG}, in linearized gravity we shall use the letters $c,d,\dots$,
instead of $a,b,\dots$,
for the indices of the energy-momentum and other tensors,
as we find it more convenient.
We shall use natural units, in which the speed of light is $1$
and $8\pi\gamma=1$,
where $\gamma$ is Newton's gravitational constant.

Let $\{ \Phi_\alpha \}$ denote a collection of fields and field components,
labeled by the index $\alpha$.
These fields are assumed to be defined on the spacetime,
or generally on a base manifold
which is not necessarily equipped with a metric.
$\Phi_\alpha$ is assumed to be real valued for each value of $\alpha$;
a complex field can be represented by a pair of real fields.
A function $f$ on the spacetime (or on the base manifold)
that is of the composite form
\begin{equation}
f(x^a) = \breve{f}(\Phi_\alpha(x^a), \partial_{a_1}\Phi_\alpha(x^a),
\partial_{a_1a_2}\Phi_\alpha(x^a), \dots,
\partial_{a_1a_2\dots a_k}\Phi_\alpha(x^a), x^a)
\end{equation}
with some fixed number $k$ and multivariate function $\breve{f}$
will be said to be a \emph{local function} of $\Phi_\alpha$.
The coordinates $x^a$ are not assumed to be orthonormal here.
$\partial_a f$ will be understood to be the partial derivative of $f$
as a composite function.
$\breve{f}$ is understood to be fixed,
therefore the value of $f$ at $x^a$ depends on
$\Phi_\alpha(x^a), \partial_{a_1}\Phi_\alpha(x^a),
\partial_{a_1a_2}\Phi_\alpha(x^a), \dots,$
$\partial_{a_1a_2\dots a_k}\Phi_\alpha(x^a)$.
This also means that $f$ is not a single function
(except in the very special case where $\breve{f}$ does not depend on
$\Phi_\alpha, \partial_{a_1}\Phi_\alpha, \dots$);
it depends on the values of $\Phi_\alpha$ throughout spacetime.
If $\partial \breve{f} / \partial x^a = 0$,
then it can be said that $f$ does not depend explicitly on $x^a$
($f$ depends on $x^a$
only through $\Phi_\alpha$ and its derivatives in this case).
Instead of $\breve{f}$, usually only $f$ will be written,
as is common in the literature.
The notation $f[\Phi_\alpha]$ will also be used for local functions.
Functions with values that have multiple components
(e.g., vector valued functions) will be said to be local
if their components are local functions.
A local function that does not depend on the derivatives of $\Phi_\alpha$
(i.e., has the form $f(x^a) = \breve{f}(\Phi_\alpha(x^a), x^a)$)
is called \emph{ultralocal}.
In the literature, local functions are introduced, for example,
in \cite{BBH}, section 4, and in \cite{Olver}.
In the latter reference the name \emph{differential function} is used.
In \cite{BBH} and \cite{Olver}, the term local function (or differential function)
refers to the multivariate function $\breve{f}$.

\newpage

The terms \emph{strongly conserved current},
\emph{weakly zero current}, \emph{trivially conserved current}
and \emph{superpotential}
will have the same meaning as in \cite{TG};
a current that is conserved for arbitrary configurations
of the basic dynamical fields in a field theory
will be called strongly conserved current\footnote{Strongly conserved currents
are also called identically conserved currents.},
a current that is zero if the fields satisfy the field equations
(especially the Euler--Lagrange equations)
will be called weakly zero current,
and a linear combination of a strongly conserved current
and a weakly zero current will be called
trivially conserved current.
If $J^a$ is a strongly conserved local current and $\Sigma^{ab}$,
which is also assumed to be a local function of the basic dynamical fields,
is an antisymmetric tensor such that
$J^a = \partial_b\Sigma^{ab}$ for arbitrary field configurations,
then $\Sigma^{ab}$ will be said to be a superpotential for $J^a$.

\section{Framework for conserved currents associated with conformal symmetry}
\label{sec.conf}

In this section a general framework
for discussing conformal symmetry
and the associated conserved currents in field theory
is introduced.
It will be used in linearized gravity in Sections \ref{sec.lg.em}, \ref{sec.gharm},
and in electrodynamics in Appendix \ref{sec.em}.
It follows the literature in many respects
(see, in particular,
\cite{FMS,Jackiw1,JM,Jackiw2,Polchinski,ForRom,Pons,Nakayama,BanRey,HobLas});
nevertheless, it has some aspects that are, to our knowledge, novel.

The first part of the framework,
which can be found in Sections \ref{sec.ctm}-\ref{sec.cft},
contains basic definitions and statements
concerning the energy-momentum tensor,
the angular momentum tensor, the dilatation current and the special conformal tensor.
In particular, the orbital and intrinsic parts of these quantities are introduced.
The content of Sections \ref{sec.ctm}-\ref{sec.orbint}
is largely independent of field theory (i.e., field variables and field equations);
it follows mostly from the structure of the conformal Lie algebra.
In Section \ref{sec.cft}, general field theoretic considerations are made,
connecting the first part of the framework to the subsequent parts,
contained by Sections \ref{sec.nthr} and \ref{sec.struJ}.

An important component of the framework,
contained by Section \ref{sec.nthr},
is an extension of Noether's theorem and its converse
to the case when additional conditions, such as gauge fixing conditions,
are imposed.
These theorems are presented in general form;
they are not tied to Minkowski spacetime, conformal symmetry and linearized gravity.
The extended Noether's theorem will be used mainly in the treatment of
the special conformal symmetry of linearized gravity,
but it will also appear in Sections \ref{sec.propjlg} and \ref{sec.bhl}.

Where a Lagrangian is assumed to be given in this section,
it is allowed in most cases
to depend on arbitrarily high derivatives of the
basic dynamical fields.
It will not be necessary to specify the form of the Lagrangian
and the nature of the fields on which it depends---e.g.,
whether the fields are scalar, vector, tensor, spinor,
or of some other type,
and how they change under conformal transformations
(an exception is Section \ref{sec.relcan}, where
a particular transformation rule does have a role).

The dimension of the spacetime is allowed to have any value greater than $1$
in this section,
moreover in Section \ref{sec.nthr}, where the extended Noether's theorem
is presented,
the base manifold is also allowed to be one-dimensional.
The case of $2$ spacetime dimensions is exceptional,
as the conformal Lie algebra is infinite dimensional
in that case.
Only the $6$-dimensional subalgebra
corresponding to global conformal transformations
will be taken into account
if the spacetime is $2$-dimensional;
with this restriction the difference between
$2$ and higher spacetime dimensions is minor.

The full conformal symmetry of linearized gravity is discussed only
in Section \ref{sec.gharm};
Section \ref{sec.lg.em} is restricted to the Poincar\'e and dilatation symmetries,
as they do not require gauge fixing.
For brevity, in the present section
Poincar\'e and dilatation symmetries
are not considered separately (i.e., without special conformal symmetry).
Nevertheless, it is not difficult to restrict the framework to these symmetries.

The conformal group itself will not have a major role in the
present and subsequent sections;
mostly only the generators of conformal transformations will have relevance.

\subsection{The generators of conformal transformations in Minkowski spacetime}
\label{sec.ctm}

An \hfill infinitesimal \hfill conformal \hfill isometry \hfill
of \hfill the \hfill Minkowski \hfill spacetime \hfill takes \hfill the \hfill form\\
$x^a \to x^a + \epsilon f^a$, where $\epsilon$ is a small parameter
and $f^a$ is the generator vector field
\begin{equation}
\label{eq.f}
f^a = e^a + \omega^{ab}x_b + \beta x^a + 2x^a c^b x_b - c^a x_b x^b.
\end{equation}
$e^a$, $\omega^{ab}$ ($\omega^{ab} = -\omega^{ba}$), $\beta$ and $c^a$
are constant parameters, which identify the elements
of the conformal Lie algebra (which is $(d+3)(d+2)/2$-dimensional).
$e^a$ and $\omega^{ab}$ are the parameters of translation
and Lorentz transformation generators,
$\beta$ is the parameter of dilatation generators,
$c^a$ is the parameter of special conformal transformation generators.
The commutation relations of the Lie algebra can be calculated
from (\ref{eq.f}), but they will not be needed.

$f^a$ satisfies the conformal Killing equation
\begin{equation}
\label{eq.ck}
\partial_a f_b + \partial_b f_a = \frac{2}{d+1}\eta_{ab} \partial_c f^c .
\end{equation}
From (\ref{eq.f}) it follows that
\begin{equation}
\partial_{ab}f_c=2(\eta_{ac}c_b+\eta_{bc}c_a-\eta_{ab}c_c),\quad
\Box f_a = 2(1-d)c_a\, ,\quad
\partial_a\partial_bf^b = 2(1+d)c_a\, ,
\end{equation}
\begin{equation}
\partial_{abc}f_d=0.
\end{equation}

$\partial_a f^a = 0$ is equivalent with $\beta = 0$, $c^a = 0$,
thus the generators of Poincar\'e transformations are
characterized by $\partial_a f^a = 0$.
The generators of Poincar\'e transformations and dilatations
are characterized by $\partial_{ab}f_c=0$
or $\partial_a\partial_bf^b = 0$
or $\Box f_a = 0$ (if $d\ne 1$),
which are equivalent with $c^a=0$.

The parameters $e^a$, $\omega^{ab}$, $\beta$, $c^a$ can be expressed
in terms of $f^a$ and $x^a$ as
\begin{eqnarray}
\label{eq.ce1}
c^a \! & = & \! \frac{1}{2(1-d)}\Box f^a \ = \ \frac{1}{2(1+d)}\partial^a\partial_bf^b  \\[1mm]
\beta \! & = & \! \frac{1}{d+1}\partial_a f^a + \frac{1}{d-1}x_a\Box f^a
\ = \ \frac{1}{d+1}(\partial_a f^a - x_a\partial^a\partial_bf^b)\\[1mm]
\omega^{ab} \! & = & \! \partial^{[b} f^{a]} + \frac{2}{d-1} x^{[a} \Box f^{b]}
\ = \ \partial^{[b} f^{a]} - \frac{2}{d+1} x^{[a} \partial^{b]}\partial_c f^c
\\[1mm]
e^a \! & = & \! f^a - \partial^{[b} f^{a]} x_b - \frac{1}{d+1}\partial_b f^b x^a
+ \frac{1}{d-1}\left(-x^a x^b +\frac{1}{2}x_c x^c \eta^{ab} \right) \Box f_b\\[1mm]
& = & \! f^a - \partial^{[b} f^{a]} x_b - \frac{1}{d+1}\partial_b f^b x^a
- \frac{1}{d+1}\left(-x^a x^b +\frac{1}{2}x_c x^c \eta^{ab} \right) \partial_b \partial_df^d.
\label{eq.ce4}
\end{eqnarray}
The formulae involving $\Box f^a$ are only valid if $d\ne 1$.

\subsection{Energy-momentum tensor, angular momentum tensor,
dilatation current, special conformal tensor}
\label{sec.eads}

A \emph{current associated with conformal transformations} will be understood to be
a current valued linear function of $e^a$, $\omega^{ab}$, $\beta$, $c^a$:
\begin{equation}
\label{eq.TMDC}
J^a = T^{ab}e_b + M^{abc}\omega_{bc} + D^a \beta + C^{ab}c_b  \qquad (M^{abc} = -M^{acb}).
\end{equation}
$J^a$ is equivalent with the collection $\{ T^{ab}, M^{abc}, D^a, C^{ab} \}$ of tensors
(which do not depend on $e^a$, $\omega^{ab}$, $\beta$, $c^a$).
$T^{ab}$ is called \emph{energy-momentum tensor},
$M^{abc}$ is called \emph{angular momentum tensor},
$D^a$ is called \emph{dilatation current},
and $C^{ab}$ is called \emph{special conformal tensor}.
$T^{ab}$ is not necessarily symmetric or traceless.
Angular momentum is understood here in relativistic sense,
as $M^{abc}$ is associated not only with rotations
but with all Lorentz transformations,
including Lorentz boosts.\footnote{In $2$ spacetime dimensions,
Lorentz boosts are the only Lorentz transformations.}

Although in field theory $T^{ab}$, $M^{abc}$, $D^a$ and $C^{ab}$
are usually defined in terms of field variables,
this is not assumed here.
$J^a$ being associated with conformal transformations is understood
merely to mean that it has the form (\ref{eq.TMDC}).
No assumptions about how $J^a$ is obtained are made.

The conservation of $J^a$ (i.e., $\partial_aJ^a = 0$)
for arbitrary values of $e^a$, $\omega^{ab}$, $\beta$, $c^a$
is equivalent with the conservation of
$T^{ab}$, $M^{abc}$, $D^a$ and $C^{ab}$, i.e.\ with
\begin{equation}
\partial_a T^{ab}=0,\quad
\partial_a M^{abc}=0,\quad \partial_a D^a=0,\quad \partial_a C^{ab}=0.
\end{equation}
In the following, the conservation of $J^a$
(and of similar currents that are
linear functions of the parameters $e^a$, $\omega^{ab}$, $\beta$, $c^a$)
will be understood to mean
that $J^a$ is conserved for all values of $e^a$, $\omega^{ab}$, $\beta$, $c^a$.
Throughout Section \ref{sec.conf},
$J^a$ (or $T^{ab}$, $M^{abc}$, $D^a$, $C^{ab}$) is not assumed
to be conserved unless explicitly stated.

Using (\ref{eq.ce1})-(\ref{eq.ce4}), $J^a$ can be written in the form
\begin{eqnarray}
\label{eq.j2}
J^a \! & = & \! \mbb{T}^{ab}f_b - S^{abc}\partial_{[b} f_{c]}
+ \frac{1}{1+d}\mbb{T}^a \partial_c f^c + \frac{1}{2(1-d)}\mbb{C}^{ab}\Box f_b\\
\label{eq.j2b}
& = & \! \mbb{T}^{ab}f_b - S^{abc}\partial_{[b} f_{c]}
+ \frac{1}{1+d}\mbb{T}^a \partial_c f^c + \frac{1}{2(1+d)}\mbb{C}^{ab}\partial_b \partial_cf^c,
\end{eqnarray}
where $\mbb{T}^{ab}$, $S^{abc}$ ($S^{abc} = -S^{acb}$),
$\mbb{T}^a$ and $\mbb{C}^{ab}$ are tensors that do not depend on
$e^a$, $\omega^{ab}$, $\beta$, $c^a$.
The right hand side of (\ref{eq.j2}) or (\ref{eq.j2b})
has the form of the most general homogeneous linear local function of $f^a$.
(The second derivative of $f^a$ can be chosen to appear
through $\Box f^a$ or $\partial^a \partial_bf^b$ if $d\ne 1$---(\ref{eq.j2}) and (\ref{eq.j2b})
correspond to these possibilities.
Linear combinations of $\Box f^a$ and $\partial^a \partial_bf^b$ could also be used.)

It is not difficult to verify that the relation between
$T^{ab}$, $M^{abc}$, $D^a$, $C^{ab}$ and
$\mbb{T}^{ab}$, $S^{abc}$, $\mbb{T}^a$, $\mbb{C}^{ab}$ is given by the equations
\begin{equation}
\label{eq.bbT1}
T^{ab}=\mbb{T}^{ab},\qquad M^{abc} = T^{a[b}x^{c]} + S^{abc},\qquad
D^a = T^{ab}x_b + \mbb{T}^a,
\end{equation}
\begin{equation}
\label{eq.bbT2}
C^{ab} = 2T^{ac}x_c x^b - T^{ab}x_c x^c
+ 2\mbb{T}^a x^b -4S^{abc}x_c
+ \mbb{C}^{ab}.
\end{equation}
From (\ref{eq.bbT1}) it is obvious that
$S^{abc} = M^{abc} - T^{a[b}x^{c]}$
and $\mbb{T}^a = D^a -  T^{ab}x_b$.
From (\ref{eq.bbT1}) and (\ref{eq.bbT2}) it follows that
\begin{equation}
\label{eq.bbT2-b}
\mbb{C}^{ab} = C^{ab} - T^{ab}x_cx^c + 2T^{ac}x_cx^b + 4M^{abc}x_c - 2D^ax^b.
\end{equation}
The latter equation gives $\mbb{C}^{ab}$ directly
in terms of $T^{ab}$, $M^{abc}$, $D^a$, $C^{ab}$.
(\ref{eq.bbT1})-(\ref{eq.bbT2-b}) show that the relation between
$\{ T^{ab}, M^{abc}, D^a, C^{ab} \}$
and $\{ \mbb{T}^{ab}, S^{abc}, \mbb{T}^a, \mbb{C}^{ab} \}$ is bijective.
It is worth noting that $d$ does not appear explicitly
in (\ref{eq.bbT1})-(\ref{eq.bbT2-b}).

\subsection{Orbital and intrinsic parts of $M^{abc}$, $D^a$, $C^{ab}$}
\label{sec.orbint}

Equations (\ref{eq.bbT1}) and (\ref{eq.bbT2}) suggest
the division of $M^{abc}$, $D^a$ and $C^{ab}$
into `orbital' and `intrinsic' parts:
\begin{equation}
\label{eq.morb}
M_{\mathrm{orb}}^{abc} = T^{a[b}x^{c]}
\end{equation}
will be called \emph{orbital angular momentum tensor},
\begin{equation}
\label{eq.dorb}
D_{\mathrm{orb}}^a = T^{ab}x_b
\end{equation}
will be called \emph{orbital dilatation current} and
\begin{equation}
\label{eq.corb}
C_{\mathrm{orb}}^{ab} = 2T^{ac}x_c x^b - T^{ab}x_c x^c
\end{equation}
will be called
\emph{orbital special conformal tensor}, whereas the remaining parts
\begin{equation}
\label{eq.mdint}
M_{\mathrm{int}}^{abc} = S^{abc},\qquad D_{\mathrm{int}}^a = \mbb{T}^a,
\end{equation}
\begin{equation}
\label{eq.cint}
C_{\mathrm{int}}^{ab} = 2\mbb{T}^a x^b -4S^{abc}x_c + \mbb{C}^{ab}
\end{equation}
will be called \emph{intrinsic angular momentum tensor},
\emph{intrinsic dilatation current} and \emph{intrinsic special conformal tensor}, respectively.
$S^{abc}$ and $\mbb{T}^a$
will also be called \emph{spin tensor} and \emph{virial current}, following the literature.
$J^a$ can be split as
\begin{equation}
\label{eq.orbint}
J^a = J^a_{\mathrm{orb}} + J^a_{\mathrm{int}}\, ,
\end{equation}
\begin{eqnarray}
\label{eq.orb}
J^a_{\mathrm{orb}} = && \hspace{-5.7mm} T^{ab}f_b
= T^{ab}e_b + M_{\mathrm{orb}}^{abc}\omega_{bc} + D_{\mathrm{orb}}^a \beta
+ C_{\mathrm{orb}}^{ab}c_b\, , \\[2mm]
\label{eq.int}
J^a_{\mathrm{int}} = && \hspace{-5.7mm} - S^{abc}\partial_{[b} f_{c]}
+ \frac{1}{1+d}\mbb{T}^a \partial_c f^c + \frac{1}{2(1-d)}\mbb{C}^{ab}\Box f_b\\[2mm]
\phantom{J^a_{\mathrm{int}}} = && \hspace{-5.7mm} - S^{abc}\partial_{[b} f_{c]}
+ \frac{1}{1+d}\mbb{T}^a \partial_c f^c + \frac{1}{2(1+d)}\mbb{C}^{ab}\partial_b \partial_cf^c\\[2mm]
\label{eq.int3}
\phantom{J^a_{\mathrm{int}}} = && \hspace{-5.7mm} S^{abc}\omega_{bc} + \mbb{T}^a \beta + C_{\mathrm{int}}^{ab}c_b\, .
\end{eqnarray}
$J^a_{\mathrm{orb}}$ comprises the orbital parts of the angular momentum tensor,
dilatation current and special conformal tensor,
whereas $J^a_{\mathrm{int}}$ comprises the intrinsic parts.

$C_{\mathrm{int}}^{ab}$ has an `extrinsic' part, $2\mbb{T}^a x^b -4S^{abc}x_c$,
which is determined by
the spin tensor and the virial current, therefore
it is possible to distinguish a \emph{`fully intrinsic' part of}
$C_{\mathrm{int}}^{ab}$, namely $\mbb{C}^{ab}$.

In the literature the relative normalization of $T^{ab}$,
$M^{abc}$ and $S^{abc}$ is usually such that
$M^{abc} = x^b T^{ac} - x^c T^{ab} + S^{abc}$,
whereas in the present paper
$M^{abc} = T^{a[b}x^{c]} + S^{abc}$ (see (\ref{eq.bbT1})).
The latter relation follows from (\ref{eq.f}), (\ref{eq.TMDC})
and (\ref{eq.j2}) (or (\ref{eq.j2b})).

Using (\ref{eq.bbT1}) and (\ref{eq.bbT2}),
one can see that the following statements are true:
\begin{prop}
Assuming that $T^{ab}$ is conserved,
the conservation of $M^{abc}$ is equivalent with
\begin{equation}
\label{eq.consm}
T^{[bc]} = \partial_a S^{abc},
\end{equation}
the conservation of $D^a$ is equivalent with
\begin{equation}
\label{eq.consd}
T_a{}^a = -\partial_a\mbb{T}^a,
\end{equation}
and the conservation of $C^{ab}$ is equivalent with
\begin{equation}
\label{eq.consc}
2(T_a{}^a + \partial_a\mbb{T}^a)x^b + 4(T^{[bc]} - \partial_a S^{abc})x_c
+ 2\mbb{T}^b+4S_a{}^{ab} = - \partial_a\mbb{C}^{ab}.
\end{equation}
\label{prop.2.2}
\end{prop}

From (\ref{eq.consm}) it can be seen that if $T^{ab}$ and $M^{abc}$ are conserved
and $T^{ab}$ is symmetric,
then $M_{\mathrm{orb}}^{abc}$ and $S^{abc}$ are conserved.
From (\ref{eq.consd}) it can be seen that
if $T^{ab}$ and $D^a$ are conserved and $T^{ab}$ is traceless,
then $D_{\mathrm{orb}}^a$ and $\mbb{T}^a$ are conserved.
Furthermore, it is not difficult to verify by direct calculation that
if $T^{ab}$ is conserved, symmetric and traceless,
then $C_{\mathrm{orb}}^{ab}$ is also conserved,
thus if $T^{ab}$, $M^{abc}$, $D^a$ and $C^{ab}$ are conserved and
$T^{ab}$ is both symmetric and traceless,
then $M_{\mathrm{orb}}^{abc}$, $S^{abc}$,
$D_{\mathrm{orb}}^a$, $\mbb{T}^a$,
$C_{\mathrm{orb}}^{ab}$ and $C_{\mathrm{int}}^{ab}$ are conserved.

(\ref{eq.consm}) also shows that if $T^{ab}$ and $M^{abc}$ are conserved,
then $S^{abc}$ cannot be zero if $T^{ab}$ is not symmetric.
Similarly, (\ref{eq.consd}) shows that
if $T^{ab}$ and $D^a$ are conserved, then
$\mbb{T}^a$ cannot be zero if $T^{ab}$ is not traceless.
The symmetry of $T^{ab}$ does not imply the vanishing of $S^{abc}$,
and the tracelessness of $T^{ab}$ does not imply the vanishing of $\mbb{T}^a$.

From Proposition \ref{prop.2.2}, it follows that
\begin{prop}
If $T^{ab}$,  $M^{abc}$ and $D^a$ are conserved,
then the conservation of $C^{ab}$ is equivalent with
\begin{equation}
\label{eq.consc2}
4S_a{}^{ab}+2\mbb{T}^b+\partial_a\mbb{C}^{ab}=0.
\end{equation}
\label{prop.2.3}
\end{prop}

\noi
From Propositions \ref{prop.2.2} and \ref{prop.2.3},
it follows that
\begin{prop}
If $T^{ab}$, $M^{abc}$, $D^a$ and $C^{ab}$ are conserved and $S_a{}^{ab}=0$,
then
\begin{equation}
T_a{}^a = \frac{1}{2}\partial_{ab}\mbb{C}^{ab}.
\label{eq.consc3}
\end{equation}
\label{prop.2.4}
\end{prop}
\noi
(\ref{eq.consc3}) can be obtained from (\ref{eq.consd})
using (\ref{eq.consc2}) and $S_a{}^{ab}=0$.
Propositions \ref{prop.2.2}, \ref{prop.2.3} and \ref{prop.2.4} show that
$T^{ab}$, $S^{abc}$ and $\mbb{T}^a$
have to satisfy some conditions if $\mbb{C}^{ab}$ vanishes.

\vspace{2mm}
In field theory the conservation of a current usually means that
it is conserved if the field equations are satisfied,
therefore in field theory equations (\ref{eq.consm})-(\ref{eq.consc2})
are usually required, or deduced\footnote{Depending on which direction
of the statements in Propositions \ref{prop.2.2} and \ref{prop.2.3} is considered.},
to hold only on the solutions of the field equations.
Similarly, in Proposition \ref{prop.2.4}
it is reasonable to require the condition $S_a{}^{ab}=0$
to hold only if the field equations are satisfied,
and equation (\ref{eq.consc3}) is only deduced to hold
on the solutions of the field equations.

\vspace{2mm}
It is well known that from $T^{ab}$ and $S^{abc}$
it is possible to construct a new energy-momentum tensor,
the Belinfante--Rosenfeld tensor,
that is symmetric \cite{Belinfante,Rosenfeld,FMS}.
Moreover, in theories that have complete conformal symmetry,
an energy-momentum tensor that is both symmetric and traceless
can also be constructed \cite{CCJ,Deser}
(see also \cite{FMS,Polchinski,GotMar,Pons,Nakayama,BGRS,KouNicSun,IKPS}
and references therein).
These tensors will not have an important role in the present paper.
Although they are favourable regarding symmetry and tracelessness,
their other properties are not necessarily desirable;
for instance, they may turn out to depend on higher than first derivatives
of the basic dynamical fields
of the theory under consideration (see, for example, \cite{BHLthe})
even when $T^{ab}$ and $S^{abc}$ do not depend on
higher than first derivatives.

\subsection{Additional properties of $J^a$ in field theory}
\label{sec.cft}

In the literature, $T^{ab}$, $M^{abc}$, $D^a$, $C^{ab}$,
and especially the intrinsic and orbital parts of these quantities,
are often introduced within a field theoretic setting,
whereas the definitions in the previous subsections
are completely disentangled from Lagrangian field theory.
This implies that assumptions about how
$T^{ab}$, $M^{abc}$, $D^a$ and $C^{ab}$ are obtained were not made
(e.g., $T^{ab}$ is not assumed to be a canonical energy-momentum tensor).

Nevertheless, in field theory
$T^{ab}$, $M^{abc}$, $D^a$, $C^{ab}$ and their orbital and intrinsic parts
usually have additional important general properties.
In particular, they are usually
local expressions in the basic dynamical fields.
Furthermore, $T^{ab}$ usually does not depend explicitly on $x^a$,
and the same is expected from $S^{abc}$, $\mbb{T}^a$ and $\mbb{C}^{ab}$---otherwise
it would not be appropriate to label the latter as `intrinsic' quantities.
The property that $T^{ab}$, $S^{abc}$, $\mbb{T}^a$ and $\mbb{C}^{ab}$
do not depend explicitly on $x^a$ implies that
the orbital quantities
$M_{\mathrm{orb}}^{abc}$, $D_{\mathrm{orb}}^a$, $C_{\mathrm{orb}}^{ab}$
and the `extrinsic' part $2\mbb{T}^a x^b -4S^{abc}x_c$ of $C_{\mathrm{int}}^{ab}$ do,
hence it is justified to label them as `orbital' and `extrinsic'.

It is not difficult to verify that the property that
$T^{ab}$, $S^{abc}$, $\mbb{T}^a$ and $\mbb{C}^{ab}$
are local functions of the basic dynamical fields and
do not depend explicitly on $x^a$
is equivalent with the following:
{\setlength{\leftmargini}{3.0em}
\begin{itemize}
\item[(FT)]
$J^a$ is a local function of the basic dynamical fields and $f^a$
with the properties that
(a) it does not depend explicitly on $x^a$
and
(b) it is homogeneous linear in $f^a$, $\partial_b f^a$ and $\partial_{bc}f^a$.
\end{itemize}}

$S^{abc}$ is determined by $M^{abc}$ and $T^{ab}$,
therefore the property that it does not depend explicitly on $x^a$
implies a restriction on the set of $M^{abc}$ tensors
that can accompany a given $T^{ab}$ tensor,
and similar statements apply to $D^a$ and $C^{ab}$ as well.

In field theory, $T^{ab}$, $M^{abc}$, $D^a$ and $C^{ab}$
are also expected to be associated with conformal transformations
(i.e., translations, Lorentz transformations, dilatations and special conformal transformations)
via Noether's theorem.
In gauge theories, conformal transformations are allowed
to be modified conformal transformations
(i.e., combinations of `simple' conformal transformations with
possibly field-dependent gauge transformations.)

These properties of $J^a$, i.e., (FT) and that $J^a$ should be
associated with conformal transformations via Noether's theorem,
will be discussed in more detail in Section \ref{sec.struJ}.

It should be remarked that
if one is merely interested in constructing conserved currents
associated with conformal symmetry,
then one may disregard
the orbital and intrinsic parts of $M^{abc}$, $D^a$ and $C^{ab}$.

\subsection{Generalization of Noether's theorem}
\label{sec.nthr}

In this subsection, we introduce a generalization of
the definition of Noether symmetry
and of Noether's theorem and its converse,
which are applicable when additional conditions,
such as gauge fixing conditions,
are imposed on the allowed configurations of the basic dynamical fields.

If additional conditions are not imposed,
then the generalized Noether's theorem reduces to
a well-known variant of the original Noether's theorem,
which we use when the generalized theorem is not needed.
Useful references
for this variant of Noether's theorem are \cite{BanRey, Pons}.
For additional references on Noether's theorem,
see \cite{Noether, BesselHagen, Olver, BCA, K-S, PKLT}.
To our knowledge, the generalized Noether's theorem
that we introduce here
has not previously appeared in the literature,
although it did appear, in essence, in a somewhat shorter and less explicit form
in Appendix A of \cite{TG2018}.
The converse of the theorem was not considered in \cite{TG2018}.

The set of allowed configurations for each basic dynamical field component
is usually the set of all smooth real functions on the spacetime\footnote{The allowed field configurations are not required to satisfy the field equations.}
(or generally on the base manifold on which the fields are defined).
The additional conditions may be any conditions that select
a subset of the set of the allowed configurations of the basic dynamical fields.
The term `additional' indicates that the conditions
are not assumed to follow from the Lagrangian.

Assuming that some additional conditions are given that select
a set $\mc{F}$ of field configurations,
we consider some other conditions to be more restrictive
if the set of field configurations $\mc{F}'$ selected by them
is a subset of $\mc{F}$.
In the case when the additional conditions are gauge fixing conditions,
$\mc{F}$ and $\mc{F}'$ are `gauges',
and one can say that $\mc{F}'$ is a \emph{subgauge} of $\mc{F}$.

Two main features of the above-mentioned variant of Noether's theorem,
which will be characteristic of the generalized theorem as well,
are that (i) transformations are taken to act on the fields in it,
leaving the spacetime coordinates unchanged,
i.e.\ spacetime and internal symmetries are dealt with on the same footing,
(ii) for a given symmetry it gives a class of conserved currents,
rather than a unique one.

Although throughout the paper it is generally assumed that
the base manifold is the Minkowski spacetime
and $x^a$ are orthonormal coordinates,
in the present exposition of Noether's theorem
no reference to any metric is made.
$x^a$ are merely fixed coordinates on the base manifold
and are not required to have any special property.
The dimension of the base manifold is allowed to be $1$,
i.e.\ the exposition extends to Lagrangian mechanics.

$\{ \Phi_\alpha \}$ will be
a collection of basic dynamical fields and field components
in the following
and $L(\Phi_\alpha, \partial_a\Phi_\alpha, \partial_{ab}\Phi_\alpha, \dots, x^a)$
will be a Lagrangian density function.
$\Phi_\alpha$ is understood to be real valued
for all values of the index $\alpha$;
complex valued fields can be taken into account
as pairs of real valued fields.
$L$ is assumed to be differentiable as many times as needed,
but otherwise it is not restricted.
It is allowed to be complex valued.

\subsubsection{Variation of the Lagrangian}

Under an infinitesimal transformation
$\Phi_\alpha\to \Phi_\alpha+\epsilon\delta\Phi_\alpha$ of $\Phi_\alpha$,
where $\epsilon$ is a small parameter
and $\delta\Phi_\alpha$ denotes the variation of $\Phi_\alpha$ that specifies the transformation,
the variation of $L$ is
\begin{eqnarray}
\label{eq.nthr1a}
\delta L & = & \frac{dL[\Phi_\alpha + \epsilon \delta \Phi_\alpha]}{d\epsilon}|_{\epsilon = 0}\\[2mm]
\label{eq.nthr1b}
& = & \frac{\partial L}{\partial \Phi_\alpha}\delta\Phi_\alpha
+ \frac{\partial L}{\partial (\partial_a \Phi_\alpha)}\partial_a \delta \Phi_\alpha
+ \frac{\partial L}{\partial (\partial_{ab} \Phi_\alpha)}\partial_{ab} \delta \Phi_\alpha
+ \dots\\[2mm]
\label{eq.nthr1c}
& = & \mr{E}(L)^\alpha \delta\Phi_\alpha + \partial_a j^a,
\end{eqnarray}
where
\begin{equation}
\mr{E}(L)^\alpha \ = \
\frac{\partial L}{\partial\Phi_\alpha}
- \partial_a\frac{\partial L}{\partial(\partial_a\Phi_\alpha)}
+ \partial_{ab}\frac{\partial L}{\partial(\partial_{ab}\Phi_\alpha)}
- \dots
\end{equation}
is the Euler--Lagrange derivative of $L$ with respect to $\Phi_\alpha$, and
\begin{eqnarray}
j^a & = &  \frac{\partial L}{\partial (\partial_a\Phi_\alpha)}\delta \Phi_\alpha +
\biggl(\frac{\partial L}{\partial (\partial_{ab}\Phi_\alpha)}\partial_b \delta \Phi_\alpha
-\partial_b\frac{\partial L}{\partial (\partial_{ab}\Phi_\alpha)} \delta \Phi_\alpha\biggr)\nonumber\\[2mm]
&& +\, \biggl(\frac{\partial L}{\partial (\partial_{abc}\Phi_\alpha)}\partial_{bc}\delta\Phi_\alpha
-\partial_b\frac{\partial L}{\partial (\partial_{abc}\Phi_\alpha)}\partial_c\delta\Phi_\alpha
+\partial_{bc}\frac{\partial L}{\partial (\partial_{abc}\Phi_\alpha)}\delta\Phi_\alpha\biggr)+\dots . \nonumber \\
\label{eq.elc}
\end{eqnarray}
Equations (\ref{eq.nthr1a})-(\ref{eq.nthr1c}) are identities
that are valid for arbitrary $\Phi_\alpha$ and $\delta\Phi_\alpha$
((\ref{eq.nthr1a}) is the definition of $\delta L$).
They are the fundamental identities from which Noether's theorem
and its generalization follows.

\subsubsection{Noether symmetry}

We generalize the definition of Noether symmetry in the following way:
\begin{defin}[Noether symmetry in the presence of additional conditions]
In the presence of additional conditions,
$\delta\Phi_\alpha$ is said to specify a Noether symmetry of a Lagrangian density $L$
if there exists a $K^a$ such that
\begin{equation}
\label{eq.defns}
\delta L = \partial_a K^a
\end{equation}
holds on all field configurations that satisfy the additional conditions.
($\delta\Phi_\alpha$ and $K^a$ are understood to be local functions of $\Phi_\alpha$.)
\label{def.nsg}
\end{defin}
In the literature, Noether symmetry is often defined as a symmetry of the action,
but it is possible to define it directly in terms of
the Lagrangian density function.

A term that vanishes if the additional conditions are satisfied
can be freely added to $\delta\Phi_\alpha$.
For fixed $L$ and $\delta\Phi_\alpha$,
$K_1^a$ and $K_2^a$ can both satisfy the equality $\delta L = \partial_a K^a$
if and only if $\partial_a (K_1^a-K_2^a)=0$
on all field configurations satisfying the additional conditions.
This implies that strongly conserved local currents can be freely added to $K^a$
regardless of additional conditions, and that
$K^a$ is less determined in the presence of additional conditions
than without such conditions.

Clearly if $\delta\Phi_\alpha$ specifies a Noether symmetry
under certain additional conditions,
then it does so under more restrictive conditions as well.
In particular, a Noether symmetry without additional conditions
(i.e., an ordinary Noether symmetry)
is also a Noether symmetry in the presence of any additional conditions.

It is important to note that Definition \ref{def.nsg}
does not require the infinitesimal symmetry transformation
specified by $\delta\Phi_\alpha$ to preserve the additional conditions.

\subsubsection{Generalization of Noether's theorem and its converse}
\label{sec.gen-Noether-th}

After Definition \ref{def.nsg}, Noether's theorem can be generalized in the following way:
\begin{thrm}[Noether's theorem in the presence of additional conditions]
If $\delta\Phi_\alpha$ specifies a Noether symmetry of a Lagrangian density $L$
under certain additional conditions,
then the current
\begin{equation}
\label{eq.nthr}
\JN^a = j^a - K^a
\end{equation}
is conserved (i.e., $\partial_a \JN^a = 0$) on those solutions of the
Euler--Lagrange equations\\
$\mr{E}(L)^\alpha = 0$
that also satisfy the additional conditions.
The equation
\begin{equation}
\label{eq.nthr2}
\partial_a \JN^a = - \mr{E}(L)^\alpha \delta\Phi_\alpha
\end{equation}
holds on all field configurations that satisfy the additional conditions.
\label{thrm.ng}
\end{thrm}
\noi Proof. According to the condition in the theorem,
$\delta L = \partial_a K^a$ holds with some suitable $K^a$
on any field configuration satisfying the additional conditions.
For $\delta L$ we also have the identity
$\delta L =  \mr{E}(L)^\alpha \delta\Phi_\alpha + \partial_a j^a$ (see (\ref{eq.nthr1c})).
Together these equalities imply that
$\partial_a \JN^a = - \mr{E}(L)^\alpha \delta\Phi_\alpha$, where $\JN^a = j^a - K^a$,
on any field configuration satisfying the additional conditions.
From this result it follows that
if a field configuration satisfies the additional conditions
and also the Euler--Lagrange equations ($\mr{E}(L)^\alpha = 0$),
then $\partial_a \JN^a = 0$.
\hfill$\blacksquare$\\

As without additional conditions, $\JN^a$ is called Noether current
and it is said to be
associated with the symmetry specified by $\delta\Phi_\alpha$.
For $\JN^a$ to be definite, a particular $K^a$ has to be chosen.
The indeterminacy of $K^a$ in (\ref{eq.defns})
is important for Definition \ref{def.nsg} to be sufficiently general.
In particular, it is important in the treatment
of the currents associated with conformal symmetry
in linearized gravity.
$\delta\Phi_\alpha$ and $K^a$ are both assumed to be local functions of $\Phi_\alpha$,
therefore $\JN^a$ is also local.

It should be noted that in contrast with strongly conserved currents,
weakly zero currents
generally cannot be freely added to a Noether current;
if a weakly zero (local) current is added to a Noether current,
the resulting current may be associated with a different variation,
or may not be a Noether current at all.

The additional conditions do not appear explicitly
in (\ref{eq.nthr}) and (\ref{eq.nthr2}).
The latter formula
will be useful for determining the divergence of the currents
associated with conformal symmetry
in linearized gravity in the presence of matter.

The converse of Noether's theorem can be generalized
to the case when additional conditions are present as follows:
\begin{thrm}[Converse of Noether's theorem in the presence of additional conditions]
Assume that a Lagrangian density $L$ and
a field variation $\delta\Phi_\alpha$
(which is assumed to be a local function of $\Phi_\alpha$)
are given.
Additional conditions on the fields may also be specified.
Then if a local current $\JN^a$ satisfies equation (\ref{eq.nthr2})
for all field configurations that satisfy the additional conditions,
then $\delta\Phi_\alpha$ specifies, under those additional conditions,
a Noether symmetry of $L$ with $K^a = j^a - \JN^a$,
and $\JN^a$ is the Noether current associated with $\delta\Phi_\alpha$,
corresponding to the choice $K^a=j^a - \JN^a$ for $K^a$.
\label{thrm.cng}
\end{thrm}
\noi Proof. Let us define $K^a$ as $K^a=j^a - \JN^a$. Then from equation (\ref{eq.nthr2}) it follows that
$\partial_a K^a = \mr{E}(L)^\alpha \delta\Phi_\alpha + \partial_a j^a$ holds on those field configurations that satisfy the additional conditions.
Taking into account the identity
$\delta L =  \mr{E}(L)^\alpha \delta\Phi_\alpha + \partial_a j^a$ (see (\ref{eq.nthr1b})),
it follows then that $\partial_a K^a = \delta L$ on those field configurations that satisfy the additional conditions.
In view of Definition \ref{def.nsg},
this means that under the additional conditions being considered
$\delta\Phi_\alpha$ specifies a Noether symmetry of $L$ with $K^a = j^a - \JN^a$.
The associated Noether current, with $K^a = j^a - \JN^a$, is $j^a-K^a = \JN^a$ (see (\ref{eq.nthr})).
\hfill$\blacksquare$\\

Clearly if (\ref{eq.nthr2}) holds under certain additional conditions,
then it also holds under more restrictive additional conditions,
thus if Theorem \ref{thrm.cng} is applicable to a current $\JN^a$
under some additional conditions,
then it is also applicable under more restrictive additional conditions.

$\delta\Phi_\alpha = 0$ specifies a Noether symmetry of any Lagrangian density
under any additional conditions,
according to Definition \ref{def.nsg}.
$j^a=0$ in this case, therefore $\JN^a = -K^a$.
$\delta L$ is also zero, thus $\JN^a$ is any local current
that is conserved on all field configurations
satisfying the additional conditions.
This means that all such currents are Noether currents
associated with $\delta\Phi_\alpha = 0$.
In the absence of additional conditions,
these currents are the strongly conserved local currents.
If additional conditions are present, then all local currents that vanish
on the field configurations satisfying the additional conditions
are Noether currents associated with $\delta\Phi_\alpha = 0$.

A linear combination $\sum_ic_i\JN^a_{(i)}$ of Noether currents
is also a Noether current,
associated with the linear combination $\sum_ic_i\delta\Phi_{(i)\alpha}$
of the respective variations $\delta\Phi_{(i)\alpha}$
with which $\JN^a_{(i)}$ are associated.
The $K^a$ quantity for $\sum_ic_i\JN^a_{(i)}$
is the linear combination $\sum_ic_iK^a_{(i)}$
of the respective $K^a$ quantities $K^a_{(i)}$ for $\JN^a_{(i)}$.

The remarks in the previous paragraphs imply that
in the presence of additional conditions
any Noether current can be freely changed
(without changing the variation with which it is associated)
by adding an arbitrary local current
that vanishes on the field configurations
satisfying the additional conditions.

A given Noether current may be associated with more than one variations,
if there exists a nonzero variation such that
the zero current is associated with it.
In the presence of additional conditions,
if a variation $\delta\Phi_\alpha$ specifies a Noether symmetry
and the associated Noether current, corresponding to some $K^a$, is $\JN^a$,
and $\delta_0\Phi_\alpha$ is a variation that vanishes
if the additional conditions are satisfied,
then $\delta\Phi_\alpha + \delta_0\Phi_\alpha$ also specifies a Noether symmetry
with the same $K^a$, and the associated Noether current is equal to $\JN^a$
on those field configurations that satisfy the additional conditions.

Definition \ref{def.nsg} and
Theorems \ref{thrm.ng} and \ref{thrm.cng}
can be extended in an obvious manner
to variations and currents having additional indices.
This extension is relevant in the treatment of the currents
associated with conformal symmetry,
as $T^{ab}$, $M^{abc}$ and $C^{ab}$ are tensorial currents.

\subsubsection{Indeterminacy of the Lagrangian density and Noether's theorem}
\label{sec.lagrind}

The Euler--Lagrange equations do not change
if a total divergence is added to the Lagrangian density function,
therefore the Lagrangian density function
is not uniquely determined by the equations of motion.
It is an important question then whether this indeterminacy
has any significance for (the generalized) Noether's theorem
(i.e., Theorem \ref{thrm.ng}).
The following statement provides an answer to this question:
\begin{prop}
If $\delta\Phi_\alpha$ specifies a Noether symmetry of $L$,
then it also specifies a Noether symmetry of any other Lagrangian density $L'$
that differs from $L$ by a total divergence.
Moreover, when $L$ is changed to $L'$,
it is possible to change $K^a$ in such a way that
the Noether current associated with $\delta\Phi_\alpha$
does not change.
\label{prop.nindet}
\end{prop}

\noi
In this statement an additional condition is understood
to be allowed to be present, although this is not explicitly mentioned.

In order to prove Proposition \ref{prop.nindet},
let us consider the variation of the Lagrangian density $L'$
due to $\Phi_\alpha\to \Phi_\alpha+\epsilon\delta\Phi_\alpha$.
The Euler--Lagrange derivative of $L_\Delta=L'-L$ is zero,
as $L_\Delta$ is a total divergence, therefore
$\delta L' = \mr{E}(L)^\alpha\delta\Phi_\alpha+\partial_aj^a+\partial_aj_\Delta^a$,
where $j^a$ comes from $L$ and $j_\Delta^a$ from $L_\Delta$.
For $\delta L$, we have
$\delta L = \mr{E}(L)^\alpha\delta\Phi_\alpha+\partial_aj^a = \partial_a K^a$
on the field configurations satisfying the additional conditions,
therefore $\delta L' = \partial_a (K^a+j_\Delta^a)$
on these field configurations.
This means that $\delta\Phi_\alpha$ specifies a Noether symmetry of $L'$ as well,
and a suitable modification of $K^a$ when $L$ is changed to $L'$ is
$K^a\to K^a+j_\Delta^a$.
If $K^a$ is modified in this way,
then the Noether current $j^a-K^a$ associated with $\delta\Phi_\alpha$
changes as $j^a-K^a \to j^a+j_\Delta^a - (K^a+j_\Delta^a)$,
i.e.\ it remains unchanged.

\vspace*{2mm}
Concerning the converse of the generalized Noether's theorem
(i.e., Theorem \ref{thrm.cng}),
equation (\ref{eq.nthr2}) depends on $L$ only through $\mr{E}(L)^\alpha$,
therefore if $\JN^a$ satisfies (\ref{eq.nthr2})
with a Lagrangian density $L$ and a field variation $\delta\Phi_\alpha$,
then it also satisfies (\ref{eq.nthr2})
with the same $\delta\Phi_\alpha$
and any Lagrangian density that differs from $L$ by a total divergence.
This means that Theorem \ref{thrm.cng} also does not impose any restriction
on the freedom of adding a total divergence to the Lagrangian density.
For fixed $\JN^a$,
$j^a$ and $K^a$ will, of course,
depend on which particular Lagrangian density is chosen.

\subsection{The structure of $J^a$ as a Noether current}
\label{sec.struJ}

In Lagrangian field theory, $J^a$ is expected to be
associated with conformal symmetry by Noether's theorem.
In the present section some basic aspects of this property of $J^a$,
in particular its relation to the property (FT) (see Section \ref{sec.cft}),
are discussed.

The discussion is based on the following general observation:
it is clear from (\ref{eq.ce1})-(\ref{eq.ce4})
and the arguments in Sections \ref{sec.eads} and \ref{sec.orbint}
that not just $J^a$,
but any quantity that is a linear function of $e^a$, $\omega^{ab}$, $\beta$ and $c^a$
has a form analogous to (\ref{eq.j2}) or (\ref{eq.j2b}),
and it is possible to define its various orbital and intrinsic parts.

\subsubsection{$\delta \Phi_\alpha$, $j^a$ and $K^a$ associated with $J^a$}
\label{sec.Jnoether1}

When $J^a$ is constructed using Noether's theorem,
not only $J^a$, but also the variation $\delta\Phi_\alpha$
with which $J^a$ is associated
is a homogeneous linear function of $e^a$, $\omega^{ab}$, $\beta$ and $c^a$.
In view of (\ref{eq.elc}) and (\ref{eq.nthr}),
$j^a$ and $K^a$ will also be of this type.
Hence,
\begin{eqnarray}
\label{eq.dPhi_lin}
\delta \Phi_\alpha = && \hspace{-5mm} \delta_T\Phi_\alpha^a e_a + \delta_M\Phi_\alpha^{ab}\omega_{ab}
+ \delta_D\Phi_\alpha \beta + \delta_C\Phi_\alpha^a c_a
\qquad (\delta_M\Phi_\alpha^{ab} = -\delta_M\Phi_\alpha^{ba}), \\[2mm]
\label{eq.K_lin}
K^a = && \hspace{-5mm} K_T^{ab} e_b + K_M^{abc}\omega_{bc}
+ K_D^a \beta + K_C^{ab}c_b
\qquad (K_M^{abc} = -K_M^{acb}), \\[2mm]
\label{eq.j_lin}
j^a = && \hspace{-5mm} j_T^{ab} e_b + j_M^{abc}\omega_{bc}
+ j_D^a \beta + j_C^{ab}c_b
\qquad (j_M^{abc} = -j_M^{acb}).
\end{eqnarray}
$\delta_T\Phi_\alpha^a e_a$, $\delta_M\Phi_\alpha^{ab}\omega_{ab}$,
$\delta_D\Phi_\alpha \beta$ and $\delta_C\Phi_\alpha^a c_a$
are the variations of $\Phi_\alpha$ under translations,
Lorentz\\
transformations, \hfill dilatations \hfill and \hfill special \hfill conformal \hfill transformations, \hfill
respectively, \hfill and\\
$K_T^{ab} e_b$, $K_M^{abc}\omega_{bc}$, $K_D^a \beta$, $K_C^{ab}c_b$ and
$j_T^{ab} e_b$, $j_M^{abc}\omega_{bc}$, $j_D^a \beta$, $j_C^{ab}c_b$
are the $K^a$ and $j^a$ quantities for the currents associated with
$\delta_T\Phi_\alpha^a e_a$, $\delta_M\Phi_\alpha^{ab}\omega_{ab}$,
$\delta_D\Phi_\alpha \beta$ and $\delta_C\Phi_\alpha^a c_a$,
which are $T^{ab} e_b$, $M^{abc}\omega_{bc}$, $D^a \beta$ and $C^{ab}c_b$
(see (\ref{eq.TMDC})).

It should be noted that the variations
$\delta_T\Phi_\alpha^a e_a$, $\delta_M\Phi_\alpha^{ab}\omega_{ab}$,
$\delta_D\Phi_\alpha \beta$ and $\delta_C\Phi_\alpha^a c_a$
do not have to satisfy the commutation relations of the conformal Lie algebra.
This is especially important in gauge theories,
where the variations
$\delta_T\Phi_\alpha^a e_a$, $\delta_M\Phi_\alpha^{ab}\omega_{ab}$,
$\delta_D\Phi_\alpha \beta$ and $\delta_C\Phi_\alpha^a c_a$
generally correspond to modified conformal transformations
and do not satisfy the commutation relations of the conformal Lie algebra.

We also stress that until the end of Section \ref{sec.depjf},
it is not necessary to specify
$\delta_T\Phi_\alpha^a$, $\delta_M\Phi_\alpha^{ab}$,
$\delta_D\Phi_\alpha$ and $\delta_C\Phi_\alpha^a$
(explicit examples of these variations
in linearized gravity and electrodynamics
can be found in Sections \ref{sec.lg.em}, \ref{sec.gharm}
and Appendix \ref{sec.em}).

\subsubsection{Orbital and intrinsic parts of $\delta \Phi_\alpha$, $j^a$ and $K^a$}
\label{sec.Jnoether2}

In view of (\ref{eq.dPhi_lin})-(\ref{eq.j_lin}),
$\delta\Phi_\alpha$, $K^a$ and $j^a$
can be written in a form similar to (\ref{eq.j2}) or (\ref{eq.j2b}),
using (\ref{eq.ce1})-(\ref{eq.ce4}):
\begin{eqnarray}
\label{eq.dPhi_orbint}
\hspace{-8mm} \delta \Phi_\alpha = && \hspace{-5mm} \delta_T\Phi_\alpha^a\, f_a
- \check{\delta}_S\Phi_\alpha^{ab}\,\partial_{[a}f_{b]}
+ \frac{1}{1+d}\,\check{\delta}_{\mbb{T}}\Phi_\alpha\,\partial_af^a
+ \frac{1}{2(1+d)}\,\check{\delta}_{\mbb{C}}\Phi_\alpha^a\, \partial_a\partial_b f^b,\\[2mm]
\label{eq.K_orbint}
\hspace{-8mm} K^a  = && \hspace{-5mm}  K_T^{ab} f_b
- \check{K}_S^{abc}\, \partial_{[b}f_{c]}
+ \frac{1}{1+d}\, \check{K}_{\mbb{T}}^a\, \partial_bf^b
+ \frac{1}{2(1+d)}\, \check{K}_{\mbb{C}}^{ab}\, \partial_b\partial_c f^c,\\[2mm]
\label{eq.j_orbint}
\hspace{-8mm} j^a  = && \hspace{-5mm}  j_T^{ab} f_b
- \check{\jmath}_S^{\,abc}\, \partial_{[b}f_{c]}
+ \frac{1}{1+d}\, \check{\jmath}_{\mbb{T}}^{\,a}\, \partial_bf^b
+ \frac{1}{2(1+d)}\, \check{\jmath}_{\mbb{C}}^{\,ab}\, \partial_b\partial_c f^c,
\end{eqnarray}
where $\check{\delta}_S\Phi_\alpha^{ab}$, $\check{K}_S^{abc}$
and $\check{\jmath}_S^{\,abc}$ are antisymmetric:
$\check{\delta}_S\Phi_\alpha^{ab} = - \check{\delta}_S\Phi_\alpha^{ba}$,
$\check{K}_S^{abc} = - \check{K}_S^{acb}$,
$\check{\jmath}_S^{\,abc} = - \check{\jmath}_S^{\,acb}$.
(The formulae in which $\Box f^a$ appears instead of $\partial_a\partial_bf^b$ are omitted.)
The check marks appearing in (\ref{eq.dPhi_orbint})-(\ref{eq.j_orbint})
and below are intended
to indicate that
the quantities marked by them are not directly related to
Noether symmetries and currents, unlike
$\delta_T\Phi_\alpha^a$, $\delta_M\Phi_\alpha^{ab}$,
$\delta_D\Phi_\alpha$, $\delta_C\Phi_\alpha^a$,
$K_T^{ab}$, $K_M^{abc}$, \dots, $j_T^{ab}$, $j_M^{abc}$, \dots .

The relations between
$\delta_T\Phi_\alpha^a$, $\delta_M\Phi_\alpha^{ab}$, $\delta_D\Phi_\alpha$, $\delta_C\Phi_\alpha^a$
and
$\check{\delta}_S\Phi_\alpha^{ab}$, $\check{\delta}_{\mbb{T}}\Phi_\alpha$, $\check{\delta}_{\mbb{C}}\Phi_\alpha^a$
take, in analogy with (\ref{eq.bbT1})-(\ref{eq.bbT2-b}), the form
\begin{eqnarray}
\label{eq.rel-dMdD}
\delta_M\Phi_\alpha^{ab} = && \hspace{-5mm} \delta_T\Phi_\alpha^{[a}x^{b]} + \check{\delta}_S\Phi_\alpha^{ab},
\qquad\quad
\delta_D\Phi_\alpha =\, \delta_T\Phi_\alpha^ax_a + \check{\delta}_{\mbb{T}}\Phi_\alpha\, ,\\[2mm]
\label{eq.rel-dC1}
\delta_C\Phi_\alpha^a = && \hspace{-5mm} 2\delta_T\Phi_\alpha^bx_bx^a - \delta_T\Phi_\alpha^ax_bx^b + 2\check{\delta}_{\mbb{T}}\Phi_\alpha x^a
-4 \check{\delta}_S\Phi_\alpha^{ab}x_b + \check{\delta}_{\mbb{C}}\Phi_\alpha^a\, , \\[2mm]
\label{eq.rel-dC2}
\check{\delta}_{\mbb{C}}\Phi_\alpha^a = && \hspace{-5mm} \delta_C\Phi_\alpha^a - \delta_T\Phi_\alpha^ax_bx^b + 2\delta_T\Phi_\alpha^bx_bx^a
+ 4\delta_M\Phi_\alpha^{ab}x_b - 2\delta_D\Phi_\alpha x^a\, .
\end{eqnarray}
Similarly, the relations between $K_T^{ab}$, $K_M^{abc}$, $K_D^a$, $K_C^{ab}$
and
$\check{K}_S^{abc}$, $\check{K}_{\mbb{T}}^a$, $\check{K}_{\mbb{C}}^{ab}$ are
\begin{eqnarray}
\label{eq.rel-KMKD}
K_M^{abc} = && \hspace{-5mm} K_T^{a[b}x_{\phantom{T}}^{c]} + \check{K}_S^{abc},\qquad\quad
K_D^a =\, K_T^{ab}x_b + \check{K}_{\mbb{T}}^a\, ,\\[2mm]
\label{eq.rel-KC1}
K_C^{ab} = && \hspace{-5mm} 2K_T^{ac}x_cx^b - K_T^{ab}x_cx^c + 2\check{K}_{\mbb{T}}^a x^b
-4 \check{K}_S^{abc}x_c + \check{K}_{\mbb{C}}^{ab}, \\[2mm]
\label{eq.rel-KC2}
\check{K}_{\mbb{C}}^{ab} = && \hspace{-5mm} K_C^{ab} - K_T^{ab}x_cx^c + 2K_T^{ac}x_cx^b
+ 4K_M^{abc}x_c - 2K_D^a x^b\, .
\end{eqnarray}
There are also completely analogous relations between
$j_T^{ab}$, $j_M^{abc}$, $j_D^a$, $j_C^{ab}$
and
$\check{\jmath}_S^{\,abc}$, $\check{\jmath}_{\mbb{T}}^{\,a}$, $\check{\jmath}_{\mbb{C}}^{\,ab}$.

Orbital and intrinsic parts of
$\delta_M\Phi_\alpha^{ab}$, $\delta_D\Phi_\alpha$, $\delta_C\Phi_\alpha^a$
and
$K_M^{abc}$, $K_D^a$, $K_C^{ab}$
can be defined,
in analogy with (\ref{eq.morb})-(\ref{eq.cint}), as
\begin{eqnarray}
\label{eq.dMorbint}
&&
\hspace{-23mm}
\makebox[15mm][r]{$\displaystyle \check{\delta}_{M,{\mathrm{orb}}}\Phi_\alpha^{ab}$}
\makebox[55mm][l]{$\displaystyle \ = \delta_T\Phi_\alpha^{[a}x^{b]},$}
\makebox[15mm][r]{$\displaystyle \check{\delta}_{M,{\mathrm{int}}}\Phi_\alpha^{ab}$}
\makebox[50mm][l]{$\displaystyle \ = \check{\delta}_S\Phi_\alpha^{ab},$}\\[2mm]
\label{eq.dDorbint}
&&
\hspace{-23mm}
\makebox[15mm][r]{$\displaystyle \check{\delta}_{D,\mathrm{orb}}\Phi_\alpha$}
\makebox[55mm][l]{$\displaystyle \ = \delta_T\Phi_\alpha^ax_a\, ,$}
\makebox[15mm][r]{$\displaystyle \check{\delta}_{D,\mathrm{int}}\Phi_\alpha$}
\makebox[50mm][l]{$\displaystyle \ = \check{\delta}_{\mbb{T}}\Phi_\alpha\, ,$}\\[2mm]
\label{eq.dCorbint}
&&
\hspace{-23mm}
\makebox[15mm][r]{$\displaystyle \check{\delta}_{C,{\mathrm{orb}}}\Phi_\alpha^a$}
\makebox[55mm][l]{$\displaystyle \ = 2\delta_T\Phi_\alpha^bx_bx^a - \delta_T\Phi_\alpha^ax_bx^b,$}
\makebox[15mm][r]{$\displaystyle \check{\delta}_{C,{\mathrm{int}}}\Phi_\alpha^a$}
\makebox[50mm][l]{$\displaystyle \ = 2\check{\delta}_{\mbb{T}}\Phi_\alpha x^a
-4 \check{\delta}_S\Phi_\alpha^{ab}x_b + \check{\delta}_{\mbb{C}}\Phi_\alpha^a\, ,$}
\end{eqnarray}
\begin{eqnarray}
\label{eq.KMorbint}
&&
\hspace{-23mm}
\makebox[15mm][r]{$\displaystyle \check{K}_{M,{\mathrm{orb}}}^{abc}$}
\makebox[55mm][l]{$\displaystyle \ = K_T^{a[b}x_{\phantom{T}}^{c]},$}
\makebox[15mm][r]{$\displaystyle \check{K}_{M,{\mathrm{int}}}^{abc}$}
\makebox[50mm][l]{$\displaystyle \ = \check{K}_S^{abc},$}\\[2mm]
&&
\hspace{-23mm}
\makebox[15mm][r]{$\displaystyle \check{K}_{D,{\mathrm{orb}}}^a$}
\makebox[55mm][l]{$\displaystyle \ = K_T^{ab}x_b\, ,$}
\makebox[15mm][r]{$\displaystyle \check{K}_{D,{\mathrm{int}}}^a$}
\makebox[50mm][l]{$\displaystyle \ = \check{K}_{\mbb{T}}^a\, ,$}\\[2mm]
&&
\hspace{-23mm}
\makebox[15mm][r]{$\displaystyle \check{K}_{C,{\mathrm{orb}}}^{ab}$}
\makebox[55mm][l]{$\displaystyle \ = 2K_T^{ac}x_cx^b - K_T^{ab}x_cx^c,$}
\makebox[15mm][r]{$\displaystyle \check{K}_{C,{\mathrm{int}}}^{ab}$}
\makebox[50mm][l]{$\displaystyle \ = 2\check{K}_{\mbb{T}}^a x^b - 4 \check{K}_S^{abc}x_c + \check{K}_{\mbb{C}}^{ab}.$}
\end{eqnarray}
$j_M^{abc}$, $j_D^a$ and $j_C^{ab}$
also have completely analogous orbital and intrinsic parts
$\check{\jmath}_{M,{\mathrm{orb}}}^{\,abc}$, $\check{\jmath}_{D,{\mathrm{orb}}}^{\,a}$, $\check{\jmath}_{C,{\mathrm{orb}}}^{\,ab}$,
$\check{\jmath}_{M,{\mathrm{int}}}^{\,abc}$, $\check{\jmath}_{D,{\mathrm{int}}}^{\,a}$, $\check{\jmath}_{C,{\mathrm{int}}}^{\,ab}$.

\vspace{2mm}
In view of (\ref{eq.nthr}), it is clear that
generally both $j^a$ and $K^a$ give rise to
contributions to $T^{ab}$, $S^{abc}$, $\mbb{T}^a$ and $\mbb{C}^{ab}$;
\begin{eqnarray}
&& T^{ab} = j_T^{ab} - K_T^{ab},\qquad
S^{abc} = \check{\jmath}_S^{\,abc} - \check{K}_S^{abc},\\
&& \mbb{T}^a = \check{\jmath}_{\mbb{T}}^{\,a} - \check{K}_{\mbb{T}}^a,\qquad
\hspace{4mm} \mbb{C}^{ab} = \check{\jmath}_{\mbb{C}}^{\,ab} - \check{K}_{\mbb{C}}^{ab}.
\end{eqnarray}
Some of the contributions
$j_T^{ab}, \check{\jmath}_S^{\,abc},\dots $,
$K_T^{ab}, \check{K}_S^{abc},\dots $ may be zero
in special cases, which are often encountered in practice.

As can be seen from (\ref{eq.elc}),
$j_T^{ab}$, $\check{\jmath}_S^{\,abc}$, $\check{\jmath}_{\mbb{T}}^{\,a}$,
$\check{\jmath}_{\mbb{C}}^{\,ab}$
are related to
$\delta_T\Phi_\alpha^a$, $\check{\delta}_S\Phi_\alpha^{ab}$,
$\check{\delta}_{\mbb{T}}\Phi_\alpha$, $\check{\delta}_{\mbb{C}}\Phi_\alpha^a$.
If $L$ does not depend on higher than first derivatives of $\Phi_\alpha$,
then the relation is simple:
\begin{eqnarray}
\label{eq.419}
&&
\makebox[8mm][r]{$\displaystyle j_T^{ab}$}
\makebox[40mm][l]{$\displaystyle \ = \frac{\partial L}{\partial(\partial_a\Phi_\alpha)} \delta_T\Phi_\alpha^b$,}
\makebox[8mm][r]{$\displaystyle \check{\jmath}_S^{\,abc}$}
\makebox[40mm][l]{$\displaystyle \ = \frac{\partial L}{\partial(\partial_a\Phi_\alpha)} \check{\delta}_S\Phi_\alpha^{bc}$,}\\
&&
\makebox[8mm][r]{$\displaystyle \check{\jmath}_{\mbb{T}}^{\,a}$}
\makebox[40mm][l]{$\displaystyle \ = \frac{\partial L}{\partial(\partial_a\Phi_\alpha)} \check{\delta}_{\mbb{T}}\Phi_\alpha$,}
\makebox[8mm][r]{$\displaystyle \check{\jmath}_{\mbb{C}}^{\,ab}$}
\makebox[40mm][l]{$\displaystyle \ = \frac{\partial L}{\partial(\partial_a\Phi_\alpha)} \check{\delta}_{\mbb{C}}\Phi_\alpha^b$.}
\end{eqnarray}
However, if $L$ depends on higher than first derivatives of $\Phi_\alpha$,
then the relation is generally more complicated.
If, for instance, $L$ also depends on the second derivatives of $\Phi_\alpha$,
but not on higher derivatives, then
\begin{eqnarray}
j_T^{ab} \!\! & = & \!\! \left(\frac{\partial L}{\partial(\partial_a\Phi_\alpha)}
- \partial_c \frac{\partial L}{\partial(\partial_{ac}\Phi_\alpha)}\right) \delta_T\Phi_\alpha^b
+ \frac{\partial L}{\partial(\partial_{ac}\Phi_\alpha)}\partial_c\delta_T\Phi_\alpha^b\, ,\\[2mm]
\check{\jmath}_S^{\,abc} \!\! & = & \!\! \left(\frac{\partial L}{\partial(\partial_a\Phi_\alpha)}
- \partial_d \frac{\partial L}{\partial(\partial_{ad}\Phi_\alpha)}\right) \check{\delta}_S\Phi_\alpha^{bc}
+ \frac{\partial L}{\partial(\partial_{ad}\Phi_\alpha)}\partial_d \check{\delta}_S\Phi_\alpha^{bc}\\[2mm]
&& \!\! -\, \frac{1}{2}\left(\frac{\partial L}{\partial(\partial_{ab}\Phi_\alpha)}\delta_T\Phi_\alpha^c
-\frac{\partial L}{\partial(\partial_{ac}\Phi_\alpha)}\delta_T\Phi_\alpha^b\right).
\end{eqnarray}
In these relations the derivatives of
$\delta_T\Phi_\alpha^a$ and $\check{\delta}_S\Phi_\alpha^{ab}$ also appear,
and in the formula for $\check{\jmath}_S^{\,abc}$
not only $\check{\delta}_S\Phi_\alpha^{ab}$,
but also $\delta_T\Phi_\alpha^a$ is present.

\subsubsection{Dependence of $J^a$ on $f^a$}
\label{sec.depjf}

Now we turn to the question of whether the property (FT) (see Section \ref{sec.cft})
and the requirement that $J^a$ should be a Noether current are compatible.
In specific theories of interest, $\delta\Phi_\alpha$ usually has
the property (FT)\footnote{Here $J^a$ is understood
to be replaced with $\delta\Phi_\alpha$ in (FT).
Similar replacements are understood in the following
for $K^a$, $j^a$ and $\delta L$.},
moreover, usually the Lagrangian density function $L$
does not depend explicitly on $x^a$,
therefore (FT) also holds for $j^a$, as can be seen from (\ref{eq.elc}).
$K^a$ has then to have the property (FT) in order for $J^a$ to have it,
since $J^a = j^a - K^a$.
This means that $K^a$ should be required to have the property (FT);
if this requirement is fulfilled, then (FT) will hold for $J^a$.

If $K^a$ has the property (FT), then $\partial_a K^a$ also has it,
therefore in view of the relation
$\delta L = \partial_a K^a$ (see (\ref{eq.defns})),
it is an interesting question whether $\delta L$ has the property (FT).
$\delta L$ usually does have this property,
because if $\delta \Phi_\alpha$
has the property (FT) and
$L$ does not depend explicitly on $x^a$
(which is the usual case),
then from (\ref{eq.nthr1b}) it follows that
$\delta L$ has the property (FT).

In the subsequent sections on linearized gravity
and electrodynamics,
the variations of the basic dynamical fields
($h_{ab}$ and $A_a$) under conformal transformations
and the corresponding $K^a$
will be homogeneous linear local functions of $f^a$
without explicit dependence on $x^a$,
and the Lagrangian densities will not depend explicitly on $x^a$,
therefore $J^a$ will have the property (FT).

\subsubsection{Relation of $J^a$ to canonical currents}
\label{sec.relcan}

The members of the $5$-parameter family of energy-momentum tensors
of the linearized gravitational field
we considered in \cite{TG}
can be written in the generic form
\begin{equation}
\label{eq.Tgen}
T^{ab} = \Tc^{ab} + T_{\mathrm{g}}^{ab} + \partial_c \Sigma^{acb},
\end{equation}
where $\Tc^{ab}$ is a canonical energy-momentum tensor,
$T_{\mathrm{g}}^{ab}$ is a term associated (via Noether's theorem) with certain
field-dependent gauge transformations,
and $\Sigma^{acb}$ is a tensor that is antisymmetric in its first two indices.
$\partial_c \Sigma^{acb}$ is obviously strongly conserved
due to the antisymmetry of $\Sigma^{acb}$
in the first two indices, whereas $T_{\mathrm{g}}^{ab}$ is trivially conserved.
The trivial conservation of $T_{\mathrm{g}}^{ab}$ is in accordance with
the well-known fact that the Noether currents associated with gauge symmetries
are trivially conserved.

The \emph{canonical energy-momentum tensor} that follows
from a Lagrangian density $L$ is given by the formula
\begin{eqnarray}
\Tc^{ab} & = &
\frac{\partial L}{\partial(\partial_a\Phi_\alpha)}\partial^b\Phi_\alpha
+ \biggl(\frac{\partial L}{\partial (\partial_{ac}\Phi_\alpha)}\partial_c \partial^b \Phi_\alpha
-\partial_c\frac{\partial L}{\partial (\partial_{ac}\Phi_\alpha)} \partial^b \Phi_\alpha\biggr)\nonumber\\[2mm]
&& +\, \biggl(\frac{\partial L}{\partial (\partial_{acd}\Phi_\alpha)}\partial_{cd}\partial^b\Phi_\alpha
-\partial_c\frac{\partial L}{\partial (\partial_{acd}\Phi_\alpha)}\partial_d\partial^b\Phi_\alpha
+\partial_{cd}\frac{\partial L}{\partial (\partial_{acd}\Phi_\alpha)}\partial^b\Phi_\alpha\biggr)+\dots \nonumber \\[2mm]
&& -\, \eta^{ab}L.
\end{eqnarray}
$\Tc^{ab}$ is conserved if $L$ does not depend explicitly on $x^a$.
In the latter case $\Tc^{ab}e_b$ is associated with space and time translations
under which the variation of $\Phi_\alpha$ is
understood to be $\delta \Phi_\alpha = e^a\partial_a\Phi_\alpha$,
and the choice for $K^a$ needed to obtain $\Tc^{ab}e_b$
from the general formula (\ref{eq.nthr}) is $K^a=e^a L$.
$\Tc^{ab}$ depends on $L$; if a total divergence,
that does not depend explicitly on $x^a$,
is added to the Lagrangian density,
then $\Tc^{ab}$ shifts by a strongly conserved term.
(In (\ref{eq.Tgen}), $\Tc^{ab}$ is understood to follow from a fixed Lagrangian density.)

$T_{\mathrm{g}}^{ab}$ and $\partial_c \Sigma^{acb}$ may be regarded as
`improvement' terms if $T^{ab}$ has
better properties in some respect than $\Tc^{ab}$.
If $T_{\mathrm{g}}^{ab}\ne 0$, then $T^{ab}e_b$ is not associated, as a Noether current,
with simple spacetime translations
(under which $\delta \Phi_\alpha = e^a\partial_a\Phi_\alpha$),
but with translations combined with (field-dependent) gauge transformations.

In the present paper, the scheme (\ref{eq.Tgen})
is generalized to conformal symmetry,
i.e.\ we construct currents associated with conformal symmetry
(see (\ref{eq.TMDC}))
that take the form
\begin{equation}
\label{eq.Jgen}
J^a = \Jc^a + J_{\mathrm{g}}^a + \partial_b \Sigma^{ab},
\end{equation}
where $\Jc^a$ is `canonical' in a suitable sense,
$J_{\mathrm{g}}^a$ is associated with suitable field-dependent gauge transformations,
and $\Sigma^{ab}$ is antisymmetric.
The three terms on the right hand side will separately have the property (FT).
$\Jc^a$ and $J_{\mathrm{g}}^a$ will be Noether currents,
therefore $J^a$ will also be a Noether current
(i.e., a Noether current valued linear function of the parameters
$e^a$, $\omega^{ab}$, $\beta$, $c^a$).
As a Noether current, $J^a$ will be associated
with the variation that is the sum of the
variations with which $\Jc^a$ and $J_{\mathrm{g}}^a$ are associated,
i.e., it will be associated with ordinary conformal transformations
combined with field-dependent gauge transformations.

\section{Linearized gravity}
\label{sec.lingrav}

In this section a brief overview of the formulation of linearized gravity
in the framework of Lagrangian field theory is presented,
focusing on those parts that will be needed in the subsequent sections.
More details can be found in \cite{TG}.
The formulae involving the Fierz tensor
and the $2$-parameter family of Lagrangian densities (\ref{eq.lg.l})
are of particular importance.

The field equations for the linearized gravitational field $h_{ab}$
are the linearized Einstein equations
\begin{equation}
\label{eq.linein}
G^{ab}  = \mc{T}^{ab},
\end{equation}
where $G^{ab} = R^{ab} - (1/2)R\eta^{ab}$
is the linearized Einstein tensor
and $\mc{T}^{ab}$ is the matter energy-momentum tensor of the linearized theory.
$\mc{T}^{ab}$ is a divergenceless symmetric tensor field.
With the exception of Sections \ref{sec.divjmat} and \ref{sec.divcmat},
$\mc{T}^{ab}=0$ will be assumed.
Expressions for the linearized Riemann tensor, Ricci tensor, Ricci scalar,
and Christoffel symbols,
which will be denoted by $R_{abcd}$, $R_{ab}$, $R$ and $\Gamma_{abc}$,
can be found in Appendix B of \cite{TG}.
$h_a{}^a$ will be denoted by $h$.

Those Lagrangian densities for the linearized Einstein equations without matter
that are Lorentz covariant, differ from the usual
Fierz--Pauli Lagrangian density $L_{\mathrm{FP}}$
(see equations (2.2) and (2.3) in Section 2 of \cite{TG}, for instance)
by a total divergence and are homogeneous quadratic expressions in $\partial_a h_{bc}$
with constant coefficients are
\begin{equation}
L=L_1+\alpha_2 L_2+\alpha_3 L_3\, ,
\label{eq.lg.l}
\end{equation}
where $\alpha_2$ and $\alpha_3$ are arbitrary constants and
\begin{eqnarray}
&& \hspace{-1cm} L_1 \ = \ \frac{1}{4}\partial_a h_{bc}\partial^a h^{bc}
-\frac{1}{4}(\partial_c h^{ba}\partial_b h^c{}_a + \partial_b h^{ba}\partial_c h^c{}_a)
+\frac{1}{2}\partial_a h^{ab} \partial_b h
-\frac{1}{4}\partial_a h \partial^a h
\label{eq.lg.l1} \\[1mm]
&& \hspace{-1cm} L_2 \ = \ \partial_c h^{ba}\partial_b h^c{}_a
- \partial_b h^{ba}\partial_c h^c{}_a
 \ = \ \partial_c [h^{ba}\partial_b h^c{}_a - h^{ca}\partial_b h^b{}_a]
\label{eq.lg.l2} \\[1mm]
&& \hspace{-1cm} L_3 \ = \ \frac{1}{4}\epsilon^{abcd}\partial_a h_{be} \partial_c h_d{}^e \ = \
\frac{1}{4}\partial_a [\epsilon^{abcd}h_{be} \partial_c h_d{}^e ].
\label{eq.lg.l3}
\end{eqnarray}
$L_2$ and $L_3$ are total divergences and the Fierz--Pauli Lagrangian density
corresponds to $\alpha_2 = - 1/4$, $\alpha_3=0$, i.e.\
$L_{\mathrm{FP}} = L_1 - (1/4)L_2$.

$G^{ab}$ is invariant under the gauge transformations
\begin{equation}
\label{eq.lg.g}
h_{ab} \to h_{ab}+ \partial_a\phi_b+\partial_b\phi_a\, ,
\end{equation}
where $\phi_a$ is an arbitrary covector field.
These gauge transformations are also Noether symmetries of $L$,
but $L$ is not strictly gauge-invariant for any value of $\alpha_2$, $\alpha_3$.
$L_1$ and $L_3$ are strictly invariant
under `scalar' gauge transformations,
which are special gauge transformations having the form
\begin{equation}
h_{ab} \to h_{ab}+ \partial_{ab}\phi\, ,
\end{equation}
where $\phi$ is an arbitrary scalar field.

It is useful to introduce the following tensors:
\begin{eqnarray}
F_{abc} & \!\! = \!\! & \partial_{[a} h_{b]c}
+\partial_d h^d{}_{[a}\eta_{b]c}
-\partial_{[a} h\hspace{0.3mm}\eta_{b]c}
\label{eq.lg.x6} \\[1mm]
\label{eq.circx}
\mathring{F}_{abc} & \!\! = \!\! &
\partial_{[a} h_{b]c} \, .
\end{eqnarray}
$F_{abc}$ is called \emph{Fierz tensor}.
It has the algebraic symmetries
\begin{equation}
F_{abc} \ = \ -F_{bac}
\label{eq.lg.x1}
\end{equation}
\begin{equation}
F_{abc} + F_{bca} + F_{cab} \ = \ 0.
\label{eq.lg.x2}
\end{equation}
$\mathring{F}_{abc}$ also has these symmetries.
(\ref{eq.lg.x2}) can also be written as
\begin{equation}
\tilde{F}_{ab}{}^b=0.
\label{eq.lg.x2b}
\end{equation}
There is a bijective algebraic relation between 
$F_{abc}$ and $\mathring{F}_{abc}$:
\begin{eqnarray}
\label{eq.circ2}
\mathring{F}_{abc}
 & \!\! = \!\! & F_{abc} - F_{[a}\eta_{b]c}\\[1mm]
F_{abc} & \!\! = \!\! & \mathring{F}_{abc} - 2\mathring{F}_{[a}\eta_{b]c}\, ,
\end{eqnarray}
where the notation $F_a=F_{ab}{}^b$, $\mathring{F}_a=\mathring{F}_{ab}{}^b$ is used.
$F_{abc}$ and $\mathring{F}_{abc}$ are invariant under scalar gauge transformations.

Further important properties of these tensors,
which hold independently of the linearized Einstein field equations, are
\begin{equation}
\partial_c F^c{}_{ab}=-G_{ab},\qquad \partial_c \tilde{\mathring{F}}^c{}_{ab}=0
\label{eq.lg.x3}
\end{equation}
\begin{equation}
\label{eq.lg.x3b}
\partial_c F_{ab}{}^c=0
\end{equation}
\begin{equation}
F_a = \partial_b h^b{}_a - \partial_a h,\qquad  \mathring{F}_a = -\frac{1}{2}F_a
\label{eq.lg.x4}
\end{equation}
\begin{equation}
\label{eq.lg.tftr}
\epsilon^{dabc}\tilde{F}_{abc}=2F^d.
\end{equation}
From (\ref{eq.lg.tftr}) it follows that $\tilde{F}_{abc}$
has the property (\ref{eq.lg.x2}) only if the trace of $F_{abc}$ is zero.

$L_1$ and $L_3$ can be expressed in the following forms
using $F_{abc}$ and $\mathring{F}_{abc} $:
\begin{eqnarray}
\label{eq.lg.x7}
&& L_1 \ = \ \frac{1}{2}( F_{abc}F^{abc} - F_a F^a )
\ = \ \frac{1}{2}\mathring{F}_{abc}\mathring{F}^{abc} - \mathring{F}_a\mathring{F}^a
\ = \ \frac{1}{2}F_{abc}\mathring{F}^{abc} \\[1mm]
\label{eq.lg.l3b}
&& L_3 \ = \ \frac{1}{4}\epsilon^{abcd}\mathring{F}_{abe}\mathring{F}_{cd}{}^e.
\end{eqnarray}
$L_2$ is not scalar gauge-invariant, therefore it cannot be expressed
in terms of $F_{abc}$ or $\mathring{F}_{abc}$.

$L_1$ is distinguished within the family of Lagrangian densities (\ref{eq.lg.l})
by scalar gauge-invariance and
the absence of $\epsilon^{abcd}$ in the expression (\ref{eq.lg.l1}).
In this paper $L_1$ will serve as the `primary' Lagrangian density
of the linearized gravitational field,
instead of the Fierz--Pauli Lagrangian density.

For calculating conserved currents, it is useful to note that
\begin{eqnarray}
\label{eq.lg.dl1ddg}
\frac{\partial L_1}{\partial (\partial_c h_{ab})} & \! = \! & F^{c(ab)} \\[1mm]
\frac{\partial L_2}{\partial (\partial_c h_{ab})} & \! = \! &
2(\partial^{(a} h^{b)c}
-\eta^{c(a}\partial_d h^{b)d}) \\[1mm]
\frac{\partial L_3}{\partial (\partial_c h_{ab})} & \! = \! &
\tilde{\mathring{F}}^{c(ab)}.
\end{eqnarray}
The Euler--Lagrange derivative of $L$ with respect to $h_{ab}$ is
\begin{equation}
\label{eq.lg.x8}
\mr{E}(L)^{ab}=
-\partial_c F^{cab} = G^{ab}.
\end{equation}
The linearized Einstein equations can thus be written as
\begin{equation}
\label{eq.lg.x9}
-\partial_c F^{cab} = \mc{T}^{ab}.
\end{equation}

\subsection{Gauge fixing conditions}
\label{sec.gfcond}

Gauge fixing has an important role in linearized gravity;
several favourable properties of $T_{\mathrm{lg}}^{cd}$ or $\tau^{cd}$, in particular, are manifested only under suitable gauge fixing conditions.
The gauges that we consider in the present paper are the
generalized harmonic gauges,
the constant trace harmonic gauge,
the traceless harmonic gauge,
the transverse traceless gauge,
the $F_a=0$, $F_{ab0}=0$ gauge,
and subgauges of these gauges.

The \emph{generalized harmonic gauge} condition is
\begin{equation}
\label{eq.ghg}
X^b(\chi)=\partial_a(h^{ab}-\chi\eta^{ab}h)=0,
\end{equation}
where $\chi$ is a parameter that can take any real value
and the first equation defines $X^a$.
$\chi=\frac{1}{2}$ corresponds to the usual \emph{harmonic gauge},
whereas in the case $\chi=1$ the gauge fixing condition can also be written as $F_a=0$.
The $\chi=1$ case is special as it preserves scalar gauge symmetry
and requires $\mc{T}_a{}^a = 0$.
In generalized harmonic gauges 
\begin{equation}
\label{eq.faghg}
F_a=(\chi-1)\partial_a h
\end{equation}
holds.

From the gauge fixing condition (\ref{eq.ghg}) it can be seen that
if a particular $h_{ab}$ field is contained by more than one
generalized harmonic gauges,
then it is contained by all generalized harmonic gauges.
The intersection of all generalized harmonic gauges is the gauge
defined by the conditions
\begin{equation}
\label{eq.sg}
\partial_a h^{ab} = 0,\qquad \partial_a h = 0.
\end{equation}
It will be called \emph{constant trace harmonic gauge} (CTH gauge).
The \emph{traceless harmonic gauge} is characterized by
the conditions $\partial_a h^{ab}=0$, $h=0$.
The CTH gauge has theoretical significance, however
a $h_{ab}$ field with constant but nonzero $h$ does not appear to be physically desirable.

The \emph{transverse traceless gauge} (TT gauge)
is defined by the conditions  $\partial_a h^{ab}=0$, $h=0$,
and $h_{i0}=0$, $i=1,2,3$. $h_{00}=0$ can also be achieved under general conditions \cite{WaldGR,FlHu},
therefore we understand the TT gauge conditions to include
the condition $h_{a0}=0$ instead of $h_{i0}=0$.

It is also useful to consider the gauge defined by the conditions $F_a=0$, $F_{ab0}=0$,
as it contains the TT gauge, respects scalar gauge symmetry,
and is sufficient for the positivity of $T_{\mathrm{lg}}^{cd}$.

In the $F_a=0$ gauge
$T_{\mathrm{lg}}^{ab}$ can be written as
\begin{equation}
\label{eq.Tlg-sym}
T_{\mathrm{lg}}^{ab} =  2\left(F^{acd}F^b{}_{cd}
- \frac{1}{4} \eta^{ab} F^{ecd}F_{ecd}\right).
\end{equation}
The tensor on the right hand side of this formula is evidently symmetric and traceless.
(\ref{eq.Tlg-sym}) is also evidently valid in the subgauges of the $F_a=0$ gauge,
in particular in the $F_a=0$, $F_{ab0}=0$ and the TT gauges.

\section{Conserved currents associated with Poincar\'e and dilatation symmetry
in linearized gravity}
\label{sec.lg.em}

In this section, the transformation properties of $h_{ab}$,
$L$ ($L$ denotes the Lagrangian density (\ref{eq.lg.l}) of
linearized gravity without matter)
and the linearized Einstein field equations
under conformal transformations are discussed first.
It is found, in agreement with other authors
(see \cite{FarHinHul}, for example),
that $L$ has dilatation symmetry,
in addition to its Poincar\'e symmetry,
but it does not appear to have complete conformal symmetry.
It is also verified that the properties of
the linearized Einstein field equations
with respect to conformal symmetry are similar.
Then, in the remaining, most extensive parts of the section,
the conserved currents
associated with Poincar\'e transformations and dilatations
are discussed.

Following the scheme outlined in Section \ref{sec.struJ}
(see, in particular, (\ref{eq.Tgen})-(\ref{eq.Jgen})),
canonical currents, correction terms related to gauge symmetry,
and further correction terms that follow from a superpotential are constructed.
As a result, a general current $J^c$
associated with Poincar\'e-dilatation symmetry,
containing several free parameters, is obtained.

Three remarkable special cases, namely the extension of
$T_{\mathrm{lg}}^{cd}$, $\tau^{cd}$ and $T_{\mathrm{LL}}^{cd}$
into conserved currents
associated with Poincar\'e-dilatation symmetry,
are also discussed (see Sections \ref{sec.gentlg},
\ref{sec.bhl}, \ref{sec.ll}).
The extension of $T_{\mathrm{lg}}^{cd}$ is a central part of the paper
and it is a step in the construction of $J^a$.

\subsection{Poincar\'e and dilatation symmetry of linearized gravity}
\label{sec.pdsym}

The Lie differentiation rule of ordinary tensor fields suggests
that the variation of $h_{ab}$ under (infinitesimal) conformal transformations
should be
\begin{equation}
\label{eq.lg.conf1}
\delta h_{ab} \ = \ \mathcal{L}_f h_{ab} \ = \
f^c \partial_c h_{ab} + h_{cb} \partial_a f^c + h_{ac} \partial_b f^c,
\end{equation}
where $f^c$ is the generator introduced in (\ref{eq.f}).
It is not difficult to verify that (\ref{eq.lg.conf1})
specifies a Noether symmetry of $L$ if $f^c$ generates Poincar\'e transformations,
but for dilatation and special conformal transformation generators
the $\delta L$ that corresponds to (\ref{eq.lg.conf1}) is not a divergence,
as can be verified by calculating the Euler--Lagrange derivative of $\delta L$ with respect to $h_{ab}$. 
Nevertheless, a somewhat more general type of variations of $h_{ab}$ may also be considered,
namely
\begin{equation}
\label{eq.lg.conf2gen}
\delta h_{ab} \ = \ f^c \partial_c h_{ab} + h_{cb} \partial_a f^c + h_{ac} \partial_b f^c + w\partial_c f^c h_{ab}\, ,
\end{equation}
where $w$ is a constant.
This type of variations is relevant to weighted tensor fields.
The commutation relations of the variations (\ref{eq.lg.conf2gen})
are not changed by the term $w\partial_c f^c h_{ab}$, i.e.,
they remain those of the conformal Lie algebra.
In the case of Poincar\'e transformations,
$w\partial_c f^c h_{ab}$ vanishes.
One finds that if $\delta h_{ab}$ is chosen, in particular, to be
\begin{equation}
\label{eq.lg.conf2}
\delta h_{ab} \ = \ f^c \partial_c h_{ab} + h_{cb} \partial_a f^c + h_{ac} \partial_b f^c - \frac{1}{4}\partial_c f^c h_{ab}\, ,
\end{equation}
then $\delta L$ is a divergence not only for Poincar\'e transformation generators,
but for dilatation generators as well.
Explicitly,
\begin{eqnarray}
\label{eq.lg.dlI}
\delta L_1 & \! = \! & \partial_c (f^c L_1) - \frac{1}{2} F_a h^{ab}\Box f_b \\[1mm]
\label{eq.lg.dlII}
\delta L_2 & \! = \! & \partial_c (f^c L_2)
- \frac{1}{2} \partial_c (h^{ab}h_{ab}\Box f^c + 4h^{ba}h^c{}_a \Box f_b - 2h h^{ca}\Box f_a)\\[1mm]
\label{eq.lg.dlIII}
\delta L_3 & \! = \! & \partial_c (f^c L_3)
\end{eqnarray}
holds for arbitrary conformal transformation generators,
and the only term on the right hand side of these equations that is not a divergence is
$\frac{1}{2} F_a h^{ab}\Box f_b$, which is zero
for Poincar\'e transformation and dilatation generators.

Regarding the conformal symmetry of the linearized Einstein equations without matter,
one finds that the variation of $G_{ab}$ due to the variation (\ref{eq.lg.conf2gen})
of $h_{ab}$ is
\begin{eqnarray}
\label{eq.dG}
\delta G_{ab} & \! = \! & f^c\partial_c G_{ab} +  G_{cb}\partial_af^c +  G_{ac}\partial_bf^c
+ \left(w+\frac{1}{2}\right)\partial_cf^c G_{ab} \nonumber\\[1mm]
&& +\,\frac{1}{2}\partial_{(a}h_{b)c}\Box f^c
+ (4w + 1)\mathring{F}_{c(ab)}\Box f^c \nonumber\\[1mm]
&& +\, (4w + 2)\mathring{F}_{(a}\Box f_{b)}
-(4w + 1)\eta_{ab}\mathring{F}_c \Box f^c
- \frac{1}{2}\eta_{ab}\partial_ch^{cd}\Box f_d\, .
\end{eqnarray}
This result shows that the linearized Einstein equations have Poincar\'e and dilatation symmetry,
i.e., if $h_{ab}$ is a solution of the linearized Einstein equations and
$f^c = e^c + \omega^{cd}x_d + \beta x^c$,
then $\delta h_{ab}$ is also a solution of the linearized Einstein equations.
Moreover, the value of $w$ can be arbitrary here.
On the other hand, (\ref{eq.dG}) shows that generally special conformal transformations
are not symmetries of the linearized Einstein equations.

Since special conformal transformations are not Noether symmetries of $L$
even if (\ref{eq.lg.conf2}) is adopted,
we shall consider only Poincar\'e transformations and dilatations
in this section,
and take (\ref{eq.lg.conf2}) to be the variation of $h_{ab}$ under these transformations.
Excluding special conformal transformations implies
$f^c = e^c + \omega^{cd}x_d + \beta x^c$
and has the consequence that $\partial_{ab}f^c = 0$,
and (\ref{eq.lg.dlI})-(\ref{eq.lg.dlIII}) reduce to the standard form
\begin{equation}
\label{eq.dLi-st}
\delta L_i = \partial_c (f^c L_i),\qquad i=1,2,3.
\end{equation}

Later (\ref{eq.lg.conf2}) will be combined with variations
corresponding to gauge transformations
(see Sections \ref{sec.gentlg}, \ref{sec.gen}, \ref{sec.ll}),
and (\ref{eq.lg.conf2}) will be referred to as `simple' variation
to distinguish it from the combined variations.

Using the conformal Killing equation (\ref{eq.ck}), (\ref{eq.lg.conf2})
can be rewritten (not excluding special conformal transformations) as
\begin{equation}
\label{eq.lg.conf2b}
\delta h_{ab} \ = \ f^c \partial_c h_{ab}
+ 2h^{[d}{}_{(a}\eta^{c]}{}_{b)}\partial_{[c}f_{d]}
+ \frac{1}{4}h_{ab}\partial_cf^c.
\end{equation}
This formula has the standard form (\ref{eq.dPhi_orbint}), so
$\delta_T h_{ab}{}^c$, $\check{\delta}_S h_{ab}{}^{cd}$ and $\check{\delta}_{\mathbb{T}}h_{ab}$
can be read off:
\begin{eqnarray}
\delta h_{ab} & \! = \! & \delta_T h_{ab}{}^c\, f_c
- \check{\delta}_S h_{ab}{}^{cd}\, \partial_{[c}f_{d]}
+ \frac{1}{4}\check{\delta}_{\mathbb{T}}h_{ab}\, \partial_cf^c
\qquad (\check{\delta}_S h_{ab}{}^{(cd)} = 0) \\[1mm]
\label{eq.hdT}
\delta_T h_{ab}{}^c & \! = \! &  \partial^c h_{ab} \\[1mm]
\label{eq.hdS}
\check{\delta}_S h_{ab}{}^{cd} & \! = \! &
2h^{[c}{}_{(a}\eta^{d]}{}_{b)} \\[1mm]
\label{eq.hdbbT}
\check{\delta}_{\mathbb{T}}h_{ab} & \! = \! & h_{ab}\, .
\end{eqnarray}
According to the general formula (\ref{eq.rel-dMdD}),
the variation of $h_{ab}$ under Lorentz transformations is
\begin{eqnarray}
\label{eq.lg.lorentz}
\delta_M h_{ab}{}^{cd} & \! = \! & \delta_T h_{ab}{}^{[c} x^{d]} + \check{\delta}_S h_{ab}{}^{cd}\\[1mm]
& \! = \! & x^{[d} \partial^{c]} h_{ab}
+ 2h^{[c}{}_{(a}\eta^{d]}{}_{b)}\, .
\end{eqnarray}
For dilatations (see again (\ref{eq.rel-dMdD})),
\begin{equation}
\label{eq.hscalingdim}
\delta_D h_{ab} \ =\ x_c \delta_T h_{ab}{}^c
+ \check{\delta}_{\mathbb{T}}h_{ab}
\ =\ x^c\partial_ch_{ab} + h_{ab}\, .
\end{equation}
In particular, the scaling weight of $h_{ab}$, which is, by definition,
the coefficient of $h_{ab}$ in (\ref{eq.hscalingdim}), is $1$.
The variations specified by (\ref{eq.hdT})-(\ref{eq.hscalingdim})
are usual
variations for a symmetric tensor field under translations,
Lorentz transformations and dilatations.
The orbital and intrinsic parts of $\delta_M h_{ab}{}^{cd}$
and $\delta_D h_{ab}$
(defined by (\ref{eq.dMorbint}) and (\ref{eq.dDorbint}))
clearly correspond to usual orbital and internal
Lorentz transformations and dilatations.

We note that from (\ref{eq.lg.conf2gen}) one gets
\begin{equation}
\label{eq.hscalingdim2}
\delta_D h_{ab} \ = \ x^c\partial_ch_{ab}+2h_{ab}+4wh_{ab}\, ,
\end{equation}
i.e.\ for general values of $w$, the scaling weight of $h_{ab}$ is $2+4w$.

The deviation of $\delta h_{ab}$ from (\ref{eq.lg.conf1}),
i.e., the presence of the last term on the right hand side of (\ref{eq.lg.conf2}),
is not a special feature of the linearized gravitational field;
a similar deviation can be found in the case of the
massless scalar field
(whose field equation is $\Box\phi = 0$)
as well.

\subsection{Canonical currents}
\label{sec.cc}

Generalizing the definition of the canonical energy-momentum tensor
mentioned in Section \ref{sec.relcan},
one defines the \emph{canonical Noether current
associated with Poincar\'e-dilatation symmetry,
following from a given Lagrangian density $L$ of linearized gravity},
to be the Noether current associated with the variation (\ref{eq.lg.conf2}) of $h_{ab}$, obtained with the choice $K^c = f^cL$ for $K^c$.
$L$ is assumed here to be a member of
the $2$-parameter family of Lagrangian densities given by (\ref{eq.lg.l}).
The choice $K^c = f^cL$ is possible according to (\ref{eq.dLi-st}).

From the Lagrangian densities (\ref{eq.lg.l}),
one obtains the $2$-parameter family of canonical Noether currents
\begin{equation}
\label{eq.lg.jc}
\Jc^c = \Jc_1^c + \alpha_2 \Jc_2^c + \alpha_3 \Jc_2^c ,
\end{equation}
where
\begin{eqnarray}
\label{eq.Jc1a}
\Jc_1^c\hspace{-5.5mm} && =  F^{cab} \delta h_{ab} -
 \frac{1}{2}f^c F^{dab}\mathring{F}_{dab} \\[2mm]
\label{eq.Jc2a}
\Jc_2^c\hspace{-5.5mm} && = 2(\partial^a h^{bc}\delta h_{ab}
- \partial_b h^{ba}\delta h_a{}^c) - f^c L_2 \\[2mm]
\label{eq.Jc3}
\Jc_3^c\hspace{-5.5mm} && = \tilde{\mathring{F}}^{cab}\delta h_{ab}
- \frac{1}{2}f^c \tilde{\mathring{F}}_{dab}\mathring{F}^{dab}.
\end{eqnarray}
$\Jc_2^c$ and $\Jc_3^c$ are the contributions following from $L_2$ and $L_3$, respectively,
and $\Jc_1^c$ is the canonical current following from $L_1$.
$\delta h_{ab}$ is the variation given by (\ref{eq.lg.conf2}).
As $L_2$ and $L_3$ are total divergences, $\Jc_2^c$ and $\Jc_3^c$ are strongly conserved
and have superpotentials:
\begin{equation}
\Jc_2^c = \partial_d\Sigma^{cd}_2,\qquad \Jc_3^c = \partial_d\Sigma^{cd}_3,
\end{equation}
with
\begin{eqnarray}
\label{eq.lg.jc1}
&& \hspace{-8mm}\Sigma^{cd}_2 = 2f^a h^{b[c}\partial_a h^{d]}{}_b
+ 2\partial^a h^{b[c} f^{d]} h_{ab}
+ 2\partial_a h^{ab} f^{[c} h^{d]}{}_b  + \Sigma^{cd}_{2'} + \Sigma^{cd}_{2''} \\[2mm]
&& \hspace{-8mm}\Sigma^{cd}_{2'} = 2\partial^a f^{[c} h^{d]b}h_{ab}\, ,\qquad
\Sigma^{cd}_{2''} = 2\partial_a f_b h^{a[c}h^{d]b}\, ,
\label{eq.sigma22}
\end{eqnarray}
and
\begin{eqnarray}
\label{eq.lg.jc2}
&& \Sigma^{cd}_3 = \tilde{\mathring{F}}^{cdb}f^a h_{ab}  +\Sigma_{3'}^{cd} \\
\label{eq.lg.sigma3p}
&& \Sigma_{3'}^{cd} = - \frac{1}{4}\epsilon^{cdab}h_b{}^f \partial_f f^e h_{ae}\, .
\end{eqnarray}

The energy-momentum tensor contained by $\Jc^c$ is
\begin{equation}
\Tc^{cd} = F^{cab}\partial^dh_{ab}-\frac{1}{2}\eta^{cd}F^{eab}\mathring{F}_{eab}
+ \alpha_2 \partial_a U^{cad}_2
+ \alpha_3\tilde{\mathring{F}}^{cab}\partial_a h_b{}^d,
\label{eq.lg.emtc}
\end{equation}
where the superpotential $U_2^{cad}$ is
\begin{equation}
U_2^{cad} = 2\partial^d h_b{}^{[a} h^{c]b}
+ 2\partial^b h^{e[c} h_{be}\eta^{a]d}
+ 2\partial_b h^{be} h^{[a}{}_e \eta^{c]d}\, .
\end{equation}

$\Jc^c$ has a nontrivial intrinsic part;
the spin tensor and the virial current contained by $\Jc^c$ are
\begin{eqnarray}
\Sc^{cde} & \!\! = \!\! & \mc{A}_{de}\bigl[ (F^{cae} + F^{cea}) h_a{}^d \nonumber\\[1mm]
&& \hspace{1cm} +\,2\alpha_2(h^{ad}\partial^e h^c{}_a  + \eta^{dc}h_b{}^e \partial_a h^{ab} 
+ \partial_a (h^{ad}h^{ce})) \nonumber\\[1mm]
&& \hspace{1cm} +\, \alpha_3(\tilde{\mathring{F}}^{cea}
+ \tilde{\mathring{F}}^{cae})h_a{}^d\,  \bigr]
\label{eq.lg.scan} \\[2mm]
\bbTc^c & \!\! = \!\! &  F^{cab}h_{ab}
+ 2\alpha_2 (h_{ab}\partial^a h^{bc} - h^{cb}\partial_a h^a{}_b)
+ \alpha_3 \tilde{\mathring{F}}^{cab}h_{ab}\, .
\label{eq.cintrdil}
\end{eqnarray}
In (\ref{eq.lg.scan}), $\mc{A}_{de}[ ... ]$ denotes the antisymmetrization (including a division by $2$)
of the tensor in the brackets with respect to the indices $d,e$.

The canonical angular momentum tensor and the dilatation current
contained by $\Jc^c$ are given by the general relations in
(\ref{eq.bbT1}), i.e.
\begin{equation}
\Mc^{cde} = \Tc^{c[d}x^{e]} + \Sc^{cde},\qquad
\Dc^c = \Tc^{cd}x_d + \bbTc^c.
\end{equation}
$\Mc^{cde}$ and $\Dc^c$ are associated with the variations (\ref{eq.lg.lorentz})-(\ref{eq.hscalingdim}).

$\Jc^c$ can also be regarded
as a $2$-parameter family of Noether currents
associated with the Poincar\'e-dilatation symmetry
of the special Lagrangian densities $L_1$,
with $K^c=f^c L_1 - \alpha_2 \partial_d\Sigma^{cd}_2 - \alpha_3 \partial_d\Sigma^{cd}_3$. 
From this point of view, however, $\Jc^c$ is a canonical current only if $\alpha_2=\alpha_3=0$,
since otherwise $K^c$ does not have the form required for a canonical current.

\subsection{Correction terms related to gauge symmetry}
\label{sec.gc}

In \cite{TG}, we found four trivially conserved quadratic correction terms
for the energy-momentum tensor that are related to the gauge symmetry of linearized gravity.
Here we generalize these terms to correction terms for currents
associated with Poincar\'e-dilatation symmetry.

We obtained the following family of Noether currents associated with the
gauge transformations (\ref{eq.lg.g}) in \cite{TG}:
\begin{eqnarray}
J^c_{\mathrm{g}} & = &  
2F^{cab} \partial_a \phi_b + \alpha_3 \tilde{\mathring{F}}^{cab} \partial_a\phi_b\\
& = & 2( \partial_a [F^{cab} \phi_b] -   \partial_a F^{cab} \phi_b)
+ \alpha_3 \partial_a [\tilde{\mathring{F}}^{cab} \phi_b].
\label{eq.lg.jg0}
\end{eqnarray}
The parameter $\alpha_3$ in this formula can take any value,
thus $J^c_{\mathrm{g}}$ comprises two independent currents
(which depend on $\phi_a$).
$J^c_{\mathrm{g}}$ is a family of trivially conserved currents as
the first and third terms in (\ref{eq.lg.jg0}) are strongly conserved
and the second term is weakly zero.
It does not depend on higher than first derivatives of $h_{ab}$.
If the Lagrangian density is taken to be $L_1$,
then the $K^c$ quantity for $J^c_{\mathrm{g}}$ is
\begin{equation}
\label{eq.Kg}
K_{\mathrm{g}}^c = -\partial_a\phi_bF^{bac} - \alpha_3 \tilde{\mathring{F}}^{cab}\partial_a\phi_b\, .
\end{equation}

The correction terms for the energy-momentum tensor
were obtained from $J^c_{\mathrm{g}}$ by the replacements
$\phi_a = h_{ab} e^b$, $\phi_a = h e_a$ in \cite{TG}.
These can be generalized as
\begin{equation}
\label{eq.lg.repl}
\phi_a = h_{ab} f^b,\qquad \phi_a = h f_a\, ;
\end{equation}
by these replacements one obtains the conserved currents
\begin{equation}
J^c_{\mathrm{g1}} = 2F^{cda}\partial_d (h_{ab} f^b) 
+  \alpha_3 \partial_d [ \tilde{\mathring{F}}^{cda} h_{ab} f^b]
\label{eq.lg.jg1}
\end{equation}
and
\begin{equation}
J^c_{\mathrm{g2}} = 2F^{cda} \partial_d (h f_a) +
\alpha_3 \partial_d [ \tilde{\mathring{F}}^{cda} h f_a],
\label{eq.lg.jg2}
\end{equation}
respectively. Since $\alpha_3$ is an arbitrary constant, $J^c_{\mathrm{g1}}$ and $J^c_{\mathrm{g2}}$
comprise four conserved currents:
\begin{equation}
\label{eq.fourc}
F^{cda}\partial_d (h_{ab} f^b),\quad \partial_d [ \tilde{\mathring{F}}^{cda} h_{ab} f^b],\quad
F^{cda} \partial_d (h f_a),\quad \partial_d [ \tilde{\mathring{F}}^{cda} h f_a].
\end{equation}
These currents are homogeneous linear local functions of $f^a$
and do not depend explicitly on $x^a$.
They depend on the derivatives of $h_{ab}$ only up to first order.

$F^{cda}\partial_d (h_{ab} f^b)$ and $F^{cda} \partial_d (h f_a)$ are trivially conserved
but not strongly conserved.
$\partial_d [ \tilde{\mathring{F}}^{cda} h_{ab} f^b]$
and $\partial_d [ \tilde{\mathring{F}}^{cda} h f_a]$
are strongly conserved currents with superpotentials
\begin{equation}
\label{eq.lg.s3}
\tilde{\mathring{F}}^{cda} h_{ab} f^b = \Sigma_3^{cd}-\Sigma_{3'}^{cd}
\end{equation}
and
\begin{equation}
\label{eq.lg.s4}
\tilde{\mathring{F}}^{cda} h f_a = \Sigma_4^{cd}.
\end{equation}
($\Sigma_3^{cd}$ and $\Sigma_{3'}^{cd}$ were defined in (\ref{eq.lg.jc2}) and (\ref{eq.lg.sigma3p}),
whereas $\Sigma_4^{cd}$ is defined by (\ref{eq.lg.s4}).)

As $J^c_{\mathrm{g1}}$ and $J^c_{\mathrm{g2}}$ are obtained from $J^c_{\mathrm{g}}$ by the replacements (\ref{eq.lg.repl}),
they are Noether currents
associated with the variations
\begin{equation}
(\delta_{\mathrm{g}}h_{ab})|_{\phi_a=h_{ab}f^b} =
\partial_a h_{bc} f^c + \partial_b h_{ac} f^c
+ h_{bc} \partial_a f^c + h_{ac} \partial_b f^c
\end{equation}
and
\begin{equation}
(\delta_{\mathrm{g}}h_{ab})|_{\phi_a=hf_a} =
\partial_a h f_b + \partial_b h f_a + h \partial_a f_b + h \partial_b f_a\, ,
\end{equation}
respectively,
where $\delta_{\mathrm{g}}h_{ab}=\partial_a\phi_b+\partial_b\phi_a$.
The $K^c$ quantities for $J^c_{\mathrm{g1}}$ and $J^c_{\mathrm{g2}}$
are also obtained from $K_{\mathrm{g}}^c$
by the replacements (\ref{eq.lg.repl}).

The energy-momentum tensor, spin tensor
and virial current type parts
of the four currents (\ref{eq.fourc}) are listed in Table \ref{tab1}.
\begin{table}[h]
\begin{center}
\begin{tabular}{l|l|lll}
\hspace{8mm} $\delta h_{ab}$ & \hspace{9mm} $J^c$ & \hspace{4mm} $T^{cd}$ & \hspace{5mm} $S^{cde}$ & \hspace{3mm} $\mbb{T}^c$\\
\hline
\vspace{-3mm}
& & & & \\
$\frac{1}{2} \delta_{\mathrm{g}}h_{ab}|_{\phi_a=h_{ab}f^b}$ & $F^{cda}\partial_d (h_{ab} f^b)$ & $F^{cab}\partial_ah_b{}^d$ & $\mc{A}_{de}[F^{cea}h_a{}^d]$ & $F^{cab}h_{ab}$ \\[1mm]
$\frac{1}{2} \delta_{\mathrm{g}}h_{ab}|_{\phi_a=hf_a}$ & $F^{cda} \partial_d (h f_a)$ & $F^{cad}\partial_a h$ & $\mc{A}_{de}[F^{ced}h]$ & $F^c h$ \\[1mm]
0 & $\partial_d [ \tilde{\mathring{F}}^{cda} h_{ab} f^b]$ & $\tilde{\mathring{F}}^{cab}\partial_a h_b{}^d$ & $\mc{A}_{de}[\tilde{\mathring{F}}^{cea}h_a{}^d]$ & $\tilde{\mathring{F}}^{cab}h_{ab}$ \\[1mm]
0 & $\partial_d [ \tilde{\mathring{F}}^{cda} h f_a]$ & $\tilde{\mathring{F}}^{cad}\partial_a h$ & $\mc{A}_{de}[\tilde{\mathring{F}}^{ced}h]$ & $0$
\end{tabular}
\caption{The $T^{cd}$, $S^{cde}$ and $\mbb{T}^c$ type parts
of the currents (\ref{eq.fourc}).
The variations with which the currents are associated are also shown.}
\label{tab1}
\end{center}
\end{table}

\newpage

\subsection{Extension of $T_{\mathrm{lg}}^{cd}$}
\label{sec.gentlg}

In this subsection, we extend $T_{\mathrm{lg}}^{cd}$
into a conserved current $J_{\mathrm{lg}}^c$
associated with Poincar\'e and dilatation symmetry,
and discuss its properties.

\subsubsection{Definition of $J_{\mathrm{lg}}^c$}
\label{sec.defjlg}

In \cite{TG}, $T_{\mathrm{lg}}^{cd}$ was obtained by subtracting the
trivially conserved tensor $F^{cab}\partial_ah_b{}^d$
from the canonical energy-momentum tensor
$\Tc^{cd}|_{\alpha_2=\alpha_3=0}$ (see (\ref{eq.lg.emtc}))
following from $L_1$.
The canonical energy-momentum tensors $\Tc^{cd}$ that follow from
the Lagrangian densities (\ref{eq.lg.l})
have been generalized to currents
associated with Poincar\'e and dilatation symmetry
in Section \ref{sec.cc} and $F^{cab}\partial_ah_b{}^d$
has also been generalized in Section \ref{sec.gc}.
The generalization of the latter is
$\frac{1}{2}J_{\mathrm{g1}}^c|_{\alpha_3=0} = F^{cda}\partial_d (h_{ab} f^b)$,
as can be seen from Table \ref{tab1} and (\ref{eq.lg.jg1}).
A straightforward generalization of $T_{\mathrm{lg}}^{cd}$ is thus
\begin{equation}
\label{eq.defjlg}
J_{\mathrm{lg}}^c = \Jc^c -\frac{1}{2}J_{\mathrm{g1}}^c\, ,\qquad \alpha_2=\alpha_3=0.
\end{equation}

\subsubsection{Properties of $J_{\mathrm{lg}}^c$}
\label{sec.propjlg}

$\Jc^c$ and $J_{\mathrm{g1}}^c$ are Noether currents,
therefore $J_{\mathrm{lg}}^c$ is a Noether current.
As discussed in more detail in Section \ref{sec.genjnoether},
it is associated with the variation
\begin{equation}
\label{eq.dhst}
\delta_{\mathrm{lg}} h_{ab} =
2\mathring{F}_{c(ab)} f^c
+ h^{[d}{}_{(a}\eta^{c]}{}_{b)}\partial_{[c}f_{d]}\, ,
\end{equation}
which corresponds to modified
Poincar\'e transformations and dilatations.
(\ref{eq.dhst}) simplifies to
$\delta_{\mathrm{lg}} h_{ab} = 2\mathring{F}_{c(ab)} f^c$
if $f^c$ is a generator of translations or dilatations (i.e., if $\omega^{ab}=0$).
The expression on the right hand side of (\ref{eq.dhst})
is not scalar gauge invariant,
but it is still true that it depends on $\partial_ch_{ab}$
only through the Fierz tensor (which is scalar gauge invariant).
In electrodynamics, the variation $\delta_{\mathrm{em}} A_a$
(see (\ref{eq.em.dAst})) with which the standard conformal current
$J_{\mathrm{em}}^a = T_{\mathrm{em}}^{ab}f_b$ (see (\ref{eq.em.jst}))
is associated is completely
gauge invariant, but $\delta_{\mathrm{lg}} h_{ab}$ could not have the same property,
as it would require $\delta_{\mathrm{lg}} h_{ab}$ to depend on at least
the second derivatives of $h_{ab}$.

\newpage

From (\ref{eq.dhst}), it can be seen that the coefficients
$\delta_{T_{\mathrm{lg}}} h_{ab}{}^c$,
$\check{\delta}_{S, \mathrm{lg}} h_{ab}{}^{cd}$ and
$\check{\delta}_{\mathbb{T}, \mathrm{lg}} h_{ab}$
(see (\ref{eq.dPhi_orbint})) are
\begin{eqnarray}
\label{eq.lg.dTh}
\delta_{T_{\mathrm{lg}}} h_{ab}{}^c & = & 2\mathring{F}^c{}_{(ab)} \\[1mm]
\label{eq.lg.dSh}
\check{\delta}_{S, \mathrm{lg}} h_{ab}{}^{cd} & = & h^{[c}{}_{(a}\eta^{d]}{}_{b)}  \\[1mm]
\check{\delta}_{\mathbb{T}, \mathrm{lg}} h_{ab} & = & 0\, .
\end{eqnarray}
These formulae should be compared with (\ref{eq.hdT})-(\ref{eq.hdbbT}).
In particular, it is worth noting that
$\check{\delta}_{\mathbb{T}, \mathrm{lg}} h_{ab} = 0$
and $\check{\delta}_{S, \mathrm{lg}} h_{ab}{}^{cd}$
differs from (\ref{eq.hdS}) by a factor $1/2$.
The corresponding formulae in electrodynamics (see Section \ref{sec.el.gentem}) are
$\delta_{T_{\mathrm{em}}} A_a{}^b = -F_a{}^b$,
$\check{\delta}_{S, \mathrm{em}} A_a{}^{bc} = 0$,
$\check{\delta}_{\mathbb{T}, \mathrm{em}} A_a = 0$.

In agreement with the fact that $T_{\mathrm{lg}}^{cd}$ is not symmetric,
and in contrast with electrodynamics (see (\ref{eq.em.jst})),
$J_{\mathrm{lg}}^c$ is not $T_{\mathrm{lg}}^{cd}f_d$.
Rather, it has the form
\begin{equation}
J_{\mathrm{lg}}^c = 
T_{\mathrm{lg}}^{cd}f_d
+ F^{ca[d}h_a{}^{e]} \partial_{[d} f_{e]}\, .
\label{eq.lg.jst}
\end{equation}
Nevertheless, the second term on the right hand side of this formula
vanishes if $f^c$ is a generator of translations or dilatations.
From (\ref{eq.lg.jst}) it can be seen that the
spin tensor and the virial current
encoded in $J_{\mathrm{lg}}^c$ are
\begin{eqnarray}
\label{eq.lgspin}
S_{\mathrm{lg}}^{cde} & \! = \! & F^{ca[e}h_a{}^{d]} \\[1mm]
\mbb{T}_{\mathrm{lg}}^c & \! = \! & 0.
\end{eqnarray}
$S_{\mathrm{lg}}^{cde}$ could not be zero, since $T_{\mathrm{lg}}^{cd}$ is not symmetric
(see Proposition \ref{prop.2.2}).

The angular momentum tensor and the dilatation current encoded in $J_{\mathrm{lg}}^c$ are given by the general relations in
(\ref{eq.bbT1}), i.e.
\begin{equation}
M_{\mathrm{lg}}^{cde} = T_{\mathrm{lg}}^{c[d}x_{\phantom{l}}^{e]}
+ S_{\mathrm{lg}}^{cde},\qquad
D_{\mathrm{lg}}^c = T_{\mathrm{lg}}^{cd}x_d\, .
\end{equation}
The variations of $h_{ab}$
with which $M_{\mathrm{lg}}^{cde}$ and $D_{\mathrm{lg}}^c$
are associated
are given by the general formulae in (\ref{eq.rel-dMdD}):
\begin{eqnarray}
\label{eq.dMlg}
\delta_{M_{\mathrm{lg}}} h_{ab}{}^{cd} & = &
\delta_{T_{\mathrm{lg}}} h_{ab}{}^{[c} x^{d]}
+ \check{\delta}_{S,\mathrm{lg}} h_{ab}{}^{cd}
\ \, = \ \, 2x^{[d} \mathring{F}^{c]}{}_{(ab)}
+ h^{[c}{}_{(a}\eta^{d]}{}_{b)} \\[1mm]
\label{eq.dDlg}
\delta_{D_{\mathrm{lg}}} h_{ab} & = &
\delta_{T_{\mathrm{lg}}} h_{ab}{}^c\, x_c \ \, = \ \,
2x_c \mathring{F}^c{}_{(ab)}\, .
\end{eqnarray}

The $K^c$ quantity for $J_{\mathrm{lg}}^c$,
assuming that the Lagrangian density is taken to be $L_1$,
is the sum of the $K^c$ quantities for the two terms
on the right hand side of (\ref{eq.defjlg}):
\begin{equation}
\label{eq.ktara}
K_{\mathrm{lg}}^c = f^c L_1 - \frac{1}{2}F^{abc}\mathring{F}_{abd}f^d
-\frac{1}{2}h_a{}^{[e} F^{d]ac} \partial_{[d} f_{e]}\, .
\end{equation}
(\ref{eq.ktara}) also takes a simpler form,
$K_{\mathrm{lg}}^c = f^c L_1 - \frac{1}{2}F^{abc}\mathring{F}_{abd}f^d$,
if $f^c$ is a generator of
translations or dilatations.
From (\ref{eq.ktara}) it can be seen that
\begin{eqnarray}
\label{eq.lg.KT}
K_{T_{\mathrm{lg}}}^{cd} & \! = \! &
\eta^{cd}L_1 - \frac{1}{2}F^{abc}\mathring{F}_{ab}{}^d \\[1mm]
\label{eq.lg.KS}
\check{K}_{S, \mathrm{lg}}^{cde} & \! = \! & \frac{1}{2}h_a{}^{[e}F^{d]ac} \\[1mm]
\label{eq.lg.KbbT}
\check{K}_{\mathbb{T}, \mathrm{lg}}^c & \! = \! & 0.
\end{eqnarray}
It is worth noting that $\check{K}_{\mathbb{T}, \mathrm{lg}}^c$ is zero,
whereas $\check{K}_{S, \mathrm{lg}}^{cde}$ is not.
$K_{\mathrm{lg}}^c$ is considerably more complicated than the $K^c$ in electrodynamics,
which is just the standard $f^c L$ (see Section \ref{sec.el.gentem}).
$K_{\mathrm{lg}}^c$ is also not scalar gauge invariant,
in agreement with the properties of $L_1$ and $\delta_{\mathrm{lg}} h_{ab}$,
nevertheless it depends on $\partial_c h_{ab}$
only through the Fierz tensor. $K_{T_{\mathrm{lg}}}^{cd}$ is scalar gauge invariant,
though it deviates from $\eta^{cd}L_1$, which would be the standard expression.
The $K^c$ quantities for $M_{\mathrm{lg}}^{cde}$ and $D_{\mathrm{lg}}^c$,
i.e.\ $K_{M_{\mathrm{lg}}}^{cde}$ and $K_{D_{\mathrm{lg}}}^c$,
are given in terms of
$K_{T_{\mathrm{lg}}}^{cd}$, $\check{K}_{S, \mathrm{lg}}^{cde}$
and $\check{K}_{\mathbb{T}, \mathrm{lg}}^c$
by the general formulae in (\ref{eq.rel-KMKD}).

The vanishing of $\mbb{T}_{\mathrm{lg}}^c$ is consistent
with the tracelessness of $T_{\mathrm{lg}}^{cd}$,
but since generally the tracelessness of the energy-momentum tensor
does not imply the vanishing of the virial current,
it is an additional notable property of $J_{\mathrm{lg}}^c$
beyond the tracelessness of $T_{\mathrm{lg}}^{cd}$.
Moreover, not only $\mbb{T}_{\mathrm{lg}}^c$,
but also $\check{\delta}_{\mathbb{T}, \mathrm{lg}} h_{ab}$
and $\check{K}_{\mathbb{T}, \mathrm{lg}}^c$ are zero,
as we have seen.

Unlike $T_{\mathrm{lg}}^{cd}$,
$S_{\mathrm{lg}}^{cde}$ is not invariant under scalar gauge transformations, nevertheless it depends on the derivatives of $h_{ab}$
only through the Fierz tensor.
The gauge $F_a=0$, $F_{ab0}=0$, which was found to be very well suited to $T_{\mathrm{lg}}^{cd}$ in \cite{TG},
does not appear to be relevant to $S_{\mathrm{lg}}^{cde}$.
This is not surprising, however,
because for dimensional reasons $S_{\mathrm{lg}}^{cde}$
has to be (as it is) an expression
consisting of terms of the type $h\partial h$,
i.e.\ it has to depend
in addition to $\partial_ch_{ab}$ on $h_{ab}$ as well.

In traceless harmonic gauge---and hence
also in TT gauge---$S_{\mathrm{lg}}^{cde}$ coincides
with $-\frac{1}{2}s^{cde}$, where $s^{cde}$ is
the gravitational spin tensor proposed in \cite{BHLang} (see (\ref{eq.bhls}))\footnote{The opposite sign is due to different conventions,
but the factor $1/2$ is a genuine difference
between $S_{\mathrm{lg}}^{cde}$ and $s^{cde}$.}.
$S_{\mathrm{lg}}^{cde}$ is different from zero even in TT gauge,
despite the fact that $T_{\mathrm{lg}}^{cd}$ is symmetric
already in the $F_a=0$ gauge.

The trace of $S_{\mathrm{lg}}^{cde}$ is
\begin{equation}
S_{\mathrm{lg}\,c}{}^{ce}=\frac{1}{2}F_a h^{ae},
\end{equation}
which vanishes in the $F_a=0$ gauge, and therefore also in TT gauge.
The tracelessness of $S_{\mathrm{lg}}^{cde}$ is desirable,
as noted in \cite{BHLang} (see Section III.B),
because some standard examples of material spin tensor
(notably the spin tensor of the Weyssenhoff fluid \cite{WeyRaa, ObKor}
and of the spin-$1/2$ field \cite{BogShir}) are traceless.

Another significant physical property of $S_{\mathrm{lg}}^{cde}$
is that it is purely spatial (i.e., $S_{\mathrm{lg}}^{c0i}=0$) in TT gauge;
this property follows directly
from the same property of $s^{cde}$ established in \cite{BHLang}
and from the aforementioned equality
$S_{\mathrm{lg}}^{cde} = -\frac{1}{2}s^{cde}$ that holds
already in traceless harmonic gauge.

We note that the symmetric
energy-momentum tensor that can be constructed from
$T_{\mathrm{lg}}^{cd}$ and $S_{\mathrm{lg}}^{cde}$
according to the Belinfante--Rosenfeld prescription
depends on the second derivatives of $h_{ab}$,
thus it is not in the class of energy-momentum tensors considered in this paper.\\[2mm]

\noi
{\bf Conservation of the intrinsic
and orbital angular momentum tensors}\\[-1mm]

\noi
Since $T_{\mathrm{lg}}^{cd}$ is symmetric in the $F_a=0$ gauge,
$S_{\mathrm{lg}}^{cde}$
and the orbital angular momentum tensor
$M_{\mathrm{lg, orb}}^{cde} =
T_{\mathrm{lg}}^{c[d}x_{\phantom{l}}^{e]}$
are separately conserved in that gauge
(see (\ref{eq.consm})).
It can be verified
(more details will be given in \cite{TG2})
that $S_{\mathrm{lg}}^{cde}$ and $M_{\mathrm{lg, orb}}^{cde}$
are associated with certain symmetries
by the generalized Noether's theorem
(i.e., Theorem \ref{thrm.ng});
$M_{\mathrm{lg, orb}}^{cde}$ is associated with
$\delta_{M_\mathrm{lg, orb}} h_{ab}{}^{de} =
\delta_{T_{\mathrm{lg}}} h_{ab}{}^{[d} x^{e]}
= \check{\delta}_{M, \mathrm{lg}, \mathrm{orb}} h_{ab}{}^{de}$,
where $\delta_{T_{\mathrm{lg}}} h_{ab}{}^d$ is given by (\ref{eq.lg.dTh}),
and $S_{\mathrm{lg}}^{cde}$ is associated with
$\delta_{S_{\mathrm{lg}}} h_{ab}{}^{de} =
\check{\delta}_{M, \mathrm{lg}, \mathrm{int}}h_{ab}{}^{de}
= \check{\delta}_{S, \mathrm{lg}} h_{ab}{}^{de}$, given by (\ref{eq.lg.dSh}).
From (\ref{eq.lg.dSh}) and (\ref{eq.hdS}) it can be seen that
$\check{\delta}_{S, \mathrm{lg}} h_{ab}{}^{de}$ corresponds to
internal Lorentz transformations, although (\ref{eq.lg.dSh}) and (\ref{eq.hdS})
differ by a factor of $2$.
The relevant $K^c$ quantities are given by (\ref{eq.KMorbint}),
with $K_{T_{\mathrm{lg}}}^{cd}$ and $\check{K}_{S, \mathrm{lg}}^{cde}$
given by (\ref{eq.lg.KT}) and (\ref{eq.lg.KS}).

\subsubsection{Concluding remarks}

In view of the above properties of $J_{\mathrm{lg}}^c$,
in particular the special properties of $S_{\mathrm{lg}}^{cde}$
and the vanishing of $\mbb{T}_{\mathrm{lg}}^c$,
$J_{\mathrm{lg}}^c$ can be regarded as
an appropriate extension of $T_{\mathrm{lg}}^{cd}$
to Poincar\'e-dilatation symmetry.
$S_{\mathrm{lg}}^{cde}$ (and the entire  $J_{\mathrm{lg}}^c$)
has been obtained in a straightforward way,
without extra efforts to ensure that it will have favourable properties.

\subsection{General family of currents associated with Poincar\'e-dilata-tion symmetry}
\label{sec.gen}

In the previous subsections we found $4$ strongly conserved currents
(the currents following from the superpotentials
$\Sigma_2^{cd}$, $\Sigma_3^{cd}$, $\Sigma_{3'}^{cd}$, $\Sigma_4^{cd}$ --- see
(\ref{eq.lg.jc1}), (\ref{eq.lg.jc2}),
(\ref{eq.lg.sigma3p}), (\ref{eq.lg.s3}), (\ref{eq.lg.s4}))
and $2$ trivially conserved currents
($F^{cda}\partial_d (h_{ab} f^b)$, $F^{cda} \partial_d (h f_a)$ --- see (\ref{eq.fourc}))
that can be used as corrections in the construction of
conserved currents associated with Poincar\'e-dilatation symmetry (see (\ref{eq.Jgen})).
In this subsection we search for further suitable
strongly conserved currents (generated by superpotentials)
in a systematic manner.
We obtain finally a conserved current $J^c$ associated with
Poincar\'e-dilatation symmetry
containing $16$ free parameters.
$J^c$ contains terms that involve $\epsilon^{abcd}$;
without these terms the number of free parameters in $J^c$ is $8$.

\subsubsection{Additional superpotentials}
\label{sec.addsp}

We consider superpotentials that are Lorentz covariant
polynomial expressions in $h_{ab}$, $f^a$,
the derivatives of $h_{ab}$ and $f^a$,
and possibly $\epsilon^{abcd}$.
More specifically, we assume that the
superpotentials are sums of terms of the types $hh\partial f$, $h\partial h f$,
$\epsilon hh\partial f$, $\epsilon h\partial h f$,
since we want the corresponding currents to be
quadratic in $\partial_c h_{ab}$ and $h_{ab}$ and linear in $f_a$ or $\partial_bf_a$.
The reason for including precisely one derivative
in the superpotential expressions
is the following:
a superpotential has two indices, therefore
an odd number of derivatives have to be present in a suitable expression,
furthermore three or more derivatives lead to currents
that contain higher than first derivatives of $h_{ab}$
(only one differentiation can act on $f^a$,
since the second derivatives of $f^a$ are zero
if $f^a$ is a generator of Poincar\'e transformations and dilatations).

Without using $\epsilon^{abcd}$, one finds $5$ additional superpotentials
(beyond $\Sigma_2^{cd}$, $\Sigma_3^{cd}$, $\Sigma_{3'}^{cd}$, $\Sigma_4^{cd}$ mentioned
at the beginning of Section \ref{sec.gen})
that have the form described above:
$\Sigma_{2'}^{cd}$, $\Sigma_{2''}^{cd}$, which were defined in (\ref{eq.sigma22}), and
\begin{eqnarray}
\Sigma_7^{cd} & \!\! = \!\! & 2h_{ab}h^{ab}\partial^{[c} f^{d]} \\[1mm]
\Sigma_8^{cd} & \!\! = \!\! & 2h^2\partial^{[c} f^{d]} \\[1mm]
\Sigma_9^{cd} & \!\! = \!\! & 2hh^{a[c}\partial_a f^{d]}\, .
\end{eqnarray}

Taking into account expressions that also contain $\epsilon^{abcd}$, $5$ further
suitable superpotentials can be constructed: 
$\tilde{\Sigma}_{2'}^{cd}$, $\tilde{\Sigma}_{2''}^{cd}$,
$\tilde{\Sigma}_7^{cd}$, $\tilde{\Sigma}_8^{cd}$, $\tilde{\Sigma}_9^{cd}$,
which are the duals of
$\Sigma_{2'}^{cd},\dots,\Sigma_9^{cd}$. 
There exist some other possibilities as well,
but they do not yield new linearly independent currents.

\subsubsection{The general parametric family of currents}
\label{sec.genjpd}

In summary, we have found $16$ possible correction terms
that can be used in the construction of conserved currents associated with
Poincar\'e-dilatation symmetry;
$2$ of them are trivially conserved currents related to gauge transformations
and $14$ are strongly conserved.
As they appear somewhat scattered throughout the previous sections,
we collect them below.

The $14$ strongly conserved terms are the currents that have the superpotentials
\begin{eqnarray}
\Sigma^{cd}_2 & = & 2f^a h^{b[c}\partial_a h^{d]}{}_b
+ 2\partial^a h^{b[c} f^{d]} h_{ab}
+ 2\partial_a h^{ab} f^{[c} h^{d]}{}_b  + \Sigma^{cd}_{2'} + \Sigma^{cd}_{2''}\\[2mm]
\Sigma^{cd}_{2'} & = & 2\partial^a f^{[c} h^{d]b}h_{ab}\\[2mm]
\Sigma^{cd}_{2''} & = & 2\partial_a f_b h^{a[c}h^{d]b}\\[2mm]
\Sigma^{cd}_3 & = & \tilde{\mathring{F}}^{cdb}f^a h_{ab}  +\Sigma_{3'}^{cd}\\[2mm]
\Sigma_{3'}^{cd} & = &  - \frac{1}{4}\epsilon^{cdab}h_b{}^f \partial_f f^e h_{ae}\\[2mm]
\Sigma_4^{cd} & = & \tilde{\mathring{F}}^{cda} h f_a \\[2mm]
\Sigma_7^{cd} & = & 2h_{ab}h^{ab}\partial^{[c} f^{d]} \\[2mm]
\Sigma_8^{cd} & = & 2h^2\partial^{[c} f^{d]} \\[2mm]
\Sigma_9^{cd} & = & 2hh^{a[c}\partial_a f^{d]}\\[2mm]
\tilde{\Sigma}_{2'}^{cd} & = & \frac{1}{2}\epsilon^{cdab} \Sigma_{2'\,ab} \\[2mm]
\tilde{\Sigma}_{2''}^{cd} & = & \frac{1}{2}\epsilon^{cdab} \Sigma_{2''\,ab} \\[2mm]
\tilde{\Sigma}_7^{cd} & = & \frac{1}{2}\epsilon^{cdab} \Sigma_{7\,ab} \\[2mm]
\tilde{\Sigma}_8^{cd} & = & \frac{1}{2}\epsilon^{cdab} \Sigma_{8\,ab} \\[2mm]
\tilde{\Sigma}_9^{cd} & = & \frac{1}{2}\epsilon^{cdab} \Sigma_{9\,ab}\, .
\end{eqnarray}
The $2$ trivially conserved currents are
\begin{eqnarray}
\label{eq.j5}
J_5^c \hspace{-4mm} && = \ \makebox[24mm][l]{$\displaystyle F^{cdb}\partial_d (h_{ba} f^a)$} \ = \ \partial_d \Sigma_5^{cd} - \partial_d F^{cdb} (h_{ba} f^a)  \\
\label{eq.j6}
J_6^c \hspace{-4mm} && = \ \makebox[24mm][l]{$\displaystyle F^{cdb}\partial_d (h f_b)$} \ = \ \partial_d \Sigma_6^{cd} - \partial_d F^{cdb} (h f_b),
\end{eqnarray}
where
$\Sigma_5^{cd} = F^{cdb} h_{ba} f^a$ and $\Sigma_6^{cd} = F^{cdb} h f_b$
are antisymmetric tensors and
the last terms $\partial_d F^{cdb} (h_{ba} f^a)$ and
$\partial_d F^{cdb} (h f_b)$
are weakly zero.

The general current containing these corrections with free coefficients
(denoted by $\lambda_2, \lambda_{2'}, \dots$) is then
\begin{eqnarray}
J^c & \! = \!  &
J_{\mathrm{lg}}^c +  \lambda_5  F^{cab}\partial_a (h_{bd} f^d)
+ \lambda_6 F^{cab}\partial_a (h f_b) \nonumber \\[1mm]
&& +\,\lambda_2\partial_d\Sigma_2^{cd} + \lambda_{2'}\partial_d\Sigma_{2'}^{cd} + \lambda_{2''}\partial_d\Sigma_{2''}^{cd} 
+ \lambda_3\partial_d\Sigma_3^{cd} + \lambda_{3'}\partial_d\Sigma_{3'}^{cd} + \lambda_4\partial_d\Sigma_4^{cd} \nonumber \\[1mm]
&& +\, \tilde{\lambda}_{2'}\partial_d\tilde{\Sigma}_{2'}^{cd} + \tilde{\lambda}_{2''}\partial_d\tilde{\Sigma}_{2''}^{cd} 
+ \sum_{i=7}^9 \lambda_i\partial_d\Sigma_i^{cd} + \sum_{i=7}^9 \tilde{\lambda}_i\partial_d\tilde{\Sigma}_i^{cd}.
\label{eq.lg.j}
\end{eqnarray}
$J^c$ clearly has the structure (\ref{eq.Jgen}), since
$J_{\mathrm{lg}}^c =
\Jc^c_1
-\frac{1}{2}J_{\mathrm{g1}}^c|_{\alpha_3=0}
=\Jc^c_1 -  F^{cab}\partial_a (h_{bd} f^d)$;
\begin{equation}
\label{eq.lg.j2}
J^c \ = \ \Jc^c_1
+ (\lambda_5-1)F^{cab}\partial_a (h_{bd} f^d)
+ \lambda_6 F^{cab}\partial_a (h f_b)
+ \lambda_{\dots} \partial_d\Sigma^{cd}_{\dots}\, ,
\end{equation}
where $\lambda_{\dots} \partial_d\Sigma^{cd}_{\dots}$ indicates
the $\lambda \partial_d\Sigma^{cd}$ type terms in (\ref{eq.lg.j}).
It is also clear that
$J^c$ contains the $2$-parameter family of canonical currents (\ref{eq.lg.jc}):
$\Jc^c =
J^c|_{\lambda_5 = 1,\lambda_2 = \alpha_2,\lambda_3=\alpha_3;\lambda_6,\lambda_{2'},...=0}$.

The energy-momentum tensor encoded in (\ref{eq.lg.j}) is
\begin{eqnarray}
T^{cd} & \! = \! & T_{\mathrm{lg}}^{cd}
+ \lambda_2 \partial_a U^{cad}_2
+ \lambda_3\tilde{\mathring{F}}^{cab}\partial_a h_b{}^d
+ \lambda_4 \tilde{\mathring{F}}^{cad}\partial_a h \nonumber \\[1mm]
&&  +\, \lambda_5 F^{cab}\partial_a h_b{}^d +  \lambda_6 F^{cad}\partial_a h .
\label{eq.lg.emt}
\end{eqnarray}
(\ref{eq.lg.emt}) is the same as the family of energy-momentum tensors we found in \cite{TG}.
From (\ref{eq.lg.emt}) it can be seen that there are many terms in (\ref{eq.lg.j})
that do not contribute to $T^{cd}$---these terms
give contribution only to the intrinsic part of $J^c$.

The virial current encoded in (\ref{eq.lg.j}) is
\begin{equation}
\mbb{T}^c = 2\lambda_2 (h_{ab}\partial^a h^{bc} - h^{cb}\partial_a h^a{}_b)
+ \lambda_3 \tilde{\mathring{F}}^{cab}h_{ab} +  \lambda_5 F^{cab}h_{ab} + \lambda_6 F^c h.
\label{eq.lg.intdcgen}
\end{equation}
This expression shows that
(a) the energy-momentum tensor determines $\mbb{T}^c$,
as the parameters of $\mbb{T}^c$ are present in (\ref{eq.lg.emt}) as well,
and (b) most of the terms in (\ref{eq.lg.j}) do not contribute to $\mbb{T}^c$.

The spin tensor encoded in (\ref{eq.lg.j}) is
\begin{eqnarray}
S^{cde} & \! = \! & \mc{A}_{de}\bigl[  F^{cae}h_a{}^d
+ \lambda_2 \bigl( h^{ac}\partial^d h^e{}_a - h^{ae}\partial^d h^c{}_a
+ \eta^{dc}(h_b{}^e \partial_a h^{ab} - h_{ab}\partial^a h^{be}) \bigr) \nonumber \\
&& +\, (\lambda_2+\lambda_{2'}) \bigl(\partial_a(h^{ab}h_b{}^e)\eta^{cd}
+\partial^e (h^{cb}h_b{}^d) \bigr)
+ 2(\lambda_2+\lambda_{2''})\partial_a (h^{ad}h^{ce})
\nonumber \\
&& -\, \tilde{\lambda}_{2'}\epsilon^{cdab}\partial_a(h_b{}^f h_f{}^e)
+ \tilde{\lambda}_{2''}\epsilon^{cfab}\partial_f (h^e{}_a h^d{}_b) \nonumber \\
&& +\,  \lambda_3\tilde{\mathring{F}}^{cea}h_a{}^d
+ (\lambda_3 + \lambda_{3'})\tilde{\mathring{F}}^{cae}h_a{}^d \nonumber \\
&& +\, \lambda_4\tilde{\mathring{F}}^{ced}h
+ \lambda_5F^{cea}h_a{}^d 
+ \lambda_6F^{ced}h \nonumber \\
&& +\, 2\lambda_7 \partial^d(h_{ab}h^{ab})\eta^{ce}
+ 2\lambda_8 \partial^d h^2 \eta^{ce}
+ \lambda_9\bigl(\partial^d(hh^{ce}) + 
\partial_a(hh^{ad}\eta^{ce}) \bigr) \nonumber \\
&& -\, \tilde{\lambda}_7 \epsilon^{cdea}\partial_a (h_{bf}h^{bf}) - \tilde{\lambda}_8 \epsilon^{cdea}\partial_a h^2
+ \tilde{\lambda}_9 \epsilon^{cdab}\partial_a (hh^e{}_b)\bigr].
\label{eq.lg.sgen}
\end{eqnarray}
(\ref{eq.lg.sgen}) contains all the parameters that are present in (\ref{eq.lg.j}).
The terms multiplied by $\lambda_{2'}$, $\lambda_{2''}$, $\tilde{\lambda}_{2'}$,
$\tilde{\lambda}_{2''}$,
$\lambda_{3'}$, $\lambda_7$, $\lambda_8$, $\lambda_9$,
$\tilde{\lambda}_7$, $\tilde{\lambda}_8$, $\tilde{\lambda}_9$
are strongly conserved.
Since the energy-momentum tensor has fewer parameters, for any fixed energy-momentum tensor
there is a significant freedom in the choice of the angular momentum tensor.
Nevertheless, the angular momentum tensor is completely determined by the energy-momentum tensor up to strongly conserved terms,
as the free parameters that remain after fixing the energy-momentum tensor are those
that we have just noted to be coefficients of strongly conserved terms in the spin tensor.

Although it is not obvious from (\ref{eq.lg.sgen}),
$S^{cde}=0$ if
$\lambda_2 = -\frac{1}{4}$, $\lambda_5 = -1$, $\lambda_6 = 1$,
$\lambda_{2''} = \frac{1}{2}$, $\lambda_8 = \frac{1}{8}$, $\lambda_9 = -\frac{1}{2}$
and the other parameters are zero---see Section \ref{sec.ll} for further details.
This implies that if $S^{cde}$ takes a value $S_\Lambda^{cde}$
for some values of the parameters,
then for an arbitrary constant $\Upsilon$
there exist other values of the parameters such that
the corresponding value of $S^{cde}$ is $\Upsilon S_\Lambda^{cde}$.

It is of some interest to look at the spatial spin density
$S^{0ij}$ ($i,j=1,2,3$) in TT gauge.
One finds that if the terms in (\ref{eq.lg.sgen}) containing $\epsilon^{abcd}$ are not considered,
then $S^{0ij}$ is unique in TT gauge up to a numerical factor:
\begin{equation}
\label{eq.S0ij}
S^{0ij} = \frac{1+\lambda_5}{4}(h^i{}_k\partial_0h^{jk}
- h^j{}_k\partial_0h^{ik}).
\end{equation}
If the energy-momentum tensor is taken to be
two times\footnote{This factor is needed for consistency
with the normalization of the energy-momentum tensor used in the present paper.}
the linearized Landau--Lifshitz pseudotensor (see Section \ref{sec.ll}),
then $\lambda_5 = -1$, i.e.\ $S^{0ij} = 0$.
In the case when $J^c = J_{\mathrm{lg}}^c$,  $\lambda_5=0$,
whereas in the case of the canonical currents $\Jc^c|_{\alpha_3=0}$,
$\lambda_5=1$
(see Sections \ref{sec.cc} and \ref{sec.bhl}).

If $J^c$ is taken to be canonical with $\alpha_3=0$
(i.e.\ $J^c=\Jc^c|_{\alpha_3=0}$, $\alpha_2$ is arbitrary),
then in TT gauge
\begin{equation}
\Tc^{0i} = F^{0ab}\partial^ih_{ab} =
\frac{1}{2}\partial_0h^{jk}\partial^ih_{jk}\, ,
\end{equation}
and for the spatial angular momentum density
\begin{equation}
\Mc^{0ij} = \frac{1}{2}(\Tc^{0i}x^j - \Tc^{0j}x^i) + \Sc^{0ij}
\end{equation}
one gets the formula
\begin{equation}
\Mc^{0ij} = -\frac{1}{4}\partial_0h^{kl}(x^i\partial^j - x^j\partial^i)h_{kl}
+\frac{1}{2}(h^i{}_k\partial_0h^{jk} - h^j{}_k\partial_0h^{ik}).
\end{equation}
The latter result is in agreement, up to normalization, with
\cite{Maggiore}, Section 2.1.3.
The same result is obtained if the energy-momentum tensor and the spin tensor
are chosen according to the proposal of
\cite{BHLemt,BHLang,BHLthe}\footnote{There are differences
between the present paper and \cite{BHLemt,BHLang,BHLthe}
in normalization factors and index ordering that should be taken into account.}
(see Section \ref{sec.bhl}).
Generally, if the terms containing $\epsilon^{abcd}$ are omitted,
\begin{equation}
T^{0i} =
\frac{1}{2}\partial_0h^{jk}\partial^ih_{jk}
- \frac{1-\lambda_5}{2}\partial_0h^{jk}\partial_jh^i{}_k
\end{equation}
and
\begin{eqnarray}
M^{0ij} & \! = \! & \frac{1}{2}(T^{0i}x^j - T^{0j}x^i) + S^{0ij} \\[1mm]
& \! = \! & - \frac{1}{4}\partial_0h^{kl}(x^i\partial^j - x^j\partial^i)h_{kl}
+ \frac{1-\lambda_5}{4}\partial_0h^{kl}(x^i \partial_k h^j{}_l - x^j \partial_k h^i{}_l)\nonumber\\[1mm]
&& +\, \frac{1+\lambda_5}{4}(h^i{}_k\partial_0h^{jk}
- h^j{}_k\partial_0h^{ik})
\end{eqnarray}
in TT gauge.
In particular, $T^{0i}$ also depends only on $\lambda_5$ in TT gauge.

Finally we note that
the total charges (i.e., the total energy, momentum, \dots)
determined by $J^0$ do not depend
on any of the free parameters $\lambda_2, \lambda_{2'}, \dots$
of $J^c$
if $h_{ab}$ and $\partial_ch_{ab}$ fall off sufficiently rapidly
at spatial infinity,
because the correction terms in (\ref{eq.lg.j}),
multiplied by these parameters,
are trivially conserved or strongly conserved,
and their superpotentials,
including $\Sigma^{cd}_5$ and $\Sigma^{cd}_6$,
are second order polynomials
in $h_{ab}$ and $\partial_c h_{ab}$.
Different values of $\lambda_2, \lambda_{2'}, \dots$
correspond to different localizations of the charges and fluxes
determined by $J^c$.

\subsubsection{$J^c$ as a Noether current}
\label{sec.genjnoether}

$J^c$ is a Noether current, as it is a sum of Noether currents,
but generally it is not associated with the variation
given by (\ref{eq.lg.conf2})
(which corresponds to `simple' Poincar\'e transformations and dilatations),
due to the presence of the correction terms (see (\ref{eq.lg.j2}))
\begin{equation}
(\lambda_5-1)F^{cab}\partial_a (h_{bd} f^d)\qquad
\mathrm{and}\qquad
\lambda_6 F^{cab}\partial_a (h f_b).
\end{equation}
The latter terms are associated with
field-dependent gauge transformations; more specifically,
they are associated with the variations
\begin{equation}
\frac{1}{2}(\lambda_5-1)\delta_{\mathrm{g}}h_{ab}|_{\phi_a=h_{ab}f^b}\qquad
\mathrm{and}\qquad
\frac{1}{2}\lambda_6 \delta_{\mathrm{g}}h_{ab}|_{\phi_a=hf_a},
\end{equation}
respectively---see Section \ref{sec.gc}.
$J^c$ is thus associated
as a Noether current with the modified variation
\begin{equation}
\label{eq.dlh1}
\delta_\lambda h_{ab} \ = \ \delta h_{ab}
+ \frac{1}{2}(\lambda_5-1)\delta_{\mathrm{g}}h_{ab}|_{\phi_a=h_{ab}f^b}
+ \frac{1}{2}\lambda_6 \delta_{\mathrm{g}}h_{ab}|_{\phi_a=hf_a}\, ,
\end{equation}
where $\delta h_{ab}$ denotes the `simple' variation
given by (\ref{eq.lg.conf2}) and
$\lambda$ denotes the pair $(\lambda_5,\lambda_6)$.
$\delta_\lambda h_{ab}$ can be written in the standard form (\ref{eq.dPhi_orbint}):
\begin{equation}
\delta_\lambda h_{ab} \ = \  \delta_{\lambda,T}h_{ab}{}^c\, f_c
- \check{\delta}_{\lambda,S}h_{ab}{}^{cd}\, \partial_{[c}f_{d]}
+ \frac{1}{4} \check{\delta}_{\lambda,\mathbb{T}}h_{ab}\, \partial_cf^c,
\label{eq.dlh2}
\end{equation}
where
\begin{eqnarray}
\label{eq.dlambdaTh}
\delta_{\lambda,T}h_{ab}{}^c & \! = \! &
2\mathring{F}^c{}_{(ab)} + \lambda_5\partial_{(a}h_{b)}{}^c
+ \lambda_6\partial_{(a}h\, \eta_{b)}{}^c \\[1mm]
\label{eq.dlambdaSh}
\check{\delta}_{\lambda,S}h_{ab}{}^{cd} & \! = \! &
(1+\lambda_5)h^{[c}{}_{(a}\eta^{d]}{}_{b)}\\[1mm]
\label{eq.dlambdabbTh}
\check{\delta}_{\lambda,\mathbb{T}}h_{ab} & \! = \! &
\lambda_5 h_{ab} + \lambda_6 h \eta_{ab}\, .
\end{eqnarray}

The case $\lambda_5 = \lambda_6 = 0$ is relevant to $J_{\mathrm{lg}}^c$.
From (\ref{eq.dlambdaSh}) and (\ref{eq.dlambdabbTh}) it can be seen that
there exist values of $\lambda_5$ and $\lambda_6$ for which
$\check{\delta}_{\lambda,S}h_{ab}{}^{cd}$
or $\check{\delta}_{\lambda,\mathbb{T}}h_{ab}$
is zero,
but at least one of $\check{\delta}_{\lambda,S}h_{ab}{}^{cd}$
and $\check{\delta}_{\lambda,\mathbb{T}}h_{ab}$ is nonzero
for any value of $\lambda_5$ and $\lambda_6$.

$\delta_{\lambda,T}h_{ab}{}^c$ above is the variation of $h_{ab}$
corresponding to (modified) translations;
for expressing the variations of $h_{ab}$
corresponding to Lorentz transformations and dilatations in terms of
$\delta_{\lambda,T} h_{ab}{}^c$, $\check{\delta}_{\lambda,S} h_{ab}{}^{cd}$
and $\check{\delta}_{\lambda,\mathbb{T}}h_{ab}$,
(\ref{eq.rel-dMdD}) is applicable, as in Section \ref{sec.pdsym}:
\begin{eqnarray}
\label{eq.dlambdaMh}
\delta_{\lambda,M} h_{ab}{}^{cd} & \! = \! &
\delta_{\lambda,T} h_{ab}{}^{[c} x^{d]}
+ \check{\delta}_{\lambda,S} h_{ab}{}^{cd}\\[1mm]
\label{eq.dlambdaDh}
\delta_{\lambda,D} h_{ab} & \! = \! & \delta_{\lambda,T} h_{ab}{}^c\, x_c
+ \check{\delta}_{\lambda,\mathbb{T}}h_{ab}\, .
\end{eqnarray}
$\delta_\lambda h_{ab} = \delta_{\lambda,T}h_{ab}{}^c e_c +
\delta_{\lambda,M} h_{ab}{}^{cd} \omega_{cd} +
\delta_{\lambda,D} h_{ab} \beta$,
in accordance with (\ref{eq.dPhi_lin}).

Assuming that the Lagrangian density is taken to be $L_1$,
the $K^c$ quantity for $J^c$ is
\begin{eqnarray}
K^c & \! = \! & f^c L_1 + \frac{1}{2}(\lambda_5-1)(F^{abc}\mathring{F}_{abd}f^d
+ F^{abc}h_{bd}\partial_a f^d)
+\frac{1}{2}\lambda_6 F^{abc}\partial_a(hf_b) \nonumber \\
&& - \,\lambda_2\partial_d\Sigma_2^{cd} - \lambda_{2'}\partial_d\Sigma_{2'}^{cd} - \lambda_{2''}\partial_d\Sigma_{2''}^{cd} 
- \lambda_3\partial_d\Sigma_3^{cd} - \lambda_{3'}\partial_d\Sigma_{3'}^{cd} - \lambda_4\partial_d\Sigma_4^{cd} \nonumber \\[1mm]
&& -\, \tilde{\lambda}_{2'}\partial_d\tilde{\Sigma}_{2'}^{cd} - \tilde{\lambda}_{2''}\partial_d\tilde{\Sigma}_{2''}^{cd} 
- \sum_{i=7}^9 \lambda_i\partial_d\Sigma_i^{cd} - \sum_{i=7}^9 \tilde{\lambda}_i\partial_d\tilde{\Sigma}_i^{cd}.
\label{eq.kcgen}
\end{eqnarray}
This is the sum of the $K^c$ quantities for the terms that appear
on the right hand side of (\ref{eq.lg.j}).
The terms multiplied by $\lambda_5-1$ and $\lambda_6$ in (\ref{eq.kcgen})
follow from
$\frac{1}{2}K_{\mathrm{g}}^c|_{\alpha_3=0}$ (see (\ref{eq.Kg}))
by the replacements (\ref{eq.lg.repl}).
In view of (\ref{eq.defjlg}),
the $K^c$ quantity for $J_{\mathrm{lg}}^c$ is the sum of the
$K^c$ quantities for $\Jc^c_1$
and $-\frac{1}{2}J_{\mathrm{g1}}^c|_{\alpha_3=0}$,
which are $f^cL_1$ (see Section \ref{sec.cc}) and
$-\frac{1}{2}(F^{abc}\mathring{F}_{abd}f^d + F^{abc}h_{bd}\partial_a f^d)$.
The latter expression arises from
$-\frac{1}{2}K_{\mathrm{g}}^c|_{\alpha_3=0}$
by the replacement $\phi_a = h_{ab}f^b$.
The $K^c$ quantity for $J_{\mathrm{lg}}^c$ is thus
\begin{equation}
K^c_{\mathrm{lg}} \ = \ f^c L_1
- \frac{1}{2}(F^{abc}\mathring{F}_{abd}f^d + F^{abc}h_{bd}\partial_a f^d),
\end{equation}
which can also be written in the form (\ref{eq.ktara}).
The second term in this expression is the origin of the $-1$ term
in the coefficient $\lambda_5-1$ in (\ref{eq.kcgen}).
For strongly conserved currents the $K^c$ quantity is just
$-1$ times the current itself;
this is the reason for the presence of the terms
$- \,\lambda_2\partial_d\Sigma_2^{cd}, \dots$ in (\ref{eq.kcgen}).

If $\lambda_5\ne 1$ or $\lambda_6\ne 0$, then the variations
comprised by $\delta_\lambda h_{ab}$ do not form a closed Lie algebra;
their commutators contain additional terms corresponding to some field-dependent gauge transformations.
Let us consider, for simplicity, the variations corresponding to
$f^a=e_1^a$ and $f^a=e_2^a$;
in this case one finds the commutator to be
\begin{eqnarray}
&&\left(\frac{\lambda_5-1}{2}\right)^2\bigl(\partial_a((e_1^ce_2^d-e_1^de_2^c)\partial_ch_{db}) + \partial_b((e_1^ce_2^d-e_1^de_2^c)\partial_ch_{da})\bigr) \nonumber\\
&&\ +\, \frac{\lambda_6^2}{2}\bigl(\partial_a((e_{1b}e_2^c - e_{2b}e_1^c)\partial_c h) + \partial_b((e_{1a}e_2^c - e_{2a}e_1^c)\partial_c h)\bigr) \nonumber\\
&&\ +\, \frac{\lambda_5-1}{2}\frac{\lambda_6}{2} \bigl(\partial_a((e_{1b}e_{2c} - e_{2b}e_{1c}) (2\partial_dh^{cd}-\partial^ch)) \nonumber\\
&& \hspace{3cm} +\, \partial_b((e_{1a}e_{2c} - e_{2a}e_{1c}) (2\partial_dh^{cd}-\partial^ch))\bigr).
\end{eqnarray}

\subsubsection{Divergence of $J^c$ in the presence of matter}
\label{sec.divjmat}

In the presence of matter the divergence of $T^{cd}$, $M^{cde}$ and $D^c$
is generally not zero.
The equations giving these divergences can be regarded as
local balance equations quantifying
the nonconservation of $T^{cd}$, $M^{cde}$ and $D^c$ due to the presence of matter.

In \cite{BHLemt,BHLang} the form of the nonconservation equations
for $T^{cd}$ and $M^{cde}$
in the presence of matter has an important role
in finding definite distinguished
energy-momentum and spin tensors
for the linearized gravitational field.

From the general identity (\ref{eq.nthr2}) for the divergence of Noether currents
it follows that
\begin{equation}
\label{eq.lg.dj}
\partial_c J^c =  - \mc{T}^{ab}\delta_\lambda h_{ab}
\end{equation}
in the presence of matter (assuming that $h_{ab}$ satisfies the linearized Einstein equations).
The divergence of $T^{cd}$, $M^{cde}$ and $D^c$
can be extracted from (\ref{eq.lg.dj});
\begin{eqnarray}
\label{eq.lg.divT-g}
\partial_c T^{cd} & \!\! = \!\! & - \mc{T}^{ab}\delta_{\lambda,T} h_{ab}{}^{d} \\
\label{eq.lg.divM-g}
\partial_c M^{cde} & \!\! = \!\! & - \mc{T}^{ab}\delta_{\lambda,M} h_{ab}{}^{de} \\
\label{eq.lg.divD-g}
\partial_c D^c & \!\! = \!\! & - \mc{T}^{ab}\delta_{\lambda,D} h_{ab}\, .
\end{eqnarray}
$\delta_{\lambda,T} h_{ab}{}^d$, $\delta_{\lambda,M} h_{ab}{}^{de}$ and $\delta_{\lambda,D} h_{ab}$
are given by (\ref{eq.dlambdaTh})-(\ref{eq.dlambdaDh}).
Explicitly,
\begin{eqnarray}
\label{eq.lg.dT}
\partial_c T^{cd} & \!\! = \!\! & - 2 \mc{T}^{ab}\mathring{F}^d{}_{ab}
- \lambda_5 \mc{T}^{ab}\partial_a h_b{}^d - \lambda_6 \mc{T}^{ad}\partial_a h \\[2mm]
\partial_c M^{cde} & \!\! = \!\! &
2\mc{T}^{ab}x^{[d}\mathring{F}^{e]}{}_{ab}
+ \mc{T}^{a[d}h_a{}^{e]} \nonumber \\[1mm]
&& +\,\lambda_5(\mc{T}^{ab}x^{[d}\partial_a h_b{}^{e]}
+ \mc{T}^{a[d}h_a{}^{e]})
- \lambda_6\mc{T}^{a[d}x^{e]}\partial_a h
\label{eq.lg.dm} \\[2mm]
\label{eq.lg.dd}
\partial_c D^c & \!\! = \!\! & -2\mc{T}^{ab}x^c\mathring{F}_{cab}
- \lambda_5 (\mc{T}^{ab}x^c\partial_a h_{bc} + \mc{T}^{ab}h_{ab})
-\lambda_6 (\mc{T}^{ab}x_a\partial_b h + \mc{T}_a{}^a h).
\end{eqnarray}
The formulae relevant to $T_{\mathrm{lg}}^{cd}$, $M_{\mathrm{lg}}^{cde}$ and $D_{\mathrm{lg}}^c$
are obtained by setting $\lambda_5$ and $\lambda_6$ to $0$.
They can be compared with (\ref{eq.em.div1})-(\ref{eq.em.div3}).

The case when (\ref{eq.lg.dT}) and (\ref{eq.lg.dm}) reduce, up to normalization,
to the forms preferred in \cite{BHLemt} and \cite{BHLang} is $\lambda_5=1$, $\lambda_6=0$;
in this case (\ref{eq.lg.dT}) and (\ref{eq.lg.dm}) take the form
\begin{eqnarray}
\label{eq.dtc}
\partial_c T^{cd}|_{\lambda_5=1,\lambda_6=0} & \!\! = \!\! & - \mc{T}^{ab}\partial^d h_{ab}\\[1mm]
\label{eq.dmc}
\partial_c M^{cde}|_{\lambda_5=1,\lambda_6=0}
 & \!\! = \!\! & \mc{T}^{ab}x^{[d}\partial^{e]} h_{ab}
+ 2\mc{T}^{a[d}h_a{}^{e]}.
\end{eqnarray}
For $\partial_c D^c|_{\lambda_5=1,\lambda_6=0}$, we have
\begin{equation}
\label{eq.ddc}
\partial_c D^c|_{\lambda_5=1,\lambda_6=0} \ = \ -\mc{T}^{ab}x^c\partial_c h_{ab} - \mc{T}^{ab}h_{ab}\, .
\end{equation}
$\delta_\lambda h_{ab}$ becomes the `simple' variation $\delta h_{ab}$
(given by (\ref{eq.lg.conf2}))
if $\lambda_5=1$ and $\lambda_6=0$.
The case $\lambda_5=1$, $\lambda_6=0$ is also the one that is relevant to the canonical currents,
thus (\ref{eq.dtc}), (\ref{eq.dmc}) and (\ref{eq.ddc})
are also the nonconservation equations for the canonical $T^{cd}$, $M^{cde}$, $D^c$.
(\ref{eq.dtc}), (\ref{eq.dmc}) and (\ref{eq.ddc}) can be compared with
(\ref{eq.em.div1c}), (\ref{eq.em.div2c}) and (\ref{eq.em.div3c})
in electrodynamics.

\subsubsection{Gauge invariance of $J^c$ up to a trivially conserved current}
\label{sec.ginfofJ}

In this subsection it is shown that $J^c$ changes
by a trivially conserved current under gauge transformations.
An important consequence of this property is that
the total charges determined by $J^c$ are invariant
under any gauge transformation whose
parameter and its derivatives up to sufficiently
high order fall off sufficiently rapidly at spatial infinity.
We note that a general derivation of the gauge invariance of total charges
in gauge theories can be found in \cite{Wald1990}.

It is easy to see that a trivially conserved current
changes by a trivially conserved current under a gauge transformation,
therefore it is sufficient to focus on $\Jc_1^c$, for instance,
instead of dealing with $J^c$.
In the following, $\delta_{\mathrm{g}}$ will denote the first order variation
under a gauge transformation (\ref{eq.lg.g}).

We start from the identity (\ref{eq.nthr2}) for $\Jc_1^c$:
\begin{equation}
\partial_c \Jc_1^c = -G^{ab}\delta h_{ab}\, ,
\end{equation}
where $\delta h_{ab}$ is given by (\ref{eq.lg.conf2}).
We take the first order variation of this identity
with respect to a gauge transformation:
\begin{equation}
\label{eq.dgJ1id}
\partial_c \delta_{\mathrm{g}} \Jc_1^c =
-\delta_{\mathrm{g}} G^{ab}\delta h_{ab}
-G^{ab}\delta_\mathrm{g} \delta h_{ab}\, .
\end{equation}
$G^{ab}$ is gauge invariant, thus $\delta_{\mathrm{g}} G^{ab} = 0$,
therefore (\ref{eq.dgJ1id}) can also be written as
$\partial_c \delta_\mathrm{g} \Jc_1^c = -G^{ab}\delta_\mathrm{g} \delta h_{ab}$.
From (\ref{eq.lg.conf2}) we get
\begin{equation}
\label{eq.dgdh}
\delta_{\mathrm{g}} \delta h_{ab} = f^c\partial_c\delta_{\mathrm{g}} h_{ab}
+\delta_{\mathrm{g}} h_{cb} \partial_a f^c
+\delta_{\mathrm{g}} h_{ac} \partial_b f^c
-\frac{1}{4}\partial_c f^c \delta_{\mathrm{g}} h_{ab}\, ,
\end{equation}
and the pieces of $G^{ab}\delta_\mathrm{g} \delta h_{ab}$
corresponding to the terms on the right hand side are
\begin{eqnarray}
\label{eq.ginvJ-1}
G^{ab}f^c\partial_c \delta_{\mathrm{g}} h_{ab} \ =
&& \hspace{-4mm} G^{ab}f^c\partial_c (\partial_a\phi_b + \partial_b\phi_a)\\
= && \hspace{-4mm} \partial_a(G^{ab} f^c\partial_c \phi_b)
- G^{ab}\partial_af^c\partial_c\phi_b \nonumber \\
\label{eq.ginvJ-2}
&& \hspace{-4mm} +\, \partial_b(G^{ab} f^c\partial_c \phi_a)
- G^{ab}\partial_bf^c\partial_c\phi_a
\end{eqnarray}
\begin{eqnarray}
\label{eq.ginvJ-3}
G^{ab}\delta_{\mathrm{g}} h_{cb}\partial_a f^c \ =
&& \hspace{-4mm} G^{ab}\partial_af^c (\partial_c\phi_b + \partial_b\phi_c)\\
\label{eq.ginvJ-4}
= && \hspace{-4mm} \partial_b(G^{ab}\partial_a f^c\phi_c)
+ G^{ab}\partial_af^c \partial_c\phi_b
\end{eqnarray}
\begin{eqnarray}
\label{eq.ginvJ-5}
G^{ab}\delta_{\mathrm{g}} h_{ac}\partial_b f^c \ =
&& \hspace{-4mm} G^{ab}\partial_bf^c (\partial_a\phi_c + \partial_c\phi_a)\\
\label{eq.ginvJ-6}
= && \hspace{-4mm} \partial_a(G^{ab}\partial_b f^c\phi_c)
+ G^{ab}\partial_bf^c \partial_c\phi_a
\end{eqnarray}
\begin{eqnarray}
\label{eq.ginvJ-7}
-\frac{1}{4}G^{ab}(\partial_cf^c \delta_{\mathrm{g}} h_{ab}) \ =
&& \hspace{-4mm} -\frac{1}{4}G^{ab}\partial_cf^c (\partial_a\phi_b + \partial_b\phi_a)\\
\label{eq.ginvJ-8}
= && \hspace{-4mm} -\frac{1}{4}\partial_a(G^{ab}\partial_cf^c \phi_b)
-\frac{1}{4}\partial_b (G^{ab}\partial_cf^c \phi_a).
\end{eqnarray}
In (\ref{eq.ginvJ-1})-(\ref{eq.ginvJ-8}),
$\partial_aG^{ab} = 0$ and $\partial_{ab}f^c = 0$
are taken into account.
Using (\ref{eq.dgdh})-(\ref{eq.ginvJ-8}) and the symmetry of $G^{ab}$
in its indices, one finds
\begin{equation}
-G^{ab}\delta_{\mathrm{g}}\delta h_{ab} =
-\partial_a(2G^{ab}f^c\partial_c\phi_b + 2G^{ab}\partial_bf^c\phi_c
- \frac{1}{2}G^{ab}\partial_cf^c\phi_b),
\end{equation}
therefore the identity
$\partial_c \delta_\mathrm{g} \Jc_1^c = -G^{ab}\delta_\mathrm{g} \delta h_{ab}$
can be written as
\begin{equation}
\partial_c(\delta_\mathrm{g} \Jc_1^c
+ 2G^{cb}f^a\partial_a\phi_b + 2G^{cb}\partial_bf^a\phi_a
- \frac{1}{2}G^{cb}\partial_af^a\phi_b) = 0.
\end{equation}
This means that
$\delta_\mathrm{g} \Jc_1^c
+ 2G^{cb}f^a\partial_a\phi_b + 2G^{cb}\partial_bf^a\phi_a
- \frac{1}{2}G^{cb}\partial_af^a\phi_b$
is strongly conserved.
Since the last three terms in the latter expression
are weakly zero, it follows that
$\delta_\mathrm{g} \Jc_1^c$ is trivially conserved.

In view of (\ref{eq.Jc1a}),
$\delta_\mathrm{g} \Jc_1^c =
\delta_\mathrm{g}F^{cab}\delta h_{ab} +
F^{cab}\delta_\mathrm{g} \delta h_{ab}
- f^c \delta_\mathrm{g} L_1$.
For $\delta_\mathrm{g} L_1$ we found
$\delta_\mathrm{g} L_1 = \partial_{da}\phi_b F^{abd}$
in \cite{TG} (see Section 4.1 of \cite{TG} and (\ref{eq.lg.x3b})).
Using these formulae,
$\delta_\mathrm{g} \Jc_1^c
+ 2G^{cb}f^a\partial_a\phi_b + 2G^{cb}\partial_bf^a\phi_a
- \frac{1}{2}G^{cb}\partial_af^a\phi_b$
can be written as
\begin{equation}
\label{eq.I}
I_a{}^c \phi^a + I_a{}^{cd}\partial_d\phi^a
+ I_a{}^{cde}\partial_{de}\phi^a
\end{equation}
with
\begin{equation}
I_a{}^c = 2G^{cb}\partial_bf_a
- \frac{1}{2}G^c{}_a\partial_bf^b,
\end{equation}
\begin{equation}
I_a{}^{cd} = (F^{cb}{}_a + F^c{}_a{}^b)\partial_b f^d
+ (F^{cbd} + F^{cdb})\partial_b f_a
- \frac{1}{4}(F^c{}_a{}^d + F^{cd}{}_a)\partial_b f^b
+ 2G^c{}_af^d,
\end{equation}
\begin{eqnarray}
I_a{}^{cde} & \! = \! & -\frac{1}{2}(F^d{}_a{}^e + F^e{}_a{}^d)f^c
+\frac{1}{2}(F^{ce}{}_a + F^c{}_a{}^e)f^d
+\frac{1}{2}(F^{cd}{}_a + F^c{}_a{}^d)f^e \nonumber \\
&& +\, \frac{1}{4}(\delta h_a{}^d\eta^{ce} + \delta h_a{}^e\eta^{cd})
-\frac{1}{2}\delta h^{de}\eta_a{}^c
+\frac{1}{2}\delta h\, \eta^{de}\eta_a{}^c
-\frac{1}{2}\delta h_a{}^c \eta^{de} \nonumber \\
&& -\,\frac{1}{4}(\delta h\, \eta_a{}^d \eta^{ce} + \delta h\, \eta_a{}^e \eta^{cd})
+\frac{1}{4}(\delta h^{ce}\eta_a{}^d + \delta h^{cd}\eta_a{}^e).
\end{eqnarray}
In view of the form (\ref{eq.I}), a superpotential for
$\delta_\mathrm{g} \Jc_1^c
+ 2G^{cb}f^a\partial_a\phi_b + 2G^{cb}\partial_bf^a\phi_a
- \frac{1}{2}G^{cb}\partial_af^a\phi_b$
can be obtained using a general theorem that can be found
in \cite{Fletcher}, Section VI, or in somewhat different form in \cite{Wald1990}.
(It is also quoted in \cite{TG2017};
see equations (2.33)-(2.36) in Section II.D.)
It can be expressed in terms of $I_a{}^c$, $I_a{}^{cd}$ and
$I_a{}^{cde}$ as
\begin{equation}
\Sigma^{cd} = \frac{1}{2}(I_a{}^{cd} - I_a{}^{dc})\phi^a
+\left[\frac{2}{3}(I_a{}^{cde} - I_a{}^{dce})\partial_e\phi^a
-\frac{1}{3}\partial_e (I_a{}^{cde} - I_a{}^{dce})\phi^a \right].
\end{equation}
$\Sigma^{cd}$ is a bilinear local function of $h_{ab}$ and $\phi_a$.
The full (i.e., not only first order) variation of $\Jc_1^c$
under a gauge transformation is
$\delta_{\mathrm{g}} \Jc_1^c + \delta_{\mathrm{g}}^{(2)} \Jc_1^c$,
where $\delta_{\mathrm{g}}^{(2)} \Jc_1^c =
(1/2) (\delta_{\mathrm{g}} \Jc_1^c)|_{h_{ab}=\partial_a\phi_b+\partial_b\phi_a}$.
$(\delta_\mathrm{g} \Jc_1^c
+ 2G^{cb}f^a\partial_a\phi_b + 2G^{cb}\partial_bf^a\phi_a
- \frac{1}{2}G^{cb}\partial_af^a\phi_b)|_{h_{ab}=\partial_a\phi_b+\partial_b\phi_a}= (\delta_{\mathrm{g}} \Jc_1^c)|_{h_{ab}=\partial_a\phi_b+\partial_b\phi_a}$,
since $G^{cd}|_{h_{ab}=\partial_a\phi_b+\partial_b\phi_a}=0$,
therefore
$(\delta_{\mathrm{g}} \Jc_1^c)|_{h_{ab}=\partial_a\phi_b+\partial_b\phi_a}
= \partial_d \Sigma^{cd}|_{h_{ab}=\partial_a\phi_b+\partial_b\phi_a}$.
For the full variation of $\Jc_1^c$ we thus have the identity
\begin{equation}
\label{eq.idfullvar}
\delta_{\mathrm{g}} \Jc_1^c + \delta_{\mathrm{g}}^{(2)} \Jc_1^c
= -2G^{cb}f^a\partial_a\phi_b - 2G^{cb}\partial_bf^a\phi_a
+ \frac{1}{2}G^{cb}\partial_af^a\phi_b
+ \partial_d (\Sigma^{cd} + \Sigma^{(2)cd}),
\end{equation}
where $\Sigma^{(2)cd} = (1/2)\Sigma^{cd}|_{h_{ab}=\partial_a\phi_b+\partial_b\phi_a}$.
$-2G^{cb}f^a\partial_a\phi_b - 2G^{cb}\partial_bf^a\phi_a
+ \frac{1}{2}G^{cb}\partial_af^a\phi_b$ is evidently weakly zero
and $\partial_d (\Sigma^{cd} + \Sigma^{(2)cd})$ is strongly conserved,
therefore (\ref{eq.idfullvar})
implies that the full variation of $\Jc_1^c$ is trivially conserved.

\subsection{The intrinsic quantities accompanying the energy-momen-tum
tensor proposed in \cite{BHLthe}}
\label{sec.bhl}

In this subsection the extension of $\tau^{cd}$
into a conserved current associated with
Poincar\'e and dilatation symmetry is discussed.
$\bar{h}_{ab}$ will denote $h_{ab}-\frac{1}{2}h\eta_{ab}$.

In \cite{BHLthe} (see equations (13a), (13b)), the energy-momentum tensor
\begin{equation}
\label{eq.bhlt}
\tau^{cd} = \frac{1}{4}\partial^c\bar{h}_{ab}\partial^dh^{ab}
- \frac{1}{2} \partial^d \bar{h}^{ca} \partial_b \bar{h}^b{}_a
- \eta^{cd}\left(\frac{1}{8}\partial_e\bar{h}_{ab}\partial^eh^{ab} - \frac{1}{4}\partial_e\bar{h}^{ea}\partial_b\bar{h}^b{}_a\right)
\end{equation}
and the spin tensor
\begin{equation}
\label{eq.bhls}
s^{cde} = \mc{A}_{de}\bigl[\bar{h}_a{}^e\partial^c\bar{h}^{da}
- \bar{h}_a{}^e\partial^a\bar{h}^{dc}
+ \eta^{ce}\bar{h}^{da}\partial_b\bar{h}^b{}_a\bigr]
\end{equation}
were found,
which are generalizations of the energy-momentum and spin tensors found in \cite{BHLemt,BHLang} in harmonic gauge. 
(In comparison with the present paper,
the indices of the energy-momentum tensor are written in the opposite order in \cite{BHLemt,BHLang,BHLthe}.
In (\ref{eq.bhlt}) the indices of $\tau^{cd}$ are already swapped.
The normalization of the energy-momentum tensor also differs by a factor $1/2$ in \cite{BHLemt,BHLang,BHLthe},
and the definition of the spin tensor differs by a sign.)
As shown in \cite{BHLemt,BHLang,BHLthe}, these tensors have several notable properties;
in particular, $\tau^{cd}$ satisfies the dominant energy condition in TT gauge.

In order to give an account of $\tau^{cd}$ and $s^{cde}$ in the framework
applied in the present paper,
we first recall that in \cite{TG} we found that $2\tau^{cd}$
is the canonical energy-momentum tensor following from the Lagrangian density
\begin{equation}
L_{\tau} = L_1+\frac{1}{4}L_2.
\end{equation}
The canonical current $\Jc^c|_{\alpha_2=1/4,\, \alpha_3=0}$
(see (\ref{eq.lg.jc}) for $\Jc^c$) following from $L_{\tau}$
is thus a straightforward extension of $2\tau^{cd}$
to Poincar\'e and dilatation symmetries.
However, the spin tensor encoded in this current is
\begin{equation}
\label{eq.spin1}
S^{cde} = \mc{A}_{de}\bigl[\bar{h}_a{}^d \partial^c \bar{h}^{ae} - \bar{h}^{cd}\partial_a\bar{h}^{ae} + \eta^{cd}\bar{h}_b{}^e\partial_a\bar{h}^{ab}\bigr],
\end{equation}
which differs from $s^{cde}$ beyond normalization.
Although (\ref{eq.spin1}) was found (up to sign) in \cite{BHLang} in harmonic gauge,
it was not accepted as the definitive result.
The favoured spin tensor
was obtained in \cite{BHLang} by adding strongly conserved terms to (\ref{eq.spin1}).
It is possible to proceed in a similar way also in the present framework:
it can be verified by direct calculation that
the spin tensor encoded in 
\begin{equation}
\label{eq.jbhl}
J_{\tau}^c = \Jc^c|_{\alpha_2=1/4,\, \alpha_3=0}
-\frac{1}{2}\partial_d\Sigma_{2''}^{cd}+\frac{1}{2}\partial_d\Sigma_9^{cd}-\frac{1}{8}\partial_d\Sigma_8^{cd}
\end{equation}
is $-s^{cde}$. The last three terms on the right hand side of (\ref{eq.jbhl})
do not contribute to the energy-momentum tensor,
therefore the energy-momentum tensor contained by $J_{\tau}^c$
is still $2\tau^{cd}$.
This means that $J_{\tau}^c$
is a Noether current associated with Poincar\'e-dilatation symmetry
with the property that the energy-momentum tensor and the spin tensor encoded in it
are those of \cite{BHLemt,BHLang,BHLthe} (i.e., $2\tau^{cd}$ and $-s^{cde}$, respectively).

Regarding the virial current encoded in $J_{\tau}^c$,
the last three terms in (\ref{eq.jbhl}) do not contribute to it,
therefore it can be obtained by setting $\alpha_2=1/4$, $\alpha_3=0$ in (\ref{eq.cintrdil}):
\begin{equation}
\label{eq.tbhl}
\mbb{T}_{\tau}^c = \frac{1}{2}\bar{h}_{ab}\partial^ch^{ab}
- \bar{h}^c{}_a\partial_b\bar{h}^{ba}.
\end{equation}
Generally $\mbb{T}_{\tau}^c\ne 0$, as $\tau^{cd}$ is not traceless, and this remains true even in TT gauge.

$J_{\tau}^c$ is the special case of the general current (\ref{eq.lg.j})
with
$\lambda_2=\frac{1}{4}$, $\lambda_5=1$, $\lambda_{2''}=-\frac{1}{2}$,
$\lambda_8=-\frac{1}{8}$, $\lambda_9=\frac{1}{2}$
being the nonzero coefficients.
It differs from a canonical current only by a strongly conserved current,
therefore it is associated with `simple' Poincar\'e transformations
and dilatations, i.e.\ with (\ref{eq.lg.conf2}).

In harmonic gauge $\tau^{cd}$ is symmetric,
therefore in addition to $M^{cde}$ the spin tensor
and the orbital angular momentum tensor are also conserved.
The orbital angular momentum tensor can be seen to be associated
by Theorem \ref{thrm.ng}
with the variation
\begin{equation}
\label{eq.4.89}
\delta_{M_{\mathrm{orb}}} h_{ab}{^{de}} =
\delta_T h_{ab}{}^{[d} x^{e]} =
x^{[e}\partial^{d]}h_{ab}\, ,
\end{equation}
whereas $-s^{cde}$ is associated with
\begin{equation}
\label{eq.4.90}
\check{\delta}_S h_{ab}{}^{de} = 2h^{[d}{}_{(a}\eta^{e]}{}_{b)}.
\end{equation}
(\ref{eq.4.89}) and (\ref{eq.4.90})
are the usual variations of a rank $2$ tensor under
orbital and internal Lorentz transformations, respectively
(see Section \ref{sec.pdsym}).

\subsection{The intrinsic quantities accompanying the linearized\\ Landau--Lifshitz pseudotensor}
\label{sec.ll}

In this subsection the extension of the
linearized Landau--Lifshitz pseudotensor $T_{\mathrm{LL}}^{cd}$
into a current associated with Poincar\'e and dilatation symmetry is discussed.
Since $T_{\mathrm{LL}}^{cd}$ is symmetric,
a particularly interesting question is whether the spin tensor
accompanying it can be taken to be zero.
As in the previous subsection, $\bar{h}_{ab}$ denotes $h_{ab}-\frac{1}{2}h\eta_{ab}$.

In \cite{TG} we noted that the linearized Landau--Lifshitz pseudotensor
$T_{\mathrm{LL}}^{cd}$ (multiplied by $2$)
corresponds to the values
$\lambda_2=-\frac{1}{4}$, $\lambda_3=\lambda_4=0$, $\lambda_5=-1$, $\lambda_6=1$
in (\ref{eq.lg.emt}),
thus a straightforward extension of $2T_{\mathrm{LL}}^{cd}$
into a current associated with Poincar\'e-dilatation symmetry
is $J^c$ (i.e., (\ref{eq.lg.j})) with the same values of
$\lambda_2, \lambda_5, \lambda_6$ and all other coefficients zero.
This current will be denoted by $J_{\mathrm{LL'}}^c$.

Since $T_{\mathrm{LL}}^{cd}$ is symmetric, the orbital angular momentum tensor
and the spin tensor
encoded in $J_{\mathrm{LL'}}^c$ are separately conserved.
Using the general formula (\ref{eq.lg.sgen}), it can be verified by direct calculation
that the spin tensor is 
\begin{equation}
\label{eq.sll}
S_{\mathrm{LL'}}^{cde}  = \frac{1}{2}\partial_a(\bar{h}^{ae}\bar{h}^{cd}
- \bar{h}^{ad}\bar{h}^{ce}).
\end{equation}

From (\ref{eq.sll}) it is obvious that $S_{\mathrm{LL'}}^{cde}$ is not only conserved,
but also strongly conserved.
The strong conservation of $S_{\mathrm{LL'}}^{cde}$ suggests that it may be possible
to set the spin tensor accompanying
$T_{\mathrm{LL}}^{cd}$ to zero by adding some strongly conserved terms
to $J_{\mathrm{LL'}}^c$.
Indeed, by comparing (\ref{eq.sll}) with the general expression (\ref{eq.lg.sgen}),
it can be seen that the spin tensor encoded in
\begin{equation}
\label{eq.jllp}
J_{\mathrm{LL}}^c = J_{\mathrm{LL'}}^c 
+\frac{1}{2}\partial_d\Sigma_{2''}^{cd}-\frac{1}{2}\partial_d\Sigma_9^{cd}+\frac{1}{8}\partial_d\Sigma_8^{cd}
\end{equation}
(which we denote by $S_{\mathrm{LL}}^{cde}$)
is zero, while the energy-momentum tensor does not receive any contribution
from the additional terms on the right hand side
(interestingly, these terms coincide with the additional terms
in (\ref{eq.jbhl})).
$J_{\mathrm{LL}}^c$ is thus an extension of $2T_{\mathrm{LL}}^{cd}$
into a Noether current associated with Poincar\'e-dilatation symmetry,
with the special feature that the spin tensor encoded in it vanishes.
It is a special case of the general parametric current (\ref{eq.lg.j}),
as can be seen from (\ref{eq.jllp})
and from the definition of $J_{\mathrm{LL'}}^c$;
the relevant nonzero coefficients in (\ref{eq.lg.j}) are
$\lambda_2=-\frac{1}{4}$, $\lambda_5=-1$, $\lambda_6=1$,
$\lambda_{2''}=\frac{1}{2}$, $\lambda_8=\frac{1}{8}$, $\lambda_9=-\frac{1}{2}$.

The possibility of choosing the spin tensor accompanying $T_{\mathrm{LL}}^{cd}$ to be zero
is an additional notable property of $T_{\mathrm{LL}}^{cd}$
beyond its symmetry, and can be paralleled with the vanishing of the virial current accompanying $T_{\mathrm{lg}}^{cd}$.

The vanishing of $S_{\mathrm{LL}}^{cde}$ appears somewhat unsatisfactory,
if one wishes to interpret the spin tensor as a tensor that quantifies
the internal spinning motion of $h_{ab}$.
(In \cite{BHLang,BHLthe}, for instance, the authors argue that the spin tensor
should admit such an interpretation.)
On the other hand, the vanishing of $S_{\mathrm{LL}}^{cde}$
does not look unnatural
if one considers that the spin part of $\delta_\lambda h_{ab}$,
i.e.\ $\check{\delta}_{\lambda,S}h_{ab}{}^{cd}$
(see (\ref{eq.dlambdaSh})) is zero if $\lambda_5=-1$.
In addition, $J_{\mathrm{lg}}^c$ also has a similar feature,
namely the vanishing of $\mbb{T}_{\mathrm{lg}}^c$
and $\check{\delta}_{\mathbb{T}, \mathrm{lg}} h_{ab}$.
Furthermore, the vanishing of the spin tensor occurs in electrodynamics as well
(see Section \ref{sec.el.gentem}).

The virial currents encoded in $J_{\mathrm{LL'}}^c$ and $J_{\mathrm{LL}}^c$ are the same,
and since $T_{\mathrm{LL}}^{cd}$ is not traceless, this current is necessarily nonzero.
Explicitly,
\begin{equation}
\mbb{T}_{\mathrm{LL}}^c = -\frac{1}{2}h_{ab}\partial^c h^{ab}
+ h^{cb}\partial_a h^a{}_b + \frac{1}{2}h\partial_a h^{ac}
-\frac{1}{2}h^{ca}\partial_a h
-\frac{1}{2}h\partial^c h.
\label{eq.tll}
\end{equation}
In TT gauge, $\mbb{T}_{\mathrm{LL}}^c = - \mbb{T}_{\tau}^c =  -\frac{1}{2}h_{ab}\partial^c h^{ab}$.

We have seen that $J_{\mathrm{LL}}^c$ and $J_{\tau}^c$ are both special cases of $J^c$.
The relevant values of the parameters differ only in their sign,
with the exception of the value of $\lambda_6$,
which is zero in the case of $J_{\tau}^c$.
As a consequence, the relation
\begin{equation}
\label{eq.lg-bhl-ll}
J_{\mathrm{lg}}^c = \frac{1}{2}(J_{\tau}^c + J_{\mathrm{LL}}^c
- F^{cab}\partial_a(hf_b))
\end{equation}
holds. This implies that there exists a simple relation between
$J_{\mathrm{lg}}^c$, $J_{\tau}^c$ and $J_{\mathrm{LL}}^c$
in those gauges where the last term is zero:
\begin{equation}
\label{eq.ttjrel}
J_{\mathrm{lg}}^c = \frac{1}{2}(J_{\tau}^c + J_{\mathrm{LL}}^c).
\end{equation}
This relation holds, in particular, in TT gauge.

\section{Conformal symmetry of linearized gravity
in generalized harmonic gauges and the associated conserved currents}
\label{sec.gharm}

In this section, the results of Section \ref{sec.lg.em}
concerning the conserved currents associated with
Poincar\'e and dilatation symmetry
are extended to the full conformal algebra.
This is made possible by the generalization of Noether's theorem
in Section \ref{sec.nthr}
(i.e., Definition \ref{def.nsg} and Theorem \ref{thrm.ng})
and the finding, presented below in Section \ref{sec.scslg},
that in generalized harmonic gauges linearized gravity
does have, in the sense of Definition \ref{def.nsg},
full conformal symmetry at the level of the Lagrangian.

\subsection{Variation of $h_{ab}$}

The (`simple') variation of $h_{ab}$ under conformal transformations
can be taken to be (\ref{eq.lg.conf2}) (see also (\ref{eq.lg.conf2b}))
with $f^c$ understood to be the complete generator
$e^c + \omega^{cd}x_d + \beta x^c + 2x^c c^a x_a - c^c x_a x^a$.
The variation of $h_{ab}$ under Poincar\'e transformations and dilatations
thus remain the same as in Section \ref{sec.lg.em},
and equations (\ref{eq.lg.conf2b})-(\ref{eq.hscalingdim}) remain valid.
From (\ref{eq.lg.conf2b}) it can be seen that
\begin{equation}
\label{eq.5.1}
\check{\delta}_{\mathbb{C}}h_{ab}{}^c \ = \ 0.
\end{equation}
For the variation of $h_{ab}$ under special conformal transformations,
i.e.\ for $\delta_C h_{ab}{}^c$, we have
\begin{eqnarray}
\label{eq.5.2}
\delta_C h_{ab}{}^c & = & 2x^cx_d\delta_T h_{ab}{}^d
- x_dx^d\delta_T h_{ab}{}^c
+ 2x^c\check{\delta}_{\mathbb{T}}h_{ab}
-4x_d\check{\delta}_S h_{ab}{}^{cd}\\[1mm]
& = & 2x^cx_d \partial^dh_{ab} - x_dx^d \partial^c h_{ab} + 2x^ch_{ab}
 - 4(x_{(a}h_{b)}{}^c - x^dh_{d(a}\eta^c{}_{b)}).
\label{eq.5.2b}
\end{eqnarray}
This equation
is the specialization of (\ref{eq.rel-dC1}) to the present case.

As in Section \ref{sec.lg.em}, (\ref{eq.lg.conf2})
will be labeled as `simple' variation
in order to distinguish it from the modified variations.

\subsection{Conformal symmetry of linearized gravity in generalized harmonic gauges}
\label{sec.scslg}

In Section \ref{sec.pdsym} it was seen that the variation of $L_1$
corresponding to the `simple' variation (\ref{eq.lg.conf2}) of $h_{ab}$
under conformal transformations is
\begin{equation}
\delta L_1  =  \partial_c (f^c L_1) - \frac{1}{2} F_a h^{ab}\Box f_b\, .
\end{equation}
Due to the second term on the right hand side,
$\delta L_1$ is not a total divergence
if $f^c$ generates special conformal transformations,
i.e.\ special conformal transformations are not Noether symmetries of $L_1$.

Nevertheless, $F_a h^{ab}\Box f_b$ is zero in the $F_a=0$ gauge,
and in other generalized harmonic gauges
it is a total divergence, as a consequence of (\ref{eq.faghg}).
Specifically, it is not difficult to verify that
\begin{equation}
\label{eq.Fhf-id}
F_a h^{ab} \Box f_b = (\chi-1)\partial_c\left(hh^{cd}\Box f_d
- \frac{\chi}{2}h^2 \Box f^c\right)
+ X^a(\chi)\bigl(h_a{}^b -(\chi-1)h\delta_a^b\bigr)\Box f_b
\end{equation}
holds identically ($X^a(\chi)$ is given by (\ref{eq.ghg})), therefore
\begin{equation}
\label{eq.adiv}
F_a h^{ab} \Box f_b = (\chi-1)\partial_c\left(hh^{cd}\Box f_d
- \frac{\chi}{2}h^2 \Box f^c\right)
\end{equation}
if $h_{ab}$ is in the generalized harmonic gauge
with the parameter $\chi$.
This means that in the generalized harmonic gauges linearized gravity
does have complete conformal symmetry at the level of the Lagrangian,
in the sense of Definition \ref{def.nsg}.
Moreover, $\delta L_1$ takes the standard form $\partial_c (f^c L_1)$
in the $F_a=0$ gauge.
This then holds in the subgauges of the $F_a=0$ gauge as well,
i.e., for instance, in the TT gauge.

In view of the conformal symmetry of linearized gravity described above,
it is possible to construct conserved currents
associated with the entire conformal symmetry algebra
in generalized harmonic gauges by applying Theorem \ref{thrm.ng}.
Accordingly, in the following subsections we extend the conserved currents
associated with Poincar\'e-dilatation symmetry
found in Section \ref{sec.lg.em} into currents associated with
full conformal symmetry.

The variation of $X^a(\chi)$ due to the variation (\ref{eq.lg.conf2}) of $h_{ab}$
under conformal transformations is
\begin{equation}
\delta X^a(\chi) = f^c\partial_c X^a(\chi) + \partial^a f_c X^c(\chi)
+ \frac{1}{4}\partial_c f^c X^a(\chi)
+\frac{3}{2}h^{ac}\Box f_c+\frac{\chi-1}{2}h\Box f^a.
\end{equation}
This formula shows that Poincar\'e transformations and dilatations preserve
the generalized harmonic gauge condition $X^a(\chi)=0$,
whereas special conformal transformations do not.
Special conformal transformations are thus examples of transformations
that are Noether symmetries in some gauge without preserving that gauge.

\subsection{Canonical currents}
\label{sec.ccghg}

From (\ref{eq.lg.dlI}), (\ref{eq.lg.dlII}), (\ref{eq.lg.dlIII}) and (\ref{eq.adiv}),
it can be seen that
\begin{equation}
\label{eq.cank5.2}
K^c = f^c L 
+\frac{1-\chi}{2}\left(hh^{ca}\Box f_a - \frac{\chi}{2}h^2 \Box f^c\right)
-\frac{\alpha_2}{2}(h^{ab}h_{ab}\Box f^c + 4h^c{}_ah^{ab} \Box f_b - 2h h^{ca}\Box f_a)
\end{equation}
can be chosen in generalized harmonic gauges in the identity
$\delta L=\partial_c K^c$
for the variation of $L$ ($L$ is given by (\ref{eq.lg.l}))
under conformal transformations
(under which the variation of $h_{ab}$
is the `simple' variation (\ref{eq.lg.conf2})).
In the $F_a=0$ gauge the second term
on the right hand side of (\ref{eq.cank5.2}) is zero,
since the $F_a=0$ gauge corresponds to $\chi=1$.

In any subgauge of a generalized harmonic gauge corresponding to
some value of $\chi$,
(\ref{eq.cank5.2}) with the same $\chi$ can be taken.
If a gauge $\mathcal{G}$ is a subgauge of more than one generalized harmonic gauges,
then it is a subgauge of all generalized harmonic gauges,
therefore in $\mathcal{G}$ any value can be chosen for $\chi$ in (\ref{eq.cank5.2}).
This holds, in particular, for the CTH gauge.
A subgauge of all generalized harmonic gauges
is obviously a subgauge of the CTH gauge,
therefore the arbitrariness of $\chi$ (in (\ref{eq.cank5.2})) in CTH gauge implies the arbitrariness of $\chi$
in $\mathcal{G}$.

A consequence of the arbitrariness of $\chi$ in (\ref{eq.cank5.2})
in the CTH gauge (and in its subgauges)
is that that $hh^{ca}\Box f_a$ and $h^2 \Box f^c$,
which appear in (\ref{eq.cank5.2})
with coefficients depending on $\chi$,
must be conserved in the CTH gauge (and hence also in its subgauges)
regardless of the linearized Einstein equations.
The conservation of $hh^{ca}\Box f_a$ and $h^2 \Box f^c$
can be verified directly as well:
in the CTH gauge (\ref{eq.sg}) holds,
and from (\ref{eq.sg}) it follows that $hh^{ca}\Box f_a$ and $h^2 \Box f^c$
are conserved,
as can be easily seen by calculating
$\partial_c(hh^{ca}\Box f_a)$ and $\partial_c (h^2 \Box f^c)$.
The linearized Einstein equations are not needed in the calculation.
$h^2 \Box f^c$ is merely a constant vector in the CTH gauge.

It must be noted that the $K^c$ given by (\ref{eq.cank5.2})
has a nonzero `fully intrinsic' part because of the terms following $f^cL$;
\begin{equation}
K_T^{cd} = \eta^{cd}L,\qquad \check{K}_S^{cde} = 0,\qquad
\check{K}_{\mathbb{T}}^c = 0,
\end{equation}
\begin{equation}
\check{K}_{\mathbb{C}}^{cd} =
2(\chi-1)\left(hh^{cd} - \frac{\chi}{2}h^2 \eta^{cd}\right)
+ 2\alpha_2 (h^{ab}h_{ab}\eta^{cd} + 4h^c{}_ah^{ad}  - 2h h^{cd}).
\end{equation}
On the other hand, the `fully intrinsic' part of $\delta h_{ab}$ is zero
(see (\ref{eq.lg.conf2}), (\ref{eq.lg.conf2b})-(\ref{eq.hdbbT})
for $\delta h_{ab}$ and $\delta_Th_{ab}{}^c$,
$\check{\delta}_Sh_{ab}{}^{cd}$, $\check{\delta}_{\mathbb{T}}h_{ab}$).

With the choice (\ref{eq.cank5.2}),
the Noether current associated with the `simple' variation (\ref{eq.lg.conf2}) is
\begin{equation}
\Jc^c = \Jc_1^c + \alpha_2 \Jc_2^c + \alpha_3 \Jc_3^c\, ,
\label{eq.lg.jcghg}
\end{equation}
where
\begin{eqnarray}
\label{eq.Jc1a-conf}
\Jc_1^c \hspace{-5.5mm} && \, = \,
F^{cab} \delta h_{ab} -
\frac{1}{2}f^c F^{dab}\mathring{F}_{dab}
+ \frac{\chi-1}{2}\biggl(hh^{ca}\Box f_a - \frac{\chi}{2}h^2 \Box f^c\biggr) \\[2mm]
\label{eq.Jc2a-conf}
\Jc_2^c \hspace{-5.5mm} && \, = \, 2(\partial^a h^{bc}\delta h_{ab}
- \partial_b h^{ba}\delta h_a{}^c) - f^c L_2 \nonumber \\[1mm]
&& \hspace{5mm} +\, \frac{1}{2}(h^{ab}h_{ab}\Box f^c
+ 4h^c{}_ah^{ab} \Box f_b - 2h h^{ca}\Box f_a) \\[1mm]
&& \, = \, \partial_d \Sigma_2^{cd} \\[2mm]
\label{eq.Jc3a-conf}
\Jc_3^c \hspace{-5.5mm} && \, = \, \tilde{\mathring{F}}^{cab}\delta h_{ab}
- \frac{1}{2}f^c \tilde{\mathring{F}}_{dab}\mathring{F}^{dab} \, = \,
\partial_d\Sigma_3^{cd}.
\end{eqnarray}
$\Jc_1^c$, $\Jc_2^c$ and $\Jc_3^c$
are the currents following from $L_1$, $L_2$ and $L_3$, respectively.
Although additional terms appear in (\ref{eq.Jc2a-conf})
in comparison with (\ref{eq.Jc2a}),
the superpotential $\Sigma_2^{cd}$ turns out to remain the same
(i.e., it is given by the formulae (\ref{eq.lg.jc1}) and (\ref{eq.sigma22}),
though, of course,
now $f^c$ is $e^c + \omega^{cd}x_d + \beta x^c + 2x^c c^a x_a - c^c x_a x^a$
instead of $e^c + \omega^{cd}x_d + \beta x^c$).
The formulae for $\Jc_3^c$ do not change in comparison with (\ref{eq.Jc3}),
(\ref{eq.lg.jc2}), (\ref{eq.lg.sigma3p}).

The current (\ref{eq.lg.jcghg}) is formally close to (\ref{eq.lg.jc})
and they coincide in the case of Poincar\'e transformations and dilatations,
therefore we take (\ref{eq.lg.jcghg})-(\ref{eq.Jc3a-conf})
to be the generalization of (\ref{eq.lg.jc})-(\ref{eq.Jc3})
to full conformal symmetry.
(\ref{eq.Jc1a-conf}) and (\ref{eq.Jc1a}) also take the same form
in the $F_a=0$ gauge, where $\chi=1$.
In the subgauges of all generalized harmonic gauges
(i.e., in the CTH gauge and in its subgauges),
an arbitrary value can be chosen for $\chi$,
in agreement with the same arbitrariness of the value of $\chi$ in (\ref{eq.cank5.2}).

The energy-momentum tensor, spin tensor and virial current encoded in (\ref{eq.lg.jcghg})
are the same as in Section \ref{sec.cc}, and the $\mbb{C}^{cd}$ tensor encoded in it is
\begin{eqnarray}
\bbCc^{cd} & \! = \! & \frac{\chi-1}{2}(-4hh^{cd} + 2\chi h^2 \eta^{cd}) \nonumber\\
&& +\, \alpha_2 (-2h^{ab}h_{ab}\eta^{cd}-8h^{ca}h_a{}^d + 4hh^{cd}).
\end{eqnarray}
It is worth noting that while the spin tensor and the virial current come
entirely from $\delta h_{ab}$ (i.e., $K^c$ does not contribute to them),
$\bbCc^{cd}$ comes entirely from $K^c$,
as the `fully intrinsic' part of $\delta h_{ab}$, and hence of $j^c$,
is zero. We also note that $\bbCc^{cd}$ is symmetric.

The special conformal tensor encoded in $\Jc^c$
is given by the general relation (\ref{eq.bbT2}), i.e.
\begin{equation}
\Cc^{cd} = 2\Tc^{ca}x_a x^d - \Tc^{cd}x_a x^a
+ 2\bbTc^c x^d -4\Sc^{cde}x_e
+ \bbCc^{cd}.
\end{equation}
$\Cc^{cd}$ is associated with the variation (\ref{eq.5.2}).

\subsection{Extension of $J_{\mathrm{lg}}^c$ to full conformal symmetry}
\label{sec.lgghg}

The definition of $J_{\mathrm{lg}}^c$ was
$J_{\mathrm{lg}}^c = \Jc^c_1
- F^{cda}\partial_d (h_{ab} f^b)$
in the case of Poincar\'e-dilatation symmetry (see (\ref{eq.defjlg})).
This definition is also suitable in the case of the full conformal symmetry;
the definition of $\Jc^c$ has been extended to full conformal symmetry
in Section \ref{sec.ccghg} (see (\ref{eq.lg.jcghg})-(\ref{eq.Jc3a-conf})),
and the current $F^{cda}\partial_d (h_{ab} f^b)$
(see (\ref{eq.fourc})), originating from
the gauge symmetry of linearized gravity,
extends to the entire conformal algebra in a straightforward way
(by simply replacing $f^c = e^c + \omega^{cd}x_d + \beta x^c$
with $f^c = e^c + \omega^{cd}x_d + \beta x^c + 2x^c c^a x_a - c^c x_a x^a$).
Taking into account (\ref{eq.Jc1a-conf}), we have
\begin{eqnarray}
J_{\mathrm{lg}}^c & \! = \! & \Jc^c_1 - F^{cda}\partial_d (h_{ab} f^b) \nonumber \\
& \! = \! & T_{\mathrm{lg}}^{cd}f_d
+ F^{ca[d}h_a{}^{e]} \partial_{[d} f_{e]}
+\frac{\chi-1}{2}\biggl(hh^{ca}\Box f_a - \frac{\chi}{2}h^2 \Box f^c\biggr).
\label{eq.lg.jstghg}
\end{eqnarray}
In comparison with (\ref{eq.lg.jst}), (\ref{eq.lg.jstghg}) contains an extra term,
namely the term with the coefficient $\chi-1$.
The formula for the variation with which $J_{\mathrm{lg}}^c$
is associated as a Noether current remains (\ref{eq.dhst});
see also Section \ref{sec.genjconf}.

In the $F_a=0$ gauge, where $\chi-1=0$, the extra term
in (\ref{eq.lg.jstghg}) becomes zero.
However, in the CTH gauge and in its subgauges $\chi$ can have any value
(see Section \ref{sec.ccghg}), thus the extra term may be nonzero.
In TT gauge $h=0$, therefore the extra term is zero
regardless of the value of $\chi$.

The energy-momentum tensor, spin tensor and virial current encoded in (\ref{eq.lg.jstghg})
are the same as in Section \ref{sec.gentlg}, and the $\mbb{C}^{cd}$ tensor,
which originates from $\Jc^c_1$, is
\begin{equation}
\label{eq.clg}
\mbb{C}_{\mathrm{lg}}^{cd} \ = \  \frac{\chi-1}{2}(-4hh^{cd} + 2\chi h^2 \eta^{cd}).
\end{equation}
$\mbb{C}_{\mathrm{lg}}^{cd} = 0$ in the $F_a=0$ gauge
and in the TT gauge (among others); in these gauges
only the spin tensor is not zero
among the basic intrinsic quantities
$S_{\mathrm{lg}}^{cde}$, $\mbb{T}_{\mathrm{lg}}^c$,
$\mbb{C}_{\mathrm{lg}}^{cd}$.
The latter property is favourable
from the point of view of similarity to electrodynamics
(where all intrinsic quantities are zero---see Section \ref{sec.el.gentem}).

The special conformal tensor encoded in $J_{\mathrm{lg}}^c$
is given by the general formula (\ref{eq.bbT2});
\begin{equation}
C_{\mathrm{lg}}^{cd} \ = \
2T_{\mathrm{lg}}^{ca}x_a x^d - T_{\mathrm{lg}}^{cd}x_a x^a
- 4S_{\mathrm{lg}}^{cde}x_e
+ \mbb{C}_{\mathrm{lg}}^{cd}.
\end{equation}
For the variation with which $C_{\mathrm{lg}}^{cd}$ is associated,
see (\ref{eq.dClg}).

The $K^c$ quantity for $J_{\mathrm{lg}}^c$,
assuming, as before, that the Lagrangian density is taken to be $L_1$,
is
\begin{equation}
\label{eq.ktaraconf}
K_{\mathrm{lg}}^c = f^c L_1
+ \frac{1-\chi}{2}\left(hh^{ca}\Box f_a - \frac{\chi}{2}h^2 \Box f^c\right)
- \frac{1}{2}F^{abc}\mathring{F}_{abd}f^d
-\frac{1}{2}h_a{}^{[e} F^{d]ac} \partial_{[d} f_{e]}\, .
\end{equation}
In comparison with (\ref{eq.ktara}),
(\ref{eq.ktaraconf}) contains an additional term, which is the term
with the coefficient $(1-\chi)/2$. This term originates from
the $K^c$ for $\Jc^c_1$ (see (\ref{eq.cank5.2})).
Due to this term, $K_{\mathrm{lg}}^c$ has a `fully intrinsic' part,
$\check{K}_{\mathbb{C},\mathrm{lg}}^{cd} =
2(\chi-1)\left(hh^{cd} - \frac{\chi}{2}h^2 \eta^{cd}\right)$.
For $K_{T_{\mathrm{lg}}}^{cd}$,
$\check{K}_{S, \mathrm{lg}}^{cde}$ and
$\check{K}_{\mathbb{T}, \mathrm{lg}}^c$,
see (\ref{eq.lg.KT})-(\ref{eq.lg.KbbT}).

\subsection{General parametric current associated with conformal symmetry}
\label{sec.genjconf}

The general current $J^c$, given by (\ref{eq.lg.j}), can be extended to full conformal symmetry
in a straightforward way: $J_{\mathrm{lg}}^c$ on the right hand side
should be understood to be given by (\ref{eq.lg.jstghg}),
while the other terms remain formally the same, though
it is understood, of course, that
$f^c = e^c + \omega^{cd}x_d + \beta x^c$ is replaced with
$f^c = e^c + \omega^{cd}x_d + \beta x^c + 2x^c c^a x_a - c^c x_a x^a$.
The energy-momentum tensor, spin tensor and virial current
encoded in the extended $J^c$ are then the same as in Section \ref{sec.genjpd}.

For the $\mbb{C}^{cd}$ tensor encoded in the extended $J^c$, one finds
\begin{eqnarray}
\mbb{C}^{cd} & \! = \! & \frac{\chi-1}{2}(-4hh^{cd} + 2\chi h^2 \eta^{cd}) \nonumber\\
&& +\, \lambda_2(-2h_{ab}h^{ab}\eta^{cd} - 8h^{ca}h_a{}^d + 4hh^{cd}) \nonumber \\
&& +\, \lambda_{2'}(-2h_{ab}h^{ab}\eta^{cd} - 4h^{ca}h_a{}^d)
+ \lambda_{2''}(-4h^{ca}h_a{}^d  + 4hh^{cd}) \nonumber \\
&& +\, 12\lambda_7 h_{ab}h^{ab}\eta^{cd} + 12\lambda_8 h^2\eta^{cd}
+ \lambda_9 (4hh^{cd} + 2h^2\eta^{cd}).
\label{eq.cghg}
\end{eqnarray}
The special conformal tensor can be obtained
in terms of the energy-momentum tensor,
spin tensor, virial current and $\mbb{C}^{cd}$ by applying (\ref{eq.bbT2}).

The variation $\delta_\lambda h_{ab}$ with which $J^c$ is associated
as a Noether current in generalized harmonic gauges
is given by the same formulae as previously in Section \ref{sec.genjnoether},
i.e.\ by equations (\ref{eq.dlh1})-(\ref{eq.dlambdabbTh}),
with $f^c = e^c + \omega^{cd}x_d + \beta x^c + 2x^c c^a x_a - c^c x_a x^a$.
In particular,
\begin{equation}
\label{eq.dlambdabbCh}
\check{\delta}_{\lambda,\mathbb{C}}h_{ab}{}^c = 0.
\end{equation}
The variation of $h_{ab}$ corresponding to
special conformal transformations is given
in terms of $\delta_{\lambda,T}h_{ab}{}^d$,
$\check{\delta}_{\lambda,S}h_{ab}{}^{cd}$
and $\check{\delta}_{\lambda,\mathbb{T}}h_{ab}$
by the following formula,
obtained by specializing (\ref{eq.rel-dC1}):
\begin{equation}
\label{eq.dlambdaCh}
\delta_{\lambda,C}h_{ab}{}^c \ = \
2x^cx_d\delta_{\lambda,T}h_{ab}{}^d
- x_dx^d\delta_{\lambda,T}h_{ab}{}^c
+ 2x^c\check{\delta}_{\lambda,\mathbb{T}}h_{ab}
- 4x_d\check{\delta}_{\lambda,S}h_{ab}{}^{cd}\, .
\end{equation}
In the case of $J_{\mathrm{lg}}^c$, this reduces to
\begin{equation}
\label{eq.dClg}
\delta_{C_{\mathrm{lg}}} h_{ab}{}^c \ = \
2x^cx_d\delta_{T_{\mathrm{lg}}} h_{ab}{}^d
- x_dx^d\delta_{T_{\mathrm{lg}}} h_{ab}{}^c
- 4x_d\check{\delta}_{S, \mathrm{lg}} h_{ab}{}^{cd} \, ,
\end{equation}
where $\delta_{T_{\mathrm{lg}}} h_{ab}{}^d$ and
$\check{\delta}_{S, \mathrm{lg}} h_{ab}{}^{cd}$
are the same as in Section \ref{sec.gentlg}.

The $K^c$ quantity for the extended $J^c$ is
\begin{eqnarray}
K^c & \! = \! & f^c L_1 + \frac{1-\chi}{2}\left(hh^{ca}\Box f_a - \frac{\chi}{2}h^2 \Box f^c\right) \nonumber\\
&& +\, \frac{1}{2}(\lambda_5-1)(F^{abc}\mathring{F}_{abd}f^d  + F^{abc}h_{bd}\partial_a f^d)
+\frac{1}{2}\lambda_6 F^{abc}\partial_a(hf_b) \nonumber \\
&& - \,\lambda_2\partial_d\Sigma_2^{cd} - \dots,
\label{eq.kcgenconf}
\end{eqnarray}
where the last terms, indicated by the dots, are the same as in (\ref{eq.kcgen}).
In comparison with (\ref{eq.kcgen}),
(\ref{eq.kcgenconf}) contains one additional term,
which is the one with the coefficient $(1-\chi)/2$.

We note that there are further possible terms
that could be included in the general expression for $J^c$,
which can be obtained by the replacement $f^a\to\Box f^a$ in the terms
$F^{cab}\partial_a (h_{bd} f^d)$, $F^{cab} \partial_a (h f_b)$,
$\partial_d \Sigma_3^{cd}$, $\partial_d \Sigma_4^{cd}$, $\partial_d \Sigma_2^{cd}$
already present in $J^c$.
They are zero for translations, Lorentz transformations and dilatations,
and contribute only to $\mbb{C}^{cd}$.
The dimension of the coefficients of these terms would be
different from the dimension of the coefficients of the terms already present,
because the dimension of $\Box f^a$ differs from that of $f^a$.

\subsubsection{Divergence of the special conformal tensor
in the presence of matter}
\label{sec.divcmat}

The divergence of the special conformal tensor
in the presence of matter can be extracted,
assuming that the generalized harmonic gauge condition holds
with some value of $\chi$,
from the formula $\partial_c J^c = - \mc{T}^{ab}\delta_\lambda h_{ab}$,
which is valid according to the generalized Noether's theorem
(Theorem \ref{thrm.ng}).
We have
\begin{equation}
\partial_cC^{cd} \ = \ -\mc{T}^{ab}\delta_{\lambda,C} h_{ab}{}^d\, ,
\end{equation}
and from this equation we obtain, using (\ref{eq.dlambdaCh})
and (\ref{eq.dlambdaTh})-(\ref{eq.dlambdabbTh}),
the explicit result
\begin{eqnarray}
\partial_cC^{cd} & \! = \! &  -\mc{T}^{ab}(4x^dx^c\mathring{F}_{cab} - 2x_cx^c\mathring{F}^d{}_{ab}) \nonumber \\
&& +\, 2\mc{T}^{ab}x_a h_b{}^d
 -2\mc{T}^{da}x^b h_{ab} \nonumber \\
&& +\, \lambda_5(-2\mc{T}^{ab}x^cx^d \partial_ah_{bc}
+\mc{T}^{ab}x_cx^c\partial_ah_b{}^d) \nonumber \\
&& -\, 2\lambda_5(\mc{T}^{ab}x^d h_{ab} + \mc{T}^{da}x^b h_{ab} - \mc{T}^{ab}x_a h_b{}^d) \nonumber \\
&& -\, \lambda_6 (2\mc{T}^{ab}x_ax^d \partial_b h - \mc{T}^{ad}x_cx^c\partial_a h + 2\mc{T}_a{}^a x^d h).
\end{eqnarray}
The case $\lambda_5=\lambda_6=0$ (which is relevant to $C_{\mathrm{lg}}^{cd}$) can be compared with (\ref{eq.em.div4}).
In the case $\lambda_5=1$, $\lambda_6=0$,
when $\delta_\lambda h_{ab}$ is the `simple' variation
given by (\ref{eq.lg.conf2}),
\begin{eqnarray}
\partial_c C^{cd}|_{\lambda_5=1,\lambda_6=0} & \! = \! &
-\mc{T}^{ab}(2x^dx^c\partial_c h_{ab}
- x_cx^c\partial^d h_{ab})\nonumber \\[1mm]
&& -\, 2\mc{T}^{ab}x^d h_{ab}
- 4\mc{T}^{da}x^b h_{ab} + 4\mc{T}^{ab}x_a h_b{}^d\, .
\end{eqnarray}
The latter formula is relevant, in particular,
to the special conformal tensors
encoded in the canonical currents (\ref{eq.lg.jcghg}).

\subsection{Extension of $J_{\tau}^c$ to full conformal symmetry}
\label{sec.kghg}

The extension of $J_{\tau}^c$ (see Section \ref{sec.bhl})
to the complete conformal algebra
is straightforward,
as all terms in the formula (\ref{eq.jbhl})
that defines $J_{\tau}^c$ have already been extended.
Accordingly, we take the extended $J_{\tau}^c$ to be the extended $J^c$
with the same values of the parameters as in Section \ref{sec.bhl}
(explicitly given below (\ref{eq.tbhl})).
The $\mbb{C}^{cd}$ tensor that is encoded
in the extended $J_{\tau}^c$ is
\begin{equation}
\label{eq.cbhl}
\mbb{C}_{\tau}^{cd} = \frac{\chi-1}{2}(-4hh^{cd} + 2\chi h^2 \eta^{cd})
-\frac{1}{2}h^{ab}h_{ab}\eta^{cd}+hh^{cd}-\frac{1}{2}h^2\eta^{cd}.
\end{equation}
In TT gauge, $\mbb{C}_{\tau}^{cd} = -\frac{1}{2}h^{ab}h_{ab}\eta^{cd}$.
From Proposition \ref{prop.2.4} it follows that $\mbb{C}_{\tau}^{cd}$
could not be zero, as $\tau_c{}^c\ne 0$ and $s_c{}^{cd}=0$ in TT gauge.

\subsection{Extension of $J_{\mathrm{LL}}^c$ to full conformal symmetry}
\label{sec.llconf}

$J_{\mathrm{LL}}^c$ (defined by (\ref{eq.jllp}))
can also be extended to the complete conformal algebra in a straightforward way:
$J_{\mathrm{LL}}^c$ is a special case of the general parametric current $J^c$ defined by (\ref{eq.lg.j})
(see Section \ref{sec.ll} for the values of the parameters),
therefore
the extended $J_{\mathrm{LL}}^c$ can be taken to be the extended $J^c$,
introduced in Section \ref{sec.genjconf},
with the same values of the parameters.

The $\mbb{C}^{cd}$ tensor encoded in the extended $J_{\mathrm{LL}}^c$ is
\begin{equation}
\mbb{C}_{\mathrm{LL}}^{cd} =  \frac{\chi-1}{2}(-4hh^{cd} + 2\chi h^2 \eta^{cd})
+\frac{1}{2}h^{ab}h_{ab}\eta^{cd}-hh^{cd}+\frac{1}{2}h^2\eta^{cd}.
\label{eq.cll}
\end{equation}
We note that the only difference between $\mbb{C}_{\mathrm{LL}}^{cd}$
and $\mbb{C}_{\tau}^{cd}$
is that the last three terms appear with opposite sign in them.
In TT gauge, $\mbb{C}_{\mathrm{LL}}^{cd} = - \mbb{C}_{\tau}^{cd}
= \frac{1}{2}h^{ab}h_{ab}\eta^{cd}$.

The relation (\ref{eq.lg-bhl-ll}) remains valid for the extended
$J_{\mathrm{lg}}^c$, $J_{\tau}^c$ and $J_{\mathrm{LL}}^c$ currents,
since these currents are special cases of the extended general current $J^c$
with values of the parameters that are the same as before the extension.
The term $F^{cab}\partial_a(hf_b)$ in (\ref{eq.lg-bhl-ll})
does not give any contribution to $\mbb{C}^{cd}$, therefore
\begin{equation}
\label{eq.ttcrel}
\mbb{C}_{\mathrm{lg}}^{cd} = \frac{1}{2}(\mbb{C}_{\tau}^{cd}
+ \mbb{C}_{\mathrm{LL}}^{cd}).
\end{equation}

\section{Summary}
\label{sec.concl}

We explored the conserved currents in linearized gravity associated with Lorentz transformations,
dilatations and special conformal transformations,
extending our previous work \cite{TG}
on the energy-momentum tensor of the linearized gravitational field.

In preparation for the investigation,
we introduced in Section \ref{sec.conf}
a general elementary framework for discussing
conserved currents associated with conformal symmetry.
This included a general definition of the orbital and
intrinsic parts of the angular momentum tensor,
dilatation current and special conformal tensor,
and also of other related quantities,
in particular the variations of fields
under conformal transformations.
In Section \ref{sec.nthr}, we provided a generalization of
Noether's theorem and its converse
that is relevant if additional conditions,
such as gauge fixing conditions,
are imposed on the dynamical fields.
This generalization is based on a corresponding
generalization of Noether symmetry (Definition \ref{def.nsg}).

We consider Section \ref{sec.conf} to be an important part of the paper.
The framework introduced there is not specialized to linearized gravity
and to $4$-dimensional spacetime,
and it can be expected that it will also be useful in further studies.
The generalized Noether's theorem
introduced in Section \ref{sec.nthr}
is a general theorem in
Lagrangian field theory and mechanics;
it is independent of linearized gravity
and conformal symmetry.

Special conformal transformations are not Noether symmetries of linearized gravity,
nevertheless we found that they are Noether symmetries
in generalized harmonic gauges
according to the generalized definition \ref{def.nsg}.
Although only the Noether symmetry property is needed
for the applicability of Noether's theorem (including its generalized version),
we mention that we found that special conformal transformations
do not preserve the generalized harmonic gauges,
whereas Poincar\'e transformations and dilatations do.
We also verified that Poincar\'e transformations and dilatations
are symmetries of the linearized Einstein field equations,
whereas special conformal transformations are not.

Regarding the primary aim of the paper,
i.e.\ finding a favourable angular momentum tensor,
dilatation current and special conformal tensor
for the linearized gravitational field,
we extended $T_{\mathrm{lg}}^{cd}$ (given by (\ref{eq.Tlg}))
into a current\footnote{More precisely, into a current valued linear function.}
\begin{equation}
J_{\mathrm{lg}}^c =
T_{\mathrm{lg}}^{cd}e_d + M_{\mathrm{lg}}^{cde}\omega_{de}
+ D_{\mathrm{lg}}^c \beta + C_{\mathrm{lg}}^{cd}c_d\, ,
\end{equation}
where $e^a$, $\omega^{ab}$, $\beta$, $c^a$ are the parameters
of the generator of the conformal isometries of the Minkowski space
(see (\ref{eq.f}), (\ref{eq.TMDC}) and (\ref{eq.dPhi_lin})), and
$M_{\mathrm{lg}}^{cde}$, $D_{\mathrm{lg}}^c$ and $C_{\mathrm{lg}}^{cd}$
are the angular momentum tensor,
dilatation current and special conformal tensor, respectively,
that we propose.
A compact expression for $J_{\mathrm{lg}}^c$
is (\ref{eq.lg.jstghg})---see also
(\ref{eq.defjlg}), (\ref{eq.lg.jst}).

We constructed $J_{\mathrm{lg}}^c$ by generalizing the method we used in \cite{TG}
to obtain $T_{\mathrm{lg}}^{cd}$.
The (generalized) method consists of three main parts:
choosing a suitable Lagrangian density,
taking the corresponding canonical current associated with conformal symmetry,
and adding a suitable correction term that is associated with gauge symmetry
(see Sections \ref{sec.relcan}, \ref{sec.defjlg}, \ref{sec.lgghg}).
It can be regarded as an adaptation of
Bessel-Hagen's method for constructing conserved quantities
associated with conformal symmetry in electrodynamics \cite{BesselHagen}
(see (\ref{eq.Jem}), (\ref{eq.em.jst}), (\ref{eq.em.MDCem})).
It produces Noether currents associated with conformal transformations
modified by (field-dependent) gauge transformations.

An adaptation of Bessel-Hagen's method was also employed
recently in linearized gravity
by Hobson, Barker and Lasenby in \cite{HoBaLa}.
There the focus was on Poincar\'e symmetry,
and the Lagrangian density chosen at the beginning
was the Fierz--Pauli Lagrangian density,
rather than $L_1$ (see (\ref{eq.lg.l1}), (\ref{eq.lg.x7})),
which we chose in this paper.
Furthermore, the strategy of \cite{HoBaLa} was to identify a
gauge-invariant equivalence class of currents,
rather than a single favourable (though gauge-dependent) local current.
Accordingly, the results of \cite{HoBaLa} differ significantly
from ours.

We obtained $J_{\mathrm{lg}}^c$ in two steps;
first we extended $T_{\mathrm{lg}}^{cd}$ only
to Poincar\'e and dilatation symmetry (see Section \ref{sec.gentlg}),
and subsequently to full conformal symmetry (see Section \ref{sec.lgghg}).
The reason for these two steps was
that dealing with the special conformal transformations
required extra efforts.

$J_{\mathrm{lg}}^c$ can also be written in the general form (see (\ref{eq.j2}))
\begin{equation}
J_{\mathrm{lg}}^c = T_{\mathrm{lg}}^{cd}f_d
- S_{\mathrm{lg}}^{cde}\partial_{[d} f_{e]}
+ \frac{1}{4}\mbb{T}_{\mathrm{lg}}^c \partial_a f^a
- \frac{1}{4}\mbb{C}_{\mathrm{lg}}^{cd}\Box f_d\, ,
\end{equation}
where $S_{\mathrm{lg}}^{cde}$ is the spin tensor,
$\mbb{T}_{\mathrm{lg}}^c$ is the virial current and
$\mbb{C}_{\mathrm{lg}}^{cd}$ is the `fully intrinsic' special conformal tensor
encoded in $J_{\mathrm{lg}}^c$.
The relations between
$T_{\mathrm{lg}}^{cd}$,
$M_{\mathrm{lg}}^{cde}$, $D_{\mathrm{lg}}^c$, $C_{\mathrm{lg}}^{cd}$
and
$S_{\mathrm{lg}}^{cde}$, $\mbb{T}_{\mathrm{lg}}^c$, $\mbb{C}_{\mathrm{lg}}^{cd}$
are given by the general formulae (\ref{eq.bbT1})-(\ref{eq.bbT2-b}).
$T_{\mathrm{lg}}^{cd}$,
$S_{\mathrm{lg}}^{cde}$, $\mbb{T}_{\mathrm{lg}}^c$ and $\mbb{C}_{\mathrm{lg}}^{cd}$
can be regarded as the basic components of
$M_{\mathrm{lg}}^{cde}$, $D_{\mathrm{lg}}^c$ and $C_{\mathrm{lg}}^{cd}$.
$\mbb{T}_{\mathrm{lg}}^c = 0$
and $S_{\mathrm{lg}}^{cde}$ and $\mbb{C}_{\mathrm{lg}}^{cd}$ are
(see Sections \ref{sec.gentlg}, \ref{sec.lgghg})
\begin{eqnarray}
S_{\mathrm{lg}}^{cde}\hspace{-6mm} && \, = \,
\frac{1}{2}(F^{cae}h^d{}_a - F^{cad}h^e{}_a), \\[2mm]
\mbb{C}_{\mathrm{lg}}^{cd}\hspace{-6mm} && \, = \,
\frac{\chi-1}{2}(-4hh^{cd} + 2\chi h^2 \eta^{cd}).
\end{eqnarray}
The vanishing of $\mbb{T}_{\mathrm{lg}}^c$
is consistent with the tracelessness of $T_{\mathrm{lg}}^{cd}$.
$S_{\mathrm{lg}}^{cde}$ could not be zero,
because $T_{\mathrm{lg}}^{cd}$ is generally not symmetric.
Interestingly, it is not zero even in TT gauge,
in spite of the property of $T_{\mathrm{lg}}^{cd}$
that it is symmetric in TT gauge.
Although $S_{\mathrm{lg}}^{cde}$ is not zero,
it has special properties:
\begin{itemize}
\setlength\itemsep{0em}
\item[(a)] $\partial_ch_{ab}$ appears only through $F_{abc}$ in it,
\item[(b)] it is conserved in the $F_a=0$ gauge and
by the generalized Noether's theorem (i.e., Theorem \ref{thrm.ng})
it is associated
with internal Lorentz transformations,
\item[(c)]
it is also traceless if $F_a=0$,
\item[(d)]
in traceless harmonic gauge it is equal, up to a factor $1/2$,
to the spin tensor proposed in \cite{BHLang},
therefore it also possesses certain
physically desirable properties pointed out in \cite{BHLang}.
\end{itemize}
$S_{\mathrm{lg}}^{cde}$ has the above properties in TT gauge as well,
as the TT gauge is contained by the $F_a=0$ gauge and by the traceless harmonic gauge.

$\mbb{C}_{\mathrm{lg}}^{cd}$, and hence also $C_{\mathrm{lg}}^{cd}$,
depends explicitly on the parameter $\chi$,
which is related to the generalized harmonic gauge condition (\ref{eq.ghg}).
Generally $C_{\mathrm{lg}}^{cd}$ is conserved only
in the generalized harmonic gauge
with the matching value of $\chi$.
$\mbb{C}_{\mathrm{lg}}^{cd} = 0$ if $\chi = 1$,
which is the value of $\chi$ corresponding to the $F_a=0$ gauge.
$\mbb{C}_{\mathrm{lg}}^{cd}$ is also zero in TT gauge
due to the vanishing of $h$ in TT gauge,
hence $C_{\mathrm{lg}}^{cd}$ does not depend on $\chi$ in TT gauge.

Without $C_{\mathrm{lg}}^{cd}$, $J_{\mathrm{lg}}^c$ is a Noether current
associated with
modified Poincaré transformations and dilatations
(see (\ref{eq.dhst}), (\ref{eq.lg.dTh}), (\ref{eq.dMlg}), (\ref{eq.dDlg})
and Section \ref{sec.genjnoether}).
In any generalized harmonic gauge with some value of $\chi$,
$C_{\mathrm{lg}}^{cd}$ with the same value of $\chi$
is associated
by the generalized Noether's theorem \ref{thrm.ng}
with modified special conformal transformations
(see (\ref{eq.dClg})).

$J_{\mathrm{lg}}^c$ can be regarded as an analogue of the standard current
$J_{\mathrm{em}}^a = T_{\mathrm{em}}^{ab}f_b$ (see Section \ref{sec.el.gentem})
associated with conformal symmetry in electrodynamics,
although its properties are less perfect than those of $J_{\mathrm{em}}^a$.
While $J_{\mathrm{em}}^a$ is gauge-invariant,
gauge fixing is important for several properties of $J_{\mathrm{lg}}^c$;
nevertheless the TT gauge, which is commonly used in linearized gravity,
is well suited to $J_{\mathrm{lg}}^c$ in this respect.
Although the intrinsic quantities encoded in $J_{\mathrm{lg}}^c$
are not completely zero,
in contrast with those encoded in $J_{\mathrm{em}}^c$,
they are not far from being zero.

$J_{\mathrm{lg}}^c$ appears,
in view of its properties (including those of $T_{\mathrm{lg}}^{cd}$),
to be a favourable choice for a current associated with
conformal symmetry in linearized gravity,
and we consider its construction to be one of the main results of the paper.
Nevertheless, there are other possibilities worth examining.
We constructed two other special currents,
$J_{\tau}^c$ (see (\ref{eq.jbhl}), (\ref{eq.cbhl}))
and $J_{\mathrm{LL}}^c$ (see (\ref{eq.jllp}), (\ref{eq.cll})).
$J_{\tau}^c$ is an extension of
the energy-momentum tensor $\tau^{cd}$ (proposed in \cite{BHLemt,BHLthe}),
whereas $J_{\mathrm{LL}}^c$ is an extension of
the linearized Landau--Lifshitz pseudotensor $T_{\mathrm{LL}}^{cd}$.
We found that in TT gauge a simple relation exists
between $J_{\mathrm{lg}}^c$, $J_{\tau}^c$ and $J_{\mathrm{LL}}^c$
(see (\ref{eq.ttjrel}), (\ref{eq.ttcrel})).

$J_{\tau}^c$ is constructed in such a way that the
spin tensor encoded in it
agrees with the spin tensor given in \cite{BHLang,BHLthe}.
The virial current and the `fully intrinsic' special conformal tensor
encoded in $J_{\tau}^c$
(see (\ref{eq.tbhl}) and (\ref{eq.cbhl})) do not vanish even in TT gauge.

Since $T_{\mathrm{LL}}^{cd}$ is symmetric,
it is an interesting question whether the spin tensor accompanying it
can be taken to be zero.
$J_{\mathrm{LL}}^c$ demonstrates that this is possible,
as the spin tensor encoded in $J_{\mathrm{LL}}^c$ is zero.
The virial current and the `fully intrinsic' special conformal tensor
encoded in $J_{\mathrm{LL}}^c$
(see (\ref{eq.tll}) and (\ref{eq.cll})) are not zero in TT gauge
(in agreement with the relation between
$J_{\mathrm{lg}}^c$, $J_{\tau}^c$ and $J_{\mathrm{LL}}^c$
mentioned above).

Concerning the vanishing of the spin tensor encoded in $J_{\mathrm{LL}}^c$,
it should be noted that
the spin part of the variation with which $J_{\mathrm{LL}}^c$
is associated as a Noether current is also zero.
Moreover, $J_{\mathrm{lg}}^c$ has a similar property;
both the virial current encoded in $J_{\mathrm{lg}}^c$
and the virial part of the variation (\ref{eq.dhst})
with which $J_{\mathrm{lg}}^c$ is associated are zero.
In electrodynamics, all intrinsic parts of $J_{\mathrm{em}}^a$
and of the variation (\ref{eq.em.dAst}) with which $J_{\mathrm{em}}^a$
is associated are zero.

A secondary aim of the paper was to explore the currents
associated with conformal symmetry more generally,
not demanding exceptionally good properties.
Regarding this direction,
we obtained a current $J^c$ (see (\ref{eq.lg.j})) associated with
the Poincar\'e and dilatation symmetries,
containing $16$ free parameters (which are distinct from
the parameters of the conformal isometry generator (\ref{eq.f})).
$J^c$ is allowed to contain terms that involve $\epsilon^{abcd}$,
it does not depend on higher than first derivatives of $h_{ab}$,
and it is quadratic in $h_{ab}$ and $\partial_c h_{ab}$.
The variation with which $J^c$ is associated
as a Noether current is given by (\ref{eq.dlh2});
it corresponds to Poincar\'e transformations and dilatations
accompanied (with the exception of special cases) by gauge transformations.
It contains $2$ of the $16$ free parameters of $J^c$.
The spin tensor (\ref{eq.lg.sgen}) encoded in $J^c$ contains all of the $16$ parameters,
whereas the energy-momentum tensor (\ref{eq.lg.emt}) contains only $5$ parameters,
and the virial current (\ref{eq.lg.intdcgen}) contains $4$ parameters.
The spin tensor thus completely determines the energy-momentum tensor
and the virial current.
In Section \ref{sec.gharm} we extended $J^c$ to the complete conformal algebra;
this extended current depends on the parameter of the generalized harmonic gauges.
We found that the `fully intrinsic' special conformal tensor encoded in $J^c$
(see (\ref{eq.cghg})) contains $6$ of the $16$ free parameters,
and it does not contain $\epsilon^{abcd}$.
In particular gauges
the number of free parameters of $J^c$ can get smaller than $16$;
this happens, for instance, in TT gauge, and more generally in those gauges where $h=0$.
$J_{\mathrm{lg}}^c$, $J_{\tau}^c$ and $J_{\mathrm{LL}}^c$
are special cases of $J^c$.
In the construction of $J_{\tau}^c$ and $J_{\mathrm{LL}}^c$, $J^c$ was utilized.

The construction of $J_{\mathrm{lg}}^c$, $J^c$ and $J_{\tau}^c$
started with the construction of canonical currents
associated with Poincar\'e-dilatation symmetry (see (\ref{eq.lg.jc})-(\ref{eq.Jc3}))
and full conformal symmetry (see (\ref{eq.lg.jcghg})-(\ref{eq.Jc3a-conf})).
These canonical currents are generalizations of
the canonical energy-momentum tensors (see (\ref{eq.lg.emtc}))
contained by the 5-parameter family of energy-momentum tensors considered in \cite{TG}
(see (\ref{eq.lg.emt})) and have $2$ free parameters,
corresponding to the parameters of the Lagrangian density (\ref{eq.lg.l}).
They are special cases of $J^c$.

The terms multiplied by the $16$ parameters in $J^c$ are
trivially conserved currents with superpotentials that are
quadratic polynomials in $h_{ab}$ and $\partial_ch_{ab}$,
therefore the total charges determined by $J^c$
do not depend on the parameters
if $h_{ab}$ and $\partial_ch_{ab}$ fall off
sufficiently rapidly at spatial infinity.

Without $C^{cd}$, $J^c$ changes by a trivially conserved current
under gauge transformations (see Section \ref{sec.ginfofJ}).
This property implies that the total charges determined by $J^c$ (without $C^{cd}$)
are invariant under gauge transformations whose parameter
satisfies suitable fall-off conditions.

If the terms containing $\epsilon^{abcd}$ are excluded, then
$J^c$ and the spin tensor encoded in it have only $8$ free parameters,
the energy-momentum tensor and
the virial current have the same $3$ parameters,
the `fully intrinsic' special conformal tensor has $6$ free parameters,
and the canonical currents form a $1$-parameter family.
In TT gauge, the spatial spin density $S^{0ij}$ is unique
up to a numerical factor,
which depends on only $1$ parameter (see (\ref{eq.S0ij})).

Using the general identity (\ref{eq.nthr2})
for the divergence of Noether currents,
we also determined the local balance equations in the presence of matter for $J^c$,
and for the energy-momentum tensors, angular momentum tensors,
dilatation currents and special conformal tensors contained by $J^c$
(see Sections \ref{sec.divjmat} and \ref{sec.divcmat}).
We compared the results with \cite{BHLemt,BHLang},
where the balance equations
for the energy-momentum tensor and the angular momentum tensor
had a fundamental role.
In the present paper and in \cite{TG}
these balance equations are allowed to have
more general forms than in \cite{BHLemt,BHLang}.

In \cite{BHLthe} it was shown that the energy-momentum tensor
and the spin tensor found and discussed in \cite{BHLemt,BHLang,BHLthe}
are a source of gravity
in a second-order perturbative expansion  of the Einstein--Cartan field equations;
it would be interesting to see if similar results can be obtained
for $T_{\mathrm{lg}}^{cd}$ and $S_{\mathrm{lg}}^{cde}$ as well.
The Belinfante--Rosenfeld tensors and
the Callan--Coleman--Jackiw
improved energy-momentum tensors \cite{CCJ}
that can be constructed in linearized gravity
could also be examined,
although these tensors can be expected to depend on
higher than first derivatives of $h_{ab}$.
Extending the investigations carried out in this paper
to theories other than
linearized Einstein gravity could be considered as well.
Concerning the question of the existence of
a gauge-invariant energy-momentum tensor,
the results of \cite{HoBaLa2,HoBaLa3} indicate
that the absence of a strictly gauge-invariant
local energy-momentum tensor is characteristic of
metric spin-2 theories more generally.
In \cite{HoBaLa4} a complementary analysis of what changes
if translations are gauged internally is provided.

\section*{Acknowledgments}

I would like to thank Michael Hobson, Will Barker
and Anthony Lasenby for their helpful explanations
concerning the results of \cite{HoBaLa3,HoBaLa4}.

\appendix

\renewcommand{\theequation}{\Alph{section}.\arabic{equation}} 
\setcounter{equation}{0}

\section*{Appendix}

\section{Conformal symmetry and the associated conserved currents in electrodynamics}
\label{sec.em}

In this appendix a brief account of the conformal symmetry of electrodynamics
and the associated conserved currents is given.
The focus is mostly on the extension of the standard energy-momentum tensor
\begin{equation}
\label{eq.Tem}
T_{\mathrm{em}}^{ab} = F^{ac}F_c{}^b + \frac{1}{4}F_{cd}F^{cd}\eta^{ab}
\end{equation}
of the electromagnetic field
into a current associated with conformal symmetry,
for comparison with $J_{\mathrm{lg}}^c$.

$A_a$ will denote the vector potential and $F_{ab}$ the electromagnetic field strength tensor, as usual.
The equations of motion for $A_a$ are the Maxwell equations
$\partial_a F^{ab}=\mc{J}^b$,
where $F_{ab} = 2\partial_{[a} A_{b]}$
and $\mc{J}^a$ denotes the electric current.
$\mc{J}^a$ is assumed to be conserved, i.e.\ $\partial_a\mc{J}^a=0$.
Until Section \ref{sec.divJem}, it is assumed that $\mc{J}^a=0$.

The standard Lagrangian density function for the electromagnetic field is
\begin{equation}
\label{eq.eml2}
L=-\frac{1}{4}F_{ab}F^{ab}.
\end{equation}
The Euler--Lagrange equations following from $L$ are the Maxwell equations
without electric current:
\begin{equation}
\label{eq.emmax}
\frac{\delta L}{\delta A_b} = -\partial_a \frac{\partial L}{\partial (\partial_a A_b)} = \partial_a F^{ab} = 0.
\end{equation}

\subsection{Conformal symmetry of the Lagrangian of electrodynamics}
\label{sec.el.conf}

In the case of electrodynamics, the variation of $A_a$ under conformal transformations
(i.e., the `simple' variation; see Section \ref{sec.pdsym})
is given by
the usual transformation rule of a covector field, i.e.\ it is the Lie derivative of
$A_a$ with respect to the generator vector field $f^a$ (which is given by (\ref{eq.f})):
\begin{equation}
\label{eq.da}
\delta A_a = f^b \partial_b A_a + A_b \partial_a f^b.
\end{equation}
It is not difficult to verify that the corresponding variation of $L$ can be written as
\begin{equation}
\label{eq.el.dl}
\delta L = \partial_a (f^a L).
\end{equation}
(\ref{eq.el.dl}) implies that conformal transformations,
acting on $A_a$ as specified by (\ref{eq.da}),
are Noether symmetries of $L$,
and shows that for $K^a$ one can make the standard choice
\begin{equation}
\label{eq.kceldin}
K^a = f^a L.
\end{equation}

The Maxwell equations also have conformal symmetry,
i.e.\ if $A_a$ satisfies the Maxwell equations (without electric current),
then so does $\delta A_a$, as can be verified by direct calculation.

(\ref{eq.da}) can be written in the general form (\ref{eq.dPhi_orbint}):
\begin{equation}
\delta A_a  =  \delta_TA_a{}^b\, f_b - \check{\delta}_SA_a{}^{bc}\, \partial_bf_c
+ \frac{1}{4}\check{\delta}_{\mathbb{T}}A_a\,\partial_bf^b
-\frac{1}{4}\check{\delta}_{\mathbb{C}}A_a{}^b\, \Box f_b\, ,
\end{equation}
with
\begin{equation}
\label{eq.C7}
\delta_TA_a{}^b  =  \partial^b A_a\, ,\qquad
\check{\delta}_SA_a{}^{bc}  = A^{[b}\eta_a{}^{c]}\, ,\qquad
\check{\delta}_{\mathbb{T}}A_a  =  A_a\, ,\qquad
\check{\delta}_{\mathbb{C}} A_a{}^b  =  0.
\end{equation}

According to (\ref{eq.rel-dMdD}) and (\ref{eq.rel-dC1}),
the variations of $A_a$ under Lorentz transformations,
dilatations and special conformal transformations are given
in terms of $\delta_TA_a{}^b$,
$\check{\delta}_SA_a{}^{bc}$,
$\check{\delta}_{\mathbb{T}}A_a$
and $\check{\delta}_{\mathbb{C}} A_a{}^b$
by the equations
\begin{eqnarray}
\label{eq.C8}
\delta_M A_a{}^{bc} \hspace{-5mm} && =\ \makebox[34mm][l]{$\displaystyle x^{[c}\delta_T A_a{}^{b]}
+ \check{\delta}_SA_a{}^{bc}$}
=\ x^{[c}\partial^{b]}A_a + A^{[b}\eta_a{}^{c]} \\[1mm]
\label{eq.C9}
\delta_D A_a \hspace{-5mm} && =\ \makebox[34mm][l]{$x_b\delta_T A_a{}^b
+ \check{\delta}_{\mathbb{T}} A_a$}
=\ x_b\partial^bA_a + A_a \\[1mm]
\label{eq.C10}
\delta_C A_a{}^b \hspace{-5mm} && =\ 2x^bx_c\delta_TA_a{}^c - x_cx^c\delta_TA_a{}^b
+2x^b\check{\delta}_{\mathbb{T}}A_a - 4x_c\check{\delta}_SA_a{}^{bc}
+ \check{\delta}_{\mathbb{C}}A_a{}^b \\
&& =\ 2x^bx_c\partial^cA_a - x_cx^c\partial^bA_a + 2x^bA_a - 2x_aA^b + 2x_cA^c\eta_a{}^b\, .
\label{eq.C11}
\end{eqnarray}
The scaling weight of $A_a$ is $1$.

\subsection{Canonical currents associated with conformal symmetry}
\label{sec.el.cc}

Since $L$ has conformal symmetry with $K^a = f^a L$, there exists an associated canonical Noether current:
\begin{equation}
\label{eq.emc}
\Jc^a =
\frac{\partial L}{\partial (\partial_a A_b)}\delta A^b - f^a L
= -F^{ab} \delta A_b +\frac{1}{4}f^a F_{bc}F^{bc}.
\end{equation}
The energy-momentum tensor encoded in $\Jc^a$ is the canonical energy-momentum tensor
\begin{equation}
\Tc^{ab} = 
-F^{ac}\partial^b A_c + \frac{1}{4}F_{cd}F^{cd}\eta^{ab}.
\label{eq.emg6}
\end{equation}
The spin tensor, virial current and $\mbb{C}^{ab}$ tensor
encoded in $\Jc^a$ are
\begin{eqnarray}
\Sc^{abc} & \! = \! & \mc{A}_{bc} \bigl[
-F^{ac}A^b \bigr]
\label{eq.em.sint1} \\
\bbTc^a & \! = \! & - F^{ab}A_b  \\
\bbCc^{ab}  & \! = \! &  0.
\end{eqnarray}
The vanishing of $\bbCc^{ab}$ is not surprising in view of
$\check{\delta}_{\mathbb{C}} A_a{}^b = 0$
and $\check{K}_{\mathbb{C}}^{ab} = 0$.

$\Mc^{abc}$, $\Dc^a$ and $\Cc^{ab}$ are given in terms of $\Tc^{ab}$, $\Sc^{abc}$ and $\bbTc^a$
by the general formulae (\ref{eq.bbT1}) and (\ref{eq.bbT2}):
\begin{equation}
\Mc^{abc} = \Tc^{a[b}x^{c]}
+ \Sc^{abc}\, ,\qquad
\Dc^a = \Tc^{ab}x_b + \bbTc^a,
\end{equation}
\begin{equation}
\Cc^{ab} = 2\Tc^{ac}x_c x^b - \Tc^{ab}x_c x^c
+ 2\bbTc^a x^b -4\Sc^{abc}x_c\, .
\end{equation}

\subsection{Correction term related to gauge symmetry}
\label{sec.el.gcorr}

The gauge symmetry of electrodynamics
allows one to construct a trivially conserved current,
which can be used as a correction term in the construction of conserved currents
associated with conformal symmetry.

First, we recall---see \cite{TG}, for instance, for more details---that
\begin{equation}
\label{eq.emg2}
J_{\mathrm{g}}^a = -F^{ab}\partial_b\phi
= -\partial_b [F^{ab}\phi ]+\partial_b F^{ab}\phi
\end{equation}
is a Noether current associated with the variation $\delta A_a = \partial_a\phi$,
which corresponds to the gauge transformation
$A_a \to A_a + \epsilon\partial_a\phi$,
with $K^a=0$.
$J_{\mathrm{g}}^a$ is trivially conserved, as $\partial_b [F^{ab}\phi ]$
is strongly conserved and
$\partial_b F^{ab}\phi$ is weakly zero.

The correction current can be obtained from $F^{ab}\partial_b\phi$
by the replacement $\phi=A_c f^c$, i.e.\ it is
\begin{equation}
J_{\mathrm{g}'}^a = F^{ab}\partial_b (A_c f^c).
\end{equation}
This is a Noether current associated with $\delta A_a = -\partial_a(A_cf^c)$, with $K^a=0$.

The $T^{ab}$, $S^{abc}$, $\mathbb{T}^a$ and $\mathbb{C}^{ab}$ tensors encoded in $F^{ab}\partial_b (A_c f^c)$
are
\begin{equation}
\label{eq.c18}
T_{\mathrm{g}'}^{ab} = F^{ac}\partial_cA^b\, ,\qquad
S_{\mathrm{g}'}^{abc} = \mc{A}_{bc}\bigl[F^{ac}A^b\bigr]\, ,\qquad
\mathbb{T}_{\mathrm{g}'}^a = F^{ab}A_b\, ,\qquad \mathbb{C}_{\mathrm{g}'}^{ab} = 0\, .
\end{equation}
It is worth noting that $S_{\mathrm{g}'}^{abc} = - \Sc^{abc}$,
$\mathbb{T}_{\mathrm{g}'}^a = - \bbTc^a$, and
$\mathbb{C}_{\mathrm{g}'}^{ab} = \bbCc^{ab} = 0$.

\subsection{Extension of $T_{\mathrm{em}}^{ab}$ to conformal symmetry}
\label{sec.el.gentem}

The standard energy-momentum tensor
\begin{equation}
\label{eq.tem}
T_{\mathrm{em}}^{ab} = F^{ac}F_c{}^b + \frac{1}{4}F_{cd}F^{cd}\eta^{ab}
\end{equation}
of the electromagnetic field can be obtained by
adding the correction term $F^{ac}\partial_c A^b$,
which is a trivially conserved tensor,
to the canonical energy-momentum tensor
$\Tc^{ab}$ that follows from the
usual Lagrangian density (\ref{eq.eml2}) of Maxwell's equations
(see (\ref{eq.emg6}) for $\Tc^{ab}$).
Hence $T_{\mathrm{em}}^{ab}$ can be generalized
to a current associated with conformal symmetry by generalizing
$\Tc^{ab}$ and $F^{ac}\partial_c A^b$.

A straightforward generalization of $\Tc^{ab}$ to conformal symmetry
is $\Jc^a$ (given by (\ref{eq.emc})).
Regarding the generalization of $F^{ac}\partial_c A^b$,
a straightforward generalization of this tensor
is the trivially conserved current $F^{ab}\partial_b (A_c f^c)$
found in Section \ref{sec.el.gcorr},
as it becomes $F^{ab}\partial_b A_c e^c$ in the case of spacetime translations
(i.e., when $f^c=e^c$).
The generalization of $T_{\mathrm{em}}^{ab}$ is then
\begin{equation}
\label{eq.Jem}
J_{\mathrm{em}}^a = \Jc^a + F^{ab}\partial_b (A_c f^c).
\end{equation}
This formula is the counterpart of (\ref{eq.defjlg}).

One finds that $J_{\mathrm{em}}^a$ can be written in the simple form
\begin{equation}
\label{eq.em.jst}
J_{\mathrm{em}}^a = T_{\mathrm{em}}^{ab}f_b\, ,
\end{equation}
i.e.\ it does not have an intrinsic part (see (\ref{eq.orbint}), (\ref{eq.orb}), (\ref{eq.int}),
and the remark below (\ref{eq.c18})).
This means that the angular momentum tensor,
dilatation current and special conformal tensor
encoded in $J_{\mathrm{em}}^a$ (we denote them by $M_{\mathrm{em}}^{abc}$, $D_{\mathrm{em}}^a$ and $C_{\mathrm{em}}^{ab}$)
are just $M_{\mathrm{orb}}^{abc}$, $D_{\mathrm{orb}}^a$, $C_{\mathrm{orb}}^{ab}$,
i.e.
\begin{equation}
M_{\mathrm{em}}^{abc} = T_{\mathrm{em}}^{a[b}x^{c]} ,\quad
D_{\mathrm{em}}^a = T_{\mathrm{em}}^{ab}x_b\, ,\quad
C_{\mathrm{em}}^{ab} = 2T_{\mathrm{em}}^{ac}x_cx^b - T_{\mathrm{em}}^{ab}x_cx^c.
\label{eq.em.MDCem}
\end{equation}
These quantities agree
with those reported in Bessel-Hagen's classic article \cite{BesselHagen}.

$J_{\mathrm{em}}^a$ is a sum of Noether currents, therefore it is a Noether current itself.
It is associated with the variation
\begin{equation}
\label{eq.em.dAst}
\delta_{\mathrm{em}} A_a = F_{ba}f^b,
\end{equation}
which is the sum of the variations with which the two terms
on the right hand side of (\ref{eq.Jem}) are associated,
i.e., it is the sum of
$f^b\partial_bA_a + A_b\partial_af^b$ and $-\partial_a(A_bf^b)$.
$J_{\mathrm{em}}^a$ is thus a Noether current
associated with modified conformal transformations.
The $K^a$ quantity for $J_{\mathrm{em}}^a$ is also
the sum of the $K^a$ quantities
for the two terms on the right hand side of (\ref{eq.Jem}),
thus it is $f^aL$.
It is worth noting that in addition to the intrinsic part of $J_{\mathrm{em}}^a$,
the intrinsic parts of $\delta_{\mathrm{em}} A_a$
and $f^aL$ are also zero.
An outstanding property of $J_{\mathrm{em}}^a$, $\delta A_a$ and $f^aL$
is that they are gauge-invariant.

Since the intrinsic parts of $\delta_{\mathrm{em}} A_a$ are zero,
the variations of $A_a$ corresponding to translations,
Lorentz transformations, dilatations and special conformal transformations are just
\begin{eqnarray}
\label{eq.em.dTA}
\delta_{T_{\mathrm{em}}} A_a{}^b \hspace{-5mm} && =\ -F_a{}^b \\[1mm]
\label{eq.em.dMA}
\delta_{M_{\mathrm{em}}} A_a{}^{bc} \hspace{-5mm} && =\ \makebox[20mm][l]{$\displaystyle \delta_{T_{\mathrm{em}}} A_a{}^{[b} x^{c]}$}
=\ x^{[b}F_a{}^{c]}  \\[1mm]
\label{eq.em.dDA}
\delta_{D_{\mathrm{em}}} A_a \hspace{-5mm} && =\ \makebox[20mm][l]{$\delta_{T_{\mathrm{em}}} A_a{}^b\, x_b$}
=\ x_bF^b{}_a \\[1mm]
\label{eq.em.dCA}
\delta_{C_{\mathrm{em}}} A_a{}^b \hspace{-5mm} && =\
2x^bx_c\delta_{T_{\mathrm{em}}}A_a{}^c - x_cx^c\delta_{T_{\mathrm{em}}}A_a{}^b\
=\ 2x^bx_cF^c{}_a - x_cx^cF^b{}_a\, .
\end{eqnarray}

The vanishing of the spin tensor encoded in $J_{\mathrm{em}}^a$ appears to be unsatisfactory
if one wishes, led by physical intuition, to have a spin tensor that can be interpreted as a tensor
that quantifies the internal spinning motion of the electromagnetic field.
(In \cite{BHLang,BHLthe}, for instance, the authors argue, in the context of linearized gravity,
that the spin tensor should admit such an interpretation.)
On the other hand, for dimensional reasons a nonzero spin tensor
would be an expression of the type $A\partial A$
(see, for example, (\ref{eq.em.sint1})), thus it would not be gauge-invariant.
Furthermore, the vanishing of the spin tensor is not unnatural
if one considers that the intrinsic part of the variation
(\ref{eq.em.dAst}) is also zero.

\subsubsection{Divergence of $J_{\mathrm{em}}^a$ and $\Jc^a$ in the presence of electric currents}
\label{sec.divJem}

In the presence of electric currents, $\partial_a J_{\mathrm{em}}^a$
can be obtained using the general identity (\ref{eq.nthr2}):
\begin{equation}
\label{eq.em.dj}
\partial_a J_{\mathrm{em}}^a \ = \ -\mc{J}^a \delta_{\mathrm{em}} A_a \ = \ -\mc{J}^a F_{ba}f^b.
\end{equation}
The divergence of $T_{\mathrm{em}}^{ab}$, $M_{\mathrm{em}}^{abc}$, $D_{\mathrm{em}}^a$ and $C_{\mathrm{em}}^{ab}$
can be extracted from (\ref{eq.em.dj}); taking into account (\ref{eq.em.dTA})-(\ref{eq.em.dCA}),
the following formulae can be obtained (in agreement with \cite{BesselHagen}):
\begin{eqnarray}
\label{eq.em.div1}
\partial_a T_{\mathrm{em}}^{ab} \hspace{-5mm} && =\ \makebox[25mm][l]{$-\mc{J}^a \delta_{T_{\mathrm{em}}} A_a{}^b$}  = \ - \mc{J}_a F^{ba} \\[1mm]
\label{eq.em.div2}
\partial_a M_{\mathrm{em}}^{abc} \hspace{-5mm} && =\ \makebox[25mm][l]{$-\mc{J}^a \delta_{M_{\mathrm{em}}} A_a{}^{bc}$}  = \ \mc{J}_a x^{[b} F^{c]a} \\[1mm]
\label{eq.em.div3}
\partial_a D_{\mathrm{em}}^a \hspace{-5mm} && =\ \makebox[25mm][l]{$-\mc{J}^a \delta_{D_{\mathrm{em}}} A_a$} = \ -\mc{J}_a x_b F^{ba} \\[1mm]
\label{eq.em.div4}
\partial_a C_{\mathrm{em}}^{ab} \hspace{-5mm} && =\ \makebox[25mm][l]{$-\mc{J}^a \delta_{C_\mathrm{em}} A_a{}^b$} = \ \mc{J}_a (-2 x_c x^b F^{ca} + x_c x^c F^{ba}).
\end{eqnarray}

For the canonical current $\Jc^a$, one obtains from (\ref{eq.nthr2}) and (\ref{eq.da}) that
\begin{equation}
\label{eq.em.djc}
\partial_a \Jc^a \ = \ -\mc{J}^a \delta A_a \ = \ -\mc{J}^a (f^b \partial_b A_a + A_b \partial_a f^b).
\end{equation}
Again,
$\partial_a \Tc^{ab}$, $\partial_a \Mc^{abc}$, $\partial_a \Dc^a$ and $\partial_a \Cc^{ab}$
can be extracted from (\ref{eq.em.djc}); taking into account (\ref{eq.C7})-(\ref{eq.C11}),
one obtains the formulae
\begin{eqnarray}
\label{eq.em.div1c}
\partial_a \Tc^{ab} \hspace{-5mm} && =\ \makebox[22mm][l]{$-\mc{J}^a \delta_T A_a{}^b$} = \ -\mc{J}_a \partial^b A^a \\[1mm]
\label{eq.em.div2c}
\partial_a \Mc^{abc} \hspace{-5mm} && =\ \makebox[22mm][l]{$-\mc{J}^a \delta_M A_a{}^{bc}$} = \ \mc{J}_a x^{[b}\partial^{c]} A^a
+ \frac{1}{2} \mc{J}^{[b} A^{c]} \\[1mm]
\label{eq.em.div3c}
\partial_a \Dc^a \hspace{-5mm} && =\ \makebox[22mm][l]{$-\mc{J}^a \delta_D A_a$} = \ -\mc{J}_a(x^b\partial_bA^a + A^a) \\[1mm]
\partial_a \Cc^{ab} \hspace{-5mm} && =\ \makebox[22mm][l]{$-\mc{J}^a \delta_C A_a{}^b$} = \ \mc{J}^a(-2x^bx^c\partial _cA_a + x_cx^c\partial^bA_a) \nonumber \\
\label{eq.div4c}
&&\hspace{35mm} -\,2\mc{J}^bx^aA_a + 2\mc{J}^a x_a A^b - 2\mc{J}^a x^b A_a\, .
\end{eqnarray}

\end{document}